\documentclass[12pt,draftclsnofoot,onecolumn]{IEEEtran}
\usepackage{graphicx}
\usepackage{cite}
\usepackage{amsmath,amssymb,placeins}
\usepackage{algorithmic}
\usepackage{graphicx}
\usepackage{textcomp}
\usepackage[mathscr]{euscript}
\usepackage{lipsum}
\usepackage{cuted}
\usepackage{bm}
\usepackage{multicol}
\usepackage{array}
\usepackage{makecell}
\usepackage{diagbox}
\usepackage[colorlinks=true,linkcolor=gray,citecolor=black]{hyperref}
\usepackage[symbol]{footmisc}
\usepackage[utf8]{inputenc}
\usepackage{caption}
\usepackage{tabularx}

\usepackage[mathscr]{euscript}
\usepackage[usenames]{color}
\usepackage[table]{xcolor}
\usepackage{amsfonts}
\usepackage{latexsym}
\usepackage{subfigure}
\usepackage{tikz}
\usetikzlibrary{arrows.meta}
\usetikzlibrary{shapes.geometric, arrows}
\usepackage{caption}
\usepackage{floatrow}
\usepackage{multirow}
\usepackage{epstopdf}
\usepackage[english]{babel}

\newcommand{\ccb}{\color{blue}}
\newcommand{\ccr}{\color{red}}
\newcommand{\ccn}{\color{black}}
\newcommand{\ggr}{\color{gray}}

\newcommand{\mbb}{\mathbb}
\newcommand{\mbf}{\mathbf}
\newcommand{\om}{\zeta_q}
\newcommand{\mal}{\mathcal}
\newcommand{\tta}{\Theta}
\newcolumntype{\arow}{\rightarrow}

\newcommand{\dis}{\displaystyle}
\newcommand{\bx}{\mathbf{x}}
\newcommand{\be}{\mathbf{e}}
\newcommand{\bZ}{\mathbf{Z}}
\newcommand{\bbZ}{\mathbb{Z}}
\newcommand{\calV}{\mathcal{V}}
\DeclareMathOperator{\wt}{wt}

\usepackage{amsthm}
\newtheorem{theorem}{Theorem}

\newtheorem{corollary}{Corollary}

\newtheorem{remark}{Remark}

\newtheorem{example}{Example}

\newtheorem{case}{\textit{Case}}
\newtheorem{scase}{{{\textit{Sub-Case}}}}
\def\BibTeX{{\rm B\kern-.05em{\sc i\kern-.025em b}\kern-.08em
		T\kern-.1667em\lower.7ex\hbox{E}\kern-.125emX}}

\def\@fnsymbol#1{\ensuremath{\ifcase#1\or *\or \dagger\or \ddagger\or
		\mathsection\or \mathparagraph\or \|\or **\or \dagger\dagger
		\or \ddagger\ddagger \else\@ctrerr\fi}}

{
\begin{document}
\title{Construction of Complete Complementary Codes over Small Alphabet} 
\author{Palash Sarkar,~Chunlei Li,~Sudhan Majhi,~and ~Zilong Liu}
\maketitle

\begin{abstract}
Complete complementary codes (CCCs) play a vital role not only in wireless communication, particularly in multicarrier systems where achieving an interference-free environment is of paramount importance, but also in the construction of other codes that necessitate appropriate functions to meet the diverse demands within today's landscape of wireless communication evaluation.
This research is focused on the area of constructing $q$-ary functions for both of {traditional and spectrally null constraint (SNC)
CCCs}\footnote{When no codes in CCCs having zero components, we call it as traditonal CCCs, else, we call it as SNC-CCCs in this pape.}  of flexible length, set size and alphabet. We construct traditional CCCs with lengths, defined as $L = \prod_{i=1}^k p_i^{m_i}$, set sizes, defined as $K = \prod_{i=1}^k p_i^{n_i+1}$, and an alphabet size of $q=\prod_{i=1}^k p_i$, such that $p_1<p_2<\cdots<p_k $.
 The parameters $m_1, m_2, \ldots, m_k$ (each greater than or equal to $2$) are positive integers, while $n_1, n_2, \ldots, n_k$ are non-negative integers satisfying $n_i \leq m_i-1$, and the variable $k$ represents a positive integer. To achieve these specific parameters, we define $q$-ary functions over a domain $\mbf{Z}_{p_1}^{m_1}\times \cdots \times \mbf{Z}_{p_k}^{m_k}$ that is considered a proper subset 
  of $\mathbb{Z}_{q}^m$ and encompasses $\prod_{i=1}^k p_i^{m_i}$ vectors, where $\mbf{Z}_{p_i}^{m_i}=\{0,1,\hdots,p_i-1\}^{m_i}$, and the value of $m$ is derived from the sum of $m_1, m_2, \ldots, m_k$. This organization of the domain allows us to encompass all conceivable integer-valued length sequences over the alphabet $\mathbb{Z}_q$. 
It has been demonstrated that by constraining a $q$-ary function that generates traditional CCCs, we can derive SNC-CCCs with identical length and alphabet, yet a smaller or equal set size compared to the traditional CCCs.
Our achieved alphabet and set size are minimized when $k\geq 2$, and $m_i$ is co-prime to $m_j$ for some $1\leq i\neq j\leq k$, in contrast to existing CCCs with the same length $L$ but possessing a set size greater than or equal to $K$.


\end{abstract}
\begin{IEEEkeywords}
	Aperiodic correlation, function, multicarrier code-division multiple-access (MC-CDMA), complete complementary code set (CCCS).
\end{IEEEkeywords}
\section{Introduction}
\label{sec:intro}
\subsection{Background of Traditional CCCs}
The study of complementary sequences has been attracting increasing research attention in recent years due to their interesting
correlation properties as well as their wide applications in engineering.  Formally, a collection of sequences with identical
length and zero aperiodic auto-correlation sum (AACS) for all non-zero time shifts are called a complementary set (CS)
and each constituent sequence is called a complementary sequence \cite{chinchong,pater2000,liug}. In particular, a CS of size two
is called a Golay complementary pair (GCP), a widely known concept proposed by Marcel J. E Golay {in the late 1940s} \cite{Thesis_1949golay}.
{In this case,} the two constituent sequences are called Golay sequences.
Furthermore, by rearranging every CS {to be an ordered set}, a collection of CSs with zero aperiodic cross-correlation sum (ACCS)
for all time shifts is called mutually orthogonal Golay complementary sets (MOGCSs) \cite{rati}. It is noted that the set
size (denoted by $K$) of MOGCSs is upper bounded by the number of constituent sequences (denoted by $M$), i.e., $K\leq M$.
When $K=M$, the resultant MOGCSs are called a set of complete complementary codes (CCCs) \cite{hator}.

So far, complementary sequences and CCCs have been widely applied in numerous applications in coding, signal processing,
and wireless communications. For example, they {have been employed} for peak-to-average power ratio (PAPR)
reduction in code-keying orthogonal frequency division multiplexing (OFDM) \cite{Davis1999},
signature sequences for interference-free multicarrier code-division multiple-access (MC-CDMA) over asynchronous wireless channels \cite{chen2007next,liumc}, training/sensing sequences in multiple-input
multiple-output (MIMO) communication/radar systems \cite{Wang2007,Pezeshiki2008,Tang2014},
information hiding in image/audio/video signals \cite{Thesis_2014Kojima},
kernel/seed sequences to construct Z-complementary code set (ZCCS) \cite{psktcom,pa_pbf} and quasi-complementary sequence set \cite{avikr_qccs, zhu_qccs,palash_qcss_tit_24} for MC-CDMA ,
and zero-correlation zone (ZCZ) sequences \cite{dfan,ltsu,appus,Tang2010,Liu_ITW2014} in quasi-synchronous direct-sequence CDMA communications.

{There have been significant research attempts} on efficient constructions of  complementary
sequences (including GCPs, MOGCSs and CCCs) with flexible lengths. With interleaving and concatenation,
Golay first constructed binary GCPs of lengths $2MN$ from two shorter GCPs with lengths $M$ and $N$,
respectively  \cite{golay1961}. Budi\v{s}in developed recursive algorithms for more polyphase and multi-level
complementary sequences through a series of sequence operations (e.g., phase rotations, shifting) associated
with permutation vectors \cite{Budisin90a,Budisin90b}. In 1999, Davis and Jedwab proposed a systematic
construction of $2^h$-ary ($h\in\mathbb{Z}^+$) GCPs with power-of-two lengths using second-order generalized
Boolean functions (GBFs) \cite{Davis1999}. Such Golay sequences are called Golay-Davis-Jedwab (GDJ) {complementary} sequences in this paper.
Paterson extended the idea of $2^h$-ary GCPs to $q$-ary (even $q$) CSs with power-of-two constituent sequences {\color{red}and} power-of-two
lengths \cite{pater2000}. His work was further generalized in 2007 by Schmidt by moving to the higher-order GBFs  \cite{Schmid2007}.
Recently, Sarkar \emph{et~al.} proposed a construction in \cite{palcs} using higher-order GBFs for complementary sequences with
lower PMEPR upper bound. For complementary sequences with non-power-of-two lengths, Chen contributed with the aid of certain truncated
Reed-Muller codes \cite{chentit}.

The constructions of MOGCSs especially CCCs are more stringent owing to the additional requirement of zero ACCS  \cite{rati,liumc,chencommlett}. In 2020, Wu \emph{et~al.} introduced a construction of MOGCSs with non-power-of-two lengths based on GBFs in \cite{swuc} where the set size is exactly half of the number of constituent sequences in a CS, i.e., $K=M/2$. Later several attemt have been made to construct non-power-of-two lengths MOGCSs with parametric restriction $K=M/2$, such as in \cite{nwmocs1,xiao2023new}. However, in a recent work \cite{pku1}, the authors successfully constructed binary CCCs of lengths in the form of $5\cdot 2^m$ and $13\cdot 2^m$, and the set size in the form of power-of-two. Still to the date, there is a large gap in the function/polynomial based construction of CCCs having flexible parameters, specially lengths and alphabet size. Another powerful tool is paraunitary (PU) generator with which a set of CCCs can be efficiently obtained by sending a pipeline of shifted impulses to multistage filterbanks whose coefficients are extracted from a group of PU matrices \cite{Budisin_QAM,Wang_SETA2016,sdas,Sdas_lett,shibu2}. Very recently, the inherent connections between the GBF generator and the PU generator in CCC construction have been uncovered by Wang \emph{et~al.} \cite{wang2020new}.
\subsection{Background of SNC-CCCs}
The conventional approach to sequence design assumes the presence of continuous spectral bands, allowing the sequences energy to be distributed across all carriers within that band. However, given the limited and valuable radio spectrum available for wireless communication and radar sensing, researchers are actively seeking more efficient utilization methods \cite{zhao2007survey}. This emerging concept is known as cognitive radio (CR) \cite{haykin2005cognitive} or cognitive radar \cite{haykin2006cognitive}. It involves a set of intelligent radio devices with the capability to sense and analyze the radio frequency (RF) environment. They identify unused spectral bands and temporarily utilize them for data transmission.
In particular, an overlay cognitive radio network continuously scans for unused frequency bands, often referred to as ``white space," within licensed spectrum ranges. This is done to accommodate new users, referred to as secondary users, while minimizing any potential impact on the primary licensed users.

Recent literature reveals, SNC complementary sequences find a lot of research attention due to it's null constraint properties and applications in wireless communications. 
More recently, the demand for OFDM or multi-carrier CDMA sequences with SNCs  (also referred to as non-contiguous) is driven by their possible use cases, such as in the Third-Generation Partnership Project (3GPP),
Long-Term Evolution (LTE) enhanced licensed-assisted access
(eLAA), and New Radio in Unlicensed (NR-U) have adopted
SNC sequences 
 \cite{Shen2022construction,aplphan_snc_aapl}. This demand is primarily motivated by the potential application scenarios of such sequences within the domain of CR communications. In the context of OFDM-based CR transmissions, secondary users are restricted to transmitting only on sub-carriers that are unoccupied by the primary user. This necessitates the presence of spectral nulls at particular positions within the code word transmitted by secondary users, aligning with the sub-carriers that the primary user utilizes. In \cite{aplphan_snc_aapl,csahin2021generic}, authors generalized traditional GCP to proposed SNC-GCP. In \cite{zhou2020new}, Zhou \emph{et al.} provide an iterative construction of SNC-MOGCS based on traditional CCCs. In \cite{ipanov2018radar}, authors used two sequences from a GCP as the seed sequence and then inserted zeros (the number of zeros is a multiple of the length of GCP) into two sequences to get new ones with zero aperiodic correlation zone. Authors in \cite{li2022spectrally}, proposed the construction of SNC zero correlation zone (ZCZ) sequences and SNC Z-periodic complementary sets (ZPCS). In \cite{shen2023}, Shen \emph{et al.}, proposed construction of SNC-CCCs based on extended Boolean functions (EBFs). While the construction \cite{shen2023} can produce CCCs of arbitrary length, its alphabet size may become as large as sequence lengths, especially when the sequence lengths take the form $\prod_{i=1}^k p_i^{m_i}$, where $m_i$ and $m_j$ are coprime to each other for some $1\leq i\neq j\leq k$, and $k>1$. In the literature, the above mentioned constructions are the only constructions for SNC-CS or SNC-CCC.
\subsection{Contributions and Approach}
{We observe that conventional direct constructions of sequences mostly rely on the tool of} functions where the sequence length and alphabet size mostly depend on its domain and co-domain, respectively. As example, in most of the GBFs based constructions, the sequence length is limited to the length power-of-two, unless any additional operations been performed on the power-of-two lengths sequences. However, one way to cross the barrier is modifying the domain in a form such that it's size can cover most of the positive integers. In \cite{shen2023}, the authors used EBFs with a domain of 
$\mathbb{Z}_q^m$ and co-doamin of $\mathbf{Z}_q$, where $q\geq 2$ is an integer. With this setting of domain and co-domain, and by choosing proper functions on them, they constructed CCCs of size $q^{n+1}$, length in the form $q^m$, and the alphabet size in the form of $q$, where $m\geq 2,~0\leq n\leq m-1$. However, this choice leads to a higher alphabet and set size when they require to generate CCCs of lengths in the form of $\prod_{i=1}^k p_i^{m_i}$ with some of $m_1,m_2,\hdots,m_k$ are co-prime to each other. In this case, the alphabet size becomes as large as $\prod_{i=1}^k p_i^{m_i}$ and also the set size. To achieve lower alphabet sizes, we introduced those functions where the domain is in the form $\mbf{Z}_{p_1}^{m_1}\times\mbf{Z}_{p_2}^{m_2}\times \cdots\times \mbf{Z}_{p_k}^{m_k}$ and co-domain in the form $\mathbb{Z}_{p_1p_2\hdots p_k}$, by doing so, we are capable of generating sequences of lengths in the form of $\prod_{i=1}^k p_i^{m_i}$ and alphabet size of $\prod_{i=1}^k p_i$. We can observe that this choices of domain and co-domain allow us to achieve all possible length sequences while maintaining a small alphabet size. \ccn

To generate CCCs with aforementioned parameters, our core idea is to 
construct $q$-ary functions having specific graphical properties as defined in Section \ref{contsec} of this paper, where $q=\prod_{i=1}^k p_i$. Specifically, we construct 
$q$-ary functions which results in a function of algebrauc degree $2$ after restricting some variables in the main function. The graph of the restricted function is a collection of Hamiltonian paths. It has been firstly shown that the restricted functions generates SNC-CCCs  
of lengths in the form $L=\prod_{i=1}^k p_i^{m_i}$, set size in the form $\prod_{i=1}^k p_i$. The alphabet size of the proposed SNC-CCCs is given by $1+\prod_{i=1}^k p_i$.  Later it has been shown that the main function generates 
traditional CCCs of same lengths $L$ and the set size in the form $\prod_{i=1}^kp_i^{n_i+1}$. The alphabet size for our proposed traditional CCCs is given by $\prod_{i=1}^k p_i$.  
 As every integers greater than $1$ can be factored
uniquely as the product {of powers of} prime numbers (by unique factorization theorem), it can be observed that the proposed $q$-ary functions is capable of
producing traditional CCCs of arbitrary length and set size. We also have derived a relationship between the proposed traditional CCCs and the SNC-CCCs.
	For $k=1$ and $p_1=2$, the GDJ sequences are obtained as a special case of our construction.
	For $k=1$, and $p_1=2$, our proposed construction is also able to produce CSs and CCCs with the same
parameters as that reported in the works by Paterson \cite{pater2000}, Schmidt  \cite{Schmid2007}, and Sarkar \textit{et~al.} \cite{palcs}. {As a byproduct of this research, we introduce a novel class of linear code which may be regarded as a generalization of RM code and its connection with the proposed complementary sequences.} 
\section{Preliminaries}
 In this paper, the following notations are used otherwise stated.
\begin{itemize}
	\item $\mathbb{Z}$ denotes the set of all integers.
	\item $\mathbf{Z}_n=\{0,1,\hdots,n-1\}\subset \mathbb{Z}$
	\item $\mathbf{a}\cdot \mathbf{b}=a_0b_0+a_1b_1+\hdots+a_{L-1}b_{L-1}$.
	\item For a finite set $S$, $|S|$ denotes the cardinality of $S$.
	\item For two finite sets $S$ and $S'$, $S\times S'$ denotes the cartesian product between $S$ and $S'$.
	\item $S^n=\underbrace{S\times S\times\cdots\times S}_{\text{n times }}$
	\item $q=\prod_{i=1}^k p_i$, where $p_1,p_1,\hdots,p_k$ are $k$ distinct primes such that $p_1<p_2<\cdots<p_k$.
	\item $\zeta_q=\exp(2\pi\sqrt{-1}/q)$ is a primitive $q$-th root of unity.
	\item $m_i$ ($\geq 2$) is a positive integer, where $i=1,2,\hdots,k$.
	\item $n_i~(\leq m_i-1)$ is a non-negative integer, where $i=1,2,\hdots,k$.
	\item $L_i=p_i^{m_i}$ $L=\prod_{i=1}^k L_i$ and $K=\prod_{i=1}^k p_i^{n_i+1}$.
	\item $m=\sum_{i=1}^km_i$.
	\item $\mal{V}_L=\mbf{Z}_{p_1}^{m_1}\times\mbf{Z}_{p_2}^{m_2}\times\cdots\times \mbf{Z}_{p_k}^{m_k}$.
\end{itemize}
\subsection{Aperiodic Auto- and Cross-Correlation}
For any two complex-valued sequences $\mathbf{a}=(a_0,a_1,\hdots,a_{L-1})$ and $\mathbf{b}=(b_0,b_1,\hdots,b_{L-1})$ of length $L$, we define the aperiodic cross-correlation function (ACCF) at a shift $\tau$, with  $0\leq |\tau|<L$ as
\begin{equation}\label{accf}
\Theta (\mathbf{a},\mathbf{b})(\tau)=
\begin{cases}
\sum_{\alpha=0}^{L-\tau-1}a_\alpha b^*_{\alpha+\tau}, & 0\leq \tau<L,\\
\sum_{\alpha=0}^{L+\tau-1}a_{\alpha-\tau} b^*_\alpha,&-L<\tau<0.
\end{cases}
\end{equation}
For $\mathbf{a}=\mathbf{b}$, the ACCF defined in (\ref{accf}) reduces to the aperiodic auto-correlation function (AACF) of 
$\mathbf{a}$, which will be denoted as $\Theta (\mathbf{a})(\tau)$ for short. The set $\{x\in\mbf{Z}_L :a_x\neq 0\}$ is called the support of $\mbf{a}$. If the set $\mathbf{Z}_L\setminus \{x\in \mbf{Z}_L :a_x\neq 0\}$ is an non-empty set, $\mbf{a}$ is called a \textit{spectrally null constrained (SNC) sequence \cite{shen2023}}; otherwise, it is called a traditional sequence. 
\ccn
Let $\mathcal{C}=\{\mathbf{C}_1,\mathbf{C}_2,\hdots,\mathbf{C}_{K-1}\}$ be a collection of $K$ codes, each containing $M$ sequences of length $L$. By arranging each code as a two-dimensional matrix, we write  $\mathbf{C}_k$ as
$$\mathbf{C}_t=\begin{bmatrix}
\mathbf{c}_t^1\\
\mathbf{c}_t^2\\
\vdots\\
\mathbf{c}_t^{M}
\end{bmatrix}_{M\times L},$$
where $t=0,1,\hdots,K-1$. The ACCF (sum) between $\mathbf{C}_{t_1}$
and $\mathbf{C}_{t_2}$ for $0\leq t_1, t_2 \leq K-1$ is defined as 
\begin{equation}\label{accfk}
\Theta (\mathbf{C}_{t_1},\mathbf{C}_{t_2})(\tau)=\sum_{j=1}^{M}\Theta   (\mathbf{c}_{t_1}^j,\mathbf{c}_{t_2}^j)(\tau),
\end{equation}
For $t_1=t_2=t$, the ACCF in (\ref{accfk}) reduces to the AACF of $\mathbf{C}_{t}$ and we denote it by $\Theta (\mathbf{C}_{t})(\tau)$. The set $\mal{C}$ is said to be a set of traditional complete complementary code (CCC) if the aperiodic correlation sum of it's codes satisfies the following properties:
\begin{equation}\label{accfc}
\begin{split}
\Theta (\mathbf{C}_{t_1},\mathbf{C}_{t_2})(\tau)=
\begin{cases}
LM,& \tau=0,~t_1=t_2,\\
0,& 0<|\tau|<L, ~t_1=t_2,\\
0,&|\tau|<L, ~t_1\neq t_2.
\end{cases} 
\end{split}
\end{equation}
In this case any of the codes in $\mal{C}$ is known as a traditional CS or complementary code (CC) \cite{sarkar2021multivariable}. If there is at least one SNC constituent sequence in any codes of $\mal{C}$ and satisfying (\ref{accfc}), the set of codes $\mal{C}$ is called SNC-CCC \cite{shen2023}. It is to be noted that both of the terms `traditional CCCs' and `CCCs' bear the same meaning in this paper\ccn.  
\subsection{Functions and Traditional Sequences}
In this section, we define necessary algebraic operations on the set $\mal{V}_L$  in order to introduce $q$-ary functions while maintaining $\mal{V}_L$ as the domain. 
We start with a given positive integer $L = L_1L_2$, define a mapping $\rho$: $\mathbf{Z}_L \rightarrow \mathbf{Z}_{L_1}\times \mathbf{Z}_{L_2}$ as 
$\rho(x):= (x_1, x_2)$ where 
\[
x_2 = x \bmod{L_2} \text{ and } x_1 = (x-x_2)/L_2.
\] It is clear that $\rho$ is bijective and its inverse $\rho^{-1}$ is given by $x = \rho^{-1}(x_1, x_2) = x_1*L_2 + x_2$.
More generally, for an integer $L=\prod_{i=1}^k L_i$,  we can define a one-to-one mapping 
$\rho$: $\mathbf{Z}_L \rightarrow \mathbf{Z}_{L_1}\times \mathbf{Z}_{L_2}\times \dots \times \mathbf{Z}_{L_k}$ as $\phi(x) = (x_1, x_2, \dots, x_k)$, where $x_i$'s is recursively derived as follows: 
\begin{equation}\label{Eq_Int2Vec}
\begin{split}
& y_k = x \\
& x_t = y_t \bmod{L_t} \text{ and } y_{t-1}= \frac{y_t-x_t}{L_t} \quad \text{ for } t =k, \dots, 2, 1. 
\end{split}
\end{equation}
From the above recursive relations, it is easy to verify that the pre-image of $(x_1,x_2,\dots, x_k)$ is given by 
\begin{equation}\label{Eq_Vec2Int}
\begin{split}
x=\rho^{-1}((x_1,x_2, \dots, x_k)) &= (y_{k-1}L_k + x_k) 
\\&= ((y_{k-2}L_{k-1} + x_{k-1})L_k + x_k) =\cdots =
\\&= ((x_1L_{2}+x_2)L_3 + x_3)L_4\cdots)L_k + x_k
\\&=x_1 L_2\cdots L_k + x_2 L_3\cdots L_k + \dots + x_{k-1}L_{k} + x_k
\\&= x_1 \Delta_1 + x_2 \Delta_2 + \dots + x_k \Delta_k
\end{split}
\end{equation} where $\Delta_i = \frac{L}{L_1\cdots L_{i}}$. Note that the above relation does not pose any restriction on the relation among $L_1, L_2,\dots, L_k$. In particular, when $L_1=L_2=\dots = L_k = p$, the image $\rho(x)=(x_1,x_2,\dots, x_k)$ corresponds to the conventional $p$-ary expansion of $x = \sum_{i=1}^{k}x_i\Delta_i$, where $\Delta_i = \frac{p^k}{\prod_{1\leq j\leq i}p} = p^{k-i}$. 
From this discussion, we can define an one-to-one mapping $\rho: \mathbf{Z}_L \rightarrow \mal{V}_L$ as 
$$\mathbf{x} := \rho(x) = (\bx_1, , \dots, \bx_k) = (x_{1,1}, \dots, x_{1,m_1}, \dots, x_{k,1}, \dots, x_{k, m_k})
$$ where $\bx_i = (x_{i,1}, \dots, x_{i, m_i}) \in \bZ_{p_i}^{m_i}$ is the $p_i$-ary expansion of the corresponding integer $x_i$ in $\bZ_{p_i^{m_i}}$, namely, $x_i = \sum\limits_{j=1}^{m_i}x_{i,j}p_i^{m_i-j}$. In the sequel we will say $\bx$ is the vectorial representation of an integer $x \in \bZ_L$ and $x$ is the integer representation of a vector $\bx \in \mal{V}_L$, which are connected via the equality $\bx = \rho(x)$.

\medskip

Now we consider functions from $\calV_L$ to $\bbZ_q$ in $m$ variables $x_{1,1}, \dots, x_{1,m_1}, \dots, x_{k,1}, \dots, x_{k, m_k}$,
where the arithmetic operations among these variables and coefficients are taken modulo $q$. 
\ccr For an element $\be = (\be_1,\dots, \be_k) \in \calV_L$, we define a monomial $\mbf{x}^\mbf{e}$ over the $m$ variables as\ccn
\[
\bx^\be = \bx_1^{\be_1} \cdots \bx_k^{\be_k} = \prod_{j=1}^{m_1} x_{1,j}^{e_{1,j}} \cdots \prod_{j=1}^{m_k} x_{k,j}^{e_{k,j}},
\] where by convention we assume $0^0=0$ and $x^0 =1$ for $x\neq 0$. The algebraic degree of $\bx^{\be}$ is given by
\[
\deg(\bx^\be) = \sum_{j=1}^{m_1} e_{1,j} + \cdots + \sum_{j=1}^{m_k} e_{k,j}=||\mbf{e}||_1,
\]
where $||\mbf{e}||_1$ denotes the $L^1$-norm of $\mbf{e}$. There are $|\calV_L|$ number of monomials and a linear combination of those monomials with $\mathbb{Z}_q$-valued coefficients represents 
a $q$-ary function $f:\, \calV_L \rightarrow \bbZ_q$ which can be expressed as
\[
f(\bx) = \sum_{\be \in \calV_L} c_{\be} \bx^\be = \sum_{\be \in \calV_L} c_{\be} \bx_1^{\be_1} \cdots \bx_k^{\be_k}.
\] 
For $k=1$ and $p_1=2$, the function $f$ reduces to a Boolean function \cite{Davis1999}, and for $p_1>2$, it reduces to a extended Boolean function \cite{shen2023}. \ccn
The algebraic degree of $f$ is defined by
$
\deg(f) = \max\{||\be||_1\,:\, c_{\be} \neq 0 \}.
$ Alternatively, we may define another notion of degree, \textit{Hamming degree}, of the above function $f$ as 
\[
\deg_H(f) = \max\{\wt(\be)\,:\, c_{\be} \neq 0 \},
\] where $\wt(\be)$ is the Hamming weight of $\be = (e_{1,1}, \dots, e_{1,m_1}, \dots, e_{k,1}, \dots, e_{k, m_k})$. 
From the definition, it is clear that any $q$-ary function $f$ has $\deg_H(f) \leq \deg(f)$.
A $q$-ary functions from $\calV_L$ to $\bbZ_q$ of Hamming degree at most $r$ can be uniquely expressed as
\[
f(\bx) = \sum_{\be \in \calV_L, \wt(\be)\leq r} c_{\be} \bx^\be.
\] 
Denote by $\mho_{L,r}$ the set of all the above $q$-ary functions with Hamming degree at most $r$. Then $\mho_{L,r}$ can be expressed as
\begin{equation}
	\mho_{L,r}=\left\{\sum_{\be \in \calV_L, \wt(\be)\leq r} c_{\be} \bx^\be: c_\be\in \mathbb{Z}_q \right\}.
\end{equation} 
Let us also denote the set of all $q$-ary functions of algebraic-degree at most $r$ by $\Omega_{L,r}$ given by 
\begin{equation}
	\Omega_{L,r}=\left\{\sum_{\be \in \calV_L, ||\be||_1\leq r} c_{\be} \bx^\be:c_\be\in \mathbb{Z}_q \right\}.
\end{equation}
which says $\Omega_{L,r}\subset \mho_{L,r}$.
\begin{example}\label{ankita008}
	Let us assume $k=2$, $p_1=2$, $p_2=3$, $m_1=3$, $m_2=2$, $q=p_1 p_2=6$, and so $L=p_1^{m_1}p_2^{m_2}=72$.   
	Table \ref{dtable} containing the $q$-ary vector representation $\mbf{x}=(x_{1,1},x_{1,2},x_{1,3},x_{2,1},x_{2,2})\in \calV_{72}$ of $x$, where $x\in \mbf{Z}_{72}$, and Table \ref{ankita700} contains all the monomials of Hamming weight at most $2$ over the $5$ variables $x_{1,1},x_{1,2},x_{1,3},x_{2,1},x_{2,2}$. There are $27$ monomials and therefore $\mho_{72,2}$ contains $6^{27}$ functions which can be obtained by taking linear combination of the monomials in Table\ref{ankita700} with coefficients from $\mathbb{Z}_6$.     
\begin{table}[H]
	\centering
	\caption{\ccr The Set $\calV_{72}$ for $p_1=2$, $p_2=3$, $m_1=3$, and $m_2=2$.}\label{dtable}
	\begin{tabular}{|l|l|l|l|l|l|l|l|l|l|l|l|}
		\hline
		$x$&$\mbf{x}$&$x$&$\mbf{x}$&$x$&$\mbf{x}$&$x$&$\mbf{x}$&$x$&$\mbf{x}$&$x$&$\mbf{x}$\\ \hline                                       
		\begin{tabular}[c]{@{}l@{}}\ggr $0$\\ \ggr $1$\\ \ggr $2$\\ \ggr $3$\\\ggr $4$\\ \ggr \ggr $5$\\\ggr $6$\\\ggr $7$\\\ggr $8$\\\ggr $9$\\\ggr $10$\\\ggr $11$\ccn \end{tabular} & \begin{tabular}[c]{@{}l@{}}$0     0     0     0     0$ \\  $0     0     0     0     1$\\  $0     0     0     0     2$\\  $0     0     0     1     0$\\  $0     0     0     1     1$\\  $ 0     0     0     1     2$\\  $0     0     0     2     0$\\  $0     0     0     2     1$\\  $0     0     0     2     2$\\  $0     0     1     0     0$\\  $0     0     1     0     1$\\  $0     0     1     0     2$\end{tabular} & \begin{tabular}[c]{@{}l@{}}\ggr $12$\\\ggr $13$\\\ggr $14$\\\ggr $15$\\\ggr $16$\\\ggr $17$\\ $18$\\ $19$\\ $20$\\ $21$\\ $22$\\  $23$\end{tabular} & \begin{tabular}[c]{@{}l@{}}$   0     0     1     1     0$\\     $ 0     0     1     1     1$\\     $ 0     0     1     1     2$\\      $0     0     1     2     0$\\      $0     0     1     2     1$\\      $0     0     1     2     2$\\      $0     1     0     0     0$\\      $0     1     0     0     1$\\      $0     1     0     0     2$\\      $0     1     0     1     0$\\      $0     1     0     1     1$\\      $0     1     0     1     2$\end{tabular} & \begin{tabular}[c]{@{}l@{}} $24$\\ $25$\\ $26$\\ $27$\\ $28$\\  $29$\\ $30$\\ $31$\\ $32$\\ $33$\\ $34$\\ $35$\end{tabular} & \begin{tabular}[c]{@{}l@{}}$ 0     1     0     2     0$\\     $ 0     1     0     2     1$\\     $ 0     1     0     2     2$\\     $ 0     1     1     0     0$\\     $ 0     1     1     0     1$\\     $ 0     1     1     0     2$\\     $ 0     1     1     1     0$\\     $ 0     1     1     1     1$\\     $ 0     1     1     1     2$\\     $ 0     1     1     2     0$\\     $ 0     1     1     2     1$\\     $ 0     1     1     2     2$\end{tabular} & \begin{tabular}[c]{@{}l@{}}\ggr $36$\\ \ggr $37$\\\ggr $38$\\\ggr $39$\\\ggr $40$\\\ggr $41$\\\ggr $42$\\\ggr $43$\\\ggr $44$\\ \ggr $45$\\\ggr $46$\\\ggr $47$\end{tabular} & \begin{tabular}[c]{@{}l@{}}$ 1     0     0     0     0$\\     $1     0     0     0     1$\\     $ 1     0     0     0     2$\\     $ 1     0     0     1     0$\\     $ 1     0     0     1     1$\\     $ 1     0     0     1     2$\\     $ 1     0     0     2     0$\\     $ 1     0     0     2     1$\\     $ 1     0     0     2     2$\\     $ 1     0     1     0     0$\\     $ 1     0     1     0     1$\\     $ 1     0     1     0     2$\end{tabular} & \begin{tabular}[c]{@{}l@{}}\ggr $48$\\\ggr $49$\\\ggr $50$\\\ggr $51$\\\ggr $52$\\ \ggr $53$\\ $54$\\ $55$\\ $56$\\ $57$\\ $58$\\ $59$\end{tabular} & \begin{tabular}[c]{@{}l@{}}
			$1     0     1     1     0$\\     $ 1     0     1     1     1$\\      $1     0     1     1     2$\\      $1     0     1     2     0$\\      $1     0     1     2     1$\\      $1     0     1     2     2$\\      $1     1     0     0     0$\\      $1     1     0     0     1$\\      $1     1     0     0     2$\\      $1     1     0     1     0$\\      $1     1     0     1     1$\\      $1     1     0     1     2$\end{tabular} & \begin{tabular}[c]{@{}l@{}}$60$\\ $61$\\ $62$\\ $63$\\ $64$\\ $65$\\ $66$\\ $67$\\ $68$\\ $69$\\ $70$\\ $71$\end{tabular} & \begin{tabular}[c]{@{}l@{}}$1     1     0     2     0$\\      $1     1     0     2     1$\\      $1     1     0     2     2$\\      $1     1     1     0     0$\\      $1     1     1     0     1$\\      $1     1     1     0     2$\\      $1     1     1     1     0$\\      $1     1     1     1     1$\\      $1     1     1     1     2$\\      $1     1     1     2     0$\\      $1     1     1     2     1$\\      $1     1     1     2     2$\end{tabular} \\ \hline
	\end{tabular}
\end{table}
	\begin{table}[H]
		\centering
		\caption{Set of Monomials of Hamming weight at most $2$.}\label{ankita700}
		\begin{tabular}{|l|l|l|l|l|l|l|l|l|l|l|l|l|}
			\hline
			$\mbf{e}\in \calV_{72}$&$\mbf{x}^\mbf{e}$&   $\mbf{e}\in \calV_{72}$  &   $\mbf{x}^\mbf{e}$  &  $\mbf{e}\in \calV_{72}$  & $\mbf{x}^\mbf{e}$   \\ \hline                                       
			\begin{tabular}[c]{@{}l@{}}$0     0     0     0     0$ \\  $0     0     0     0     1$\\  $0     0     0     0     2$\\  $0     0     0     1     0$\\  $0     0     0     1     1$\\  $ 0     0     0     1     2$\\  $0     0     0     2     0$\\  $0     0     0     2     1$\\  $0     0     0     2     2$\end{tabular} & 
			\begin{tabular}[c]{@{}l@{}} $1$\\ $x_{2,2}$\\ $x_{2,2}^2$\\ $x_{2,1}$\\ $x_{2,1}x_{2,2}$\\ $x_{2,1}x_{2,2}^2$\\ $x_{2,1}^2$\\ $x_{2,1}^2x_{2,2}$\\ $x_{2,1}^2 x_{2,2}^2$\end{tabular} 
			
			& 
			
			\begin{tabular}[c]{@{}l@{}} $0     0     1     0     0$\\  $0     0     1     0     1$\\  $0     0     1     0     2$\\$   0     0     1     1     0$\\  $0     0     1     2     0$\\           $0     1     0     0     0$\\      $0     1     0     0     1$\\      $0     1     0     0     2$\\      $0     1     0     1     0$ \end{tabular} & \begin{tabular}[c]{@{}l@{}} $x_{1,3}$\\ $x_{1,3}x_{2,2}$\\  $x_{1,3}x_{2,2}^2$\\ $x_{1,3}x_{2,1}$\\ $x_{1,3}x_{2,1}^2$\\ $x_{1,2}$\\ $x_{1,2}x_{2,2}$\\ $x_{1,2}x_{2,2}^2$\\ $x_{1,2}x_{2,1}$\end{tabular} &
			\begin{tabular}[c]{@{}l@{}}$ 0     1     0     2     0$\\ $ 0     1     1     0     0$\\$ 1     0     0     0     0$\\     $1     0     0     0     1$\\     $ 1     0     0     0     2$\\     $ 1     0     0     1     0$\\ $ 1     0     0     2     0$\\ $ 1     0     1     0     0$\\ $1     1     0     0     0$ \end{tabular} 
			&
			\begin{tabular}[c]{@{}l@{}}
				$x_{1,2}x_{2,1}^2$\\ $x_{1,2}x_{1,3}$ \\ $x_{1,1}$\\ $x_{1,1} x_{2,2}$\\ $x_{1,1}x_{2,2}^2$\\ $x_{1,1}x_{2,1}$\\ $x_{1,1}x_{2,1}^2$\\ $x_{1,1}x_{1,3}$\\$x_{1,1}x_{1,2}$
			\end{tabular} \\
			\hline
		\end{tabular}
	\end{table}
\end{example}
For any function $f\in \mho_{L,2}$, we can 
represent it by an undirected graph $G(f)$ as follows: $G(f)$ contains $m$ vertices representing the $m$ variable $x_{i, j}$;
if $f$ contains a nonzero term $cx_{i,j}^e$ with $e\geq 1$, then there are $e-1$ edges from vertex $x_{i,j}$ to itself with a common label $c$ as shown in Figure \ref{fig_somvf121};
if $f$ contains a nonzero term in two variables, say $cx_{i_1,j_1}^{e_1}x_{i_2,j_2}^{e_2}$, then there is an edge between the two variables labeled with weight $c$, 
and there are corresponding number of edges from each vertex to itself as presented in Figure \ref{fig_somvf12}.
\ccn

\begin{figure}[H]
	\centering
	\begin{tikzpicture}[line width=.6pt]
	\filldraw 
	(0, 0) circle [radius=5pt];
	\draw(0, .5) node {$x_{i,j}$}; 	
	\draw[densely dotted] (0, -0.3) ellipse [x radius=5 pt, y radius=10pt];
	\draw[densely dotted] (0, -0.4) ellipse [x radius=6 pt, y radius=12pt];
	\draw[densely dotted] (0, -0.5) ellipse [x radius=7 pt, y radius=14pt];
	\draw(0, -1.5) node {$c$};	
	\end{tikzpicture}	
	\caption{Graph of $cx_{i,j}^{e}$}
	\label{fig_somvf121}
\end{figure}

\begin{figure}[H]
	\centering
	\begin{tikzpicture}[line width=.6pt]
	\filldraw 
	(0, 0) circle [radius=5pt]
	(3, 0) circle [radius=5pt];
	\draw(0, .5) node {$x_{i_1,j_1}$}; 
	\draw(1.5, 0.3) node {$c $};
	\draw(3, 0.5) node {$x_{i_2,j_2}$};
	\draw(0,0) -- (3,0); 
	
	\draw[densely dotted] (0, -0.3) ellipse [x radius=5 pt, y radius=10pt];
	\draw[densely dotted] (0, -0.4) ellipse [x radius=6 pt, y radius=12pt];
	\draw[densely dotted] (0, -0.5) ellipse [x radius=7 pt, y radius=14pt];
	
	\draw[densely dotted] (3, -0.3) ellipse [x radius=5 pt, y radius=10pt];
	\draw[densely dotted] (3, -0.4) ellipse [x radius=6 pt, y radius=12pt];
	\draw[densely dotted] (3, -0.5) ellipse [x radius=7 pt, y radius=14pt];
	\end{tikzpicture}
	\caption{Graph of $c x_{i_1,j_1}^{e_1} x_{i_2,j_2}^{e_2}$}
	\label{fig_somvf12}
\end{figure}
%
%
%
%
%
%
%
%
%
%
%
%
\ccn
\ccn
For a $q$-ary function $f:\calV_L\rightarrow \mathbb{Z}_q$, we define the $\mbb{Z}_q$-valued sequence of length $L$ as follows:
\begin{equation}\label{ajkit1}
	\begin{split}
	\eta(f)=(f_0,f_1,\hdots,f_{L-1}),
	\end{split}
\end{equation}
where $f_x=f(\mbf{x})$ and $x\in\mbf{Z}_L$.
We define the complex-valued sequence over the alphabet $\mbb{Z}_q$ as 
\begin{equation}\label{seq007}
	\psi(f)=(\om^{f_0},\om^{f_1},\hdots,\om^{f_{L-1}}).
\end{equation}
As all the components of $\psi(f)$ is non-zero, $\psi(f)$ is a tradition sequence over the alphabet $\mathbb{Z}_q$, and (\ref{seq007}) establish a relationship between $f$ with a length-$L$ traditonal complex-valued sequence $\psi(f)$. 
\begin{example}\label{exmo1}
In Example \ref{ankita008}, let us consider the following $6$-ary function $f\in \mho_{72,2}$: 
\begin{equation}\label{ankby1}
f(x_{1,1},x_{1,2},x_{1,3},x_{2,1},x_{2,2})=
2x_{1,1}x_{1,2}+4x_{1,2}x_{1,3}+x_{1,2}x_{2,1}+x_{1,2}x_{2,2}+3x_{1,1}x_{1,3}+2x_{2,1}x_{2,2}
+x_{1,2}+2.
\end{equation}
In Figure (\ref{fig_somvf11}), we present the graph corresponding to $f$.
From (\ref{ajkit1}), the $\mathbb{Z}_6$-valued sequence corresponding to $f$ is of length, $L=2^3 3^2=72$, and appears as 
$$\eta(f)=(2     2     2     2     3     4     2     4     0     2     2     2     2     3     4     2     4     0     2     2     2     2     3     4     2     4 0     3     3     3     3     4     5     3     5     1     3     4     5     4     0     2     5     2     5     4     5     0     5     1     3     0  3     0     4     5     0     5     1     3     0     3     0     0     1     2     1     3     5     2     5     2),$$
and the complex-valued sequence corresponding to $f$ is 
\begin{equation}\nonumber
\begin{split}
\psi(f)=(      \omega_6^2     \omega_6^2     \omega_6^2     \omega_6^2     \omega_6^3     \omega_6^4     \omega_6^2     \omega_6^4     \omega_6^0     \omega_6^2     \omega_6^2     \omega_6^2     \omega_6^2     \omega_6^3     \omega_6^4     \omega_6^2     \omega_6^4     \omega_6^0     \omega_6^2     \omega_6^2     \omega_6^2     \omega_6^2     \omega_6^3     \omega_6^4     \omega_6^2     \omega_6^4     \omega_6^0     \omega_6^3     \omega_6^3     \omega_6^3     \omega_6^3     \omega_6^4     \omega_6^5     \omega_6^3     \omega_6^5 \\    \omega_6^1     \omega_6^3     \omega_6^4     \omega_6^5     \omega_6^4     \omega_6^0     \omega_6^2     \omega_6^5     \omega_6^2     \omega_6^5     \omega_6^4     \omega_6^5     \omega_6^0     \omega_6^5     \omega_6^1     \omega_6^3     \omega_6^0     \omega_6^3     \omega_6^0     \omega_6^4     \omega_6^5     \omega_6^0     \omega_6^5     \omega_6^1     \omega_6^3     \omega_6^0     \omega_6^3     \omega_6^0     \omega_6^0     \omega_6^1     \omega_6^2     \omega_6^1     \omega_6^3     \omega_6^5     \omega_6^2     \omega_6^5     
\omega_6^2).
\end{split}
\end{equation}
%
\begin{figure}[H]
	\centering
	\begin{tikzpicture}[line width=.6pt]
	\filldraw 
	(0, 0) circle [radius=5pt]
	(2, 0) circle [radius=5pt]
	(1, -1.5) circle [radius=5pt]
	(0, -3) circle [radius=5pt]
	(2, -3) circle [radius=5pt]
	;
	
	
	\draw(0, .5) node {$x_{1,1}$}; 
	\draw(1, 0.3) node {$3$};
	\draw(2, 0.5) node {$x_{1,3}$};
	\draw(-0.2,-1.5) node {$x_{1,2}$};
	\draw(0,-3.5) node {$x_{2,1}$};
	\draw(2,-3.5) node {$x_{2,2}$};
	\draw(1,-3.3) node {$2$};
	
	\draw(0,0) -- (2,0); 
	\draw(1,-1.5) -- (0,-3);
	\draw(1,-1.5) -- (2,-3);
	\draw(1,-1.5) -- (0,0);
	\draw(1,-1.5) -- (2,0);
	\draw(0,-3) -- (2,-3);
	
	\end{tikzpicture}
	\caption{Graph corresponding to $f$ in (\ref{ankby1})}
	\label{fig_somvf11}
\end{figure}

\end{example}
In the below section, we shall also draw a connection between $q$-ary functions $f$ and SNC sequences along with SNC position sets.
\subsection{Functions and SNC Sequences}
Similarly as $\calV_L$, we define another set $\calV_{L'}$ containing the vector representations of $0,1,\hdots,\prod_{i=1}^kp_i^{n_i}-1$ of length $L'=\prod_{i=1}^kp_i^{n_i}$. Then $\calV_{L'}$
can be expressed as
$$\calV_{L'}=\left\{\mathbf{c}=(\mathbf{c}_1,\mathbf{c}_2,\hdots,\mathbf{c}_k):  \mathbf{c}_i\in\mathbf{Z}_{p_i}^{n_i},~i=1,2,\hdots,k\right\}=\mathbf{Z}_{p_1}^{n_1}\times \mathbf{Z}_{p_2}^{n_2}\times \cdots\times\mathbf{Z}_{p_k}^{n_k}$$
Let $J=\{J_1,J_2,\hdots,J_k\}$, where $J_i=\{j_{i,1},j_{i,2},\hdots,j_{i,n_i}\}\subset \{1,2,\hdots,m_i\}$, and $\mathbf{x}_{J}=({\mathbf{x}_1}_{J_1},{\mathbf{x}_2}_{J_2},\hdots,{\mathbf{x}_k}_{J_k})$, where ${\mathbf{x}_i}_{J_i}=({x}_{i,j_{i,1}},{x}_{i,j_{i,2}},\hdots,{x}_{i,j_{i,n_i})}$, $i=1,2,\hdots,k$. 
When $\mathbf{x}_{J}=\mbf{c}$ for some $\mbf{c}\in \calV_{L'}$ in $f$, we call it a restriction of $f$ and we denote it by
$f\arrowvert_{\mbf{x}_J=\mbf{c}}$. To define sequences corresponding to a $q$-ary function having restriction on some variables, let us first define the following sets
$$\calV_{L}^\mbf{c}=\{\mbf{x}\in \calV_{L}: {\mathbf{x}_1}_{J_1}=\mbf{c}_1,{\mathbf{x}_2}_{J_2}=\mbf{c}_2,\hdots,{\mathbf{x}_k}_{J_k}=\mbf{c}_k\}\subset \calV_{L},$$ and $$\mal{N}_\mbf{c}=\{x\in \mbf{Z}_L:x=\varphi(\mbf{x}), \mbf{x}\in\calV_{L}^\mbf{c}\}\subset \mbf{Z}_L.$$
Then
\begin{equation}\label{nul1}
	\begin{split}
\calV_{L}=\displaystyle{\cup_{\mbf{c}\in \calV_{L'}}} {\calV_{L}^\mbf{c}},~
\text{and}~
\mathbf{Z}_L=\displaystyle{\cup_{\mbf{c}\in \calV_{L'}}} {\mal{N}_\mbf{c}}
	\end{split}
\end{equation}
Also for any $\mbf{c}\neq\mbf{c}'$ in $\calV_{L'}$,
\begin{equation}\label{nul2}
	\begin{split}
	\calV_{L}^\mbf{c}\cap \calV_{L}^{\mbf{c}'}=\emptyset, ~\text{and}~\mal{N}_{\mbf{c}}\cap \mal{N}_{\mbf{c}'}=\emptyset.
	\end{split}
\end{equation}
We define a length-$L$ SNC complex-valued sequence corresponding to $f\arrowvert_{\mbf{x}_J=\mbf{c}}$ as follows:
\begin{equation}\label{lbl1}
\psi(f\arrowvert_{\mbf{x}_J=\mbf{c}})=
\begin{cases}
\zeta_q^{f_x}, & x\in \mal{N}_{\mbf{c}},\\
0,& \text{otherwise},
\end{cases}
\end{equation}
where $f_x=f(\mbf{x})$, and $\mbf{x}\in \calV_{L}^\mbf{c}$. In (\ref{lbl1}), the elements in $\mal{N}_\mbf{c}$ represent the positions of non-zero componets of $\psi(f\arrowvert_{\mbf{x}_J=\mbf{c}})$. Therefore,  
$\mbf{Z}_L\setminus \mal{N}_\mbf{c}$ represent the positions of null components. As $|\mal{N}_\mbf{c}|=|\calV_{L}^\mbf{c}|=\prod_{i=1}^k p_i^{m_i-n_i}$, $$|\mbf{Z}_L\setminus \mal{N}_\mbf{c}|=L-\prod_{i=1}^k p_i^{m_i-n_i}.$$ 
Therefore, it is clear that the set $\mbf{Z}_L\setminus \mal{N}_\mbf{c}$ is non-empty only when $n_i\neq 0$ for some values of $i$, $1\leq i\leq k$, and in this case, $\psi(f\arrowvert_{\mbf{x}_J=\mbf{c}})$ forms a SNC sequence with support $\mal{N}_\mbf{c}$. It is to be noted that the alphabet size for $\psi(f\arrowvert_{\mbf{x}_J=\mbf{c}})$ is of $q+1$ as it contains some null components along with $|\mal{N}_\mbf{c}|$ non-zero components from the $L$ components of $\psi(f)$. \ccn 
From (\ref{nul1}), (\ref{nul2}), and (\ref{lbl1}), we have
\begin{equation}\label{slit1}
	\begin{split}
\psi(f)=\sum_{\mbf{c}\in\mal{D}'} \psi(f\arrowvert_{\mbf{x}_J=\mbf{c}}),
	\end{split}
\end{equation}
which says a $q$-ary function $f$ can be expressed as the sum of $\prod_{i=1}^k p_i^{n_i}$
$L$-length complex-valued SNC sequences each having $L-\prod_{i=1}^k p_i^{m_i-n_i}$ null or zero components. 
In the later presentation, whenever we use the notation $\tau$, we consider it as non-negative integer. 
\subsection{SNC Sequences and Their Correlation Distribution}
For two $\mathbb{Z}_q$-valued functions $f$ and $g$, from (\ref{lbl1}), the ACCF between $\psi(f\arrowvert_{\mbf{x}_J=\mbf{c}})$ and $\psi(g\arrowvert_{\mbf{x}_J=\mbf{c}})$ at  $\tau$ can be expressed as 
\begin{equation}\label{lbl2}
\begin{split}
	\Theta(\psi(f\arrowvert_{\mbf{x}_J=\mbf{c}}),\psi(g\arrowvert_{\mbf{x}_J=\mbf{c}}))(\tau)=\begin{cases}
\dis\sum_{x=0}^{L-1-\tau}	\zeta_q^{f_x-g_{x+\tau}}, & (x,x+\tau)\in \mal{N}_\mbf{c}\times \mal{N}_\mbf{c}, ~0\leq \tau<L,\\
	0,& \text{otherwise.}
	\end{cases}
\end{split}
\end{equation}
To simplify (\ref{lbl2}) and for later presentation, let us define the following set to determine correlation functions and their distributions for SNC sequences:
\begin{equation}\label{corr_restric_c}
\mal{A}_\tau(\mathbf{c})=\{(x,y)\in\mal{N}_\mbf{c}\times \mal{N}_\mbf{c}:0\leq x\leq L-\tau-1,y=x+\tau\}. 
\end{equation}
Let us also define the set of all $\tau$ for which $\mal{A}_\tau\neq\emptyset$ as
\begin{equation}\label{tures}
\mal{T}_\mbf{c}=\{\tau\in \mbf{Z}_L:|\mal{A}_\tau(\mbf{c})|\neq\emptyset\}.
\end{equation}
As  $|\mal{N}_\mbf{c}|=|\calV_{L}^\mbf{c}|=\prod_{i=1}^k p_i^{m_i-n_i}$, $$ |\mal{T}_\mbf{c}|\geq \prod_{i=1}^k p_i^{m_i-n_i},$$ 
which says there exist at least $\prod_{i=1}^{k}p_i^{m_i-n_i}$ number of $\tau\in \mbf{
	Z}_L$ for whcih $\mal{A}_\tau(\mbf{c})$ is non-empty, however, it doesn't gurantee the ACCF in (\ref{lbl2}) to be zero. Then $|\mbf{Z}_L \setminus \mal{T}_\mbf{c}|\leq L-\prod_{i=1}^k p_i^{m_i-n_i} $, and so there exist at most $ L-\prod_{i=1}^k p_i^{m_i-n_i} $ many $\tau$'s for which $\mal{A}_\tau (\mbf{c})=\emptyset $.  
Then (\ref{lbl2}) can be represented as 
\begin{equation}\label{lbl3}
\begin{split}
\Theta(\psi(f\arrowvert_{\mbf{x}_J=\mbf{c}}),\psi(g\arrowvert_{\mbf{x}_J=\mbf{c}}))(\tau)=\begin{cases}
\displaystyle\sum_{(x,y)\in \mal{A}_\tau(\mathbf{c})}	\zeta_q^{f_x-g_{x+\tau}}, & \tau\in \mal{T}_\mbf{c},~ \tau\geq 0,\\
0,& \text{otherwise.}
\end{cases}
\end{split}
\end{equation}
Below, we present another example to explain the above-described matters which we have mainly defined after Example \ref{exmo1}.
\ccn
\begin{example}
Let us consider the same function $f$, as defined in (\ref{ankby1}), which is presented below:
$$f(x_{1,1},x_{1,2},x_{1,3},x_{2,1},x_{2,2})=2x_{1,1}x_{1,2}+4x_{1,2}x_{1,3}+x_{1,2}x_{2,1}+x_{1,2}x_{2,2}+3x_{1,1}x_{1,3}+2x_{2,1}x_{2,2}+x_{1,2}+2.$$
Let us consider that $n_1=1$, $n_2=0$, and $J=\{J_1,J_2\}=\{J_{1,2}\}=\{2\}$, where $J_2=\emptyset$. Hence 
$\mbf{x}_J=({\mathbf{x}_1}_{J_1},{\mathbf{x}_2}_{J_2})=(x_{1,2})$. Then 
\begin{equation}
\calV_{72}^0=\left\{(x_{1,1},x_{1,2},x_{1,3},x_{2,1},x_{2,2})\in \calV_{72}:x_{1,2}=0 \right\},
\end{equation}
and 
\begin{equation}
\begin{split}
\mal{N}_0 & =\left\{x\in \mbf{Z}_{72}:x=9((x_{1,1},0,x_{1,3})\cdot (4,2,1))+(x_{2,1},x_{2,1})\cdot (3,1), x_{1,1}, x_{1,3}\in \mbf{Z}_2,~x_{2,1},x_{2,2}\in \mbf{Z}_3\right\}\\
&=[0,17] \cup [36,53]. 
\end{split}
\end{equation}
Similarly, we can obtain $\calV_{72}^1$ and $\mal{N}_1$ given below:
\begin{equation}
\calV_{72}^1=\left\{(x_{1,1},x_{1,2},x_{1,3},x_{2,1},x_{2,2})\in \calV_{72}:x_{1,2}=1 \right\},
\end{equation}
and 
\begin{equation}
\begin{split}
\mal{N}_1 & =\left\{x\in \mbf{Z}_{72}:x=9\left((x_{1,1},1,x_{1,3})\cdot (4,2,1)\right)+(x_{2,1},x_{2,1})\cdot (3,1), x_{1,1}, x_{1,3}\in \mbf{Z}_2,~x_{2,1},x_{2,2}\in \mbf{Z}_3\right\}\\
&=[18,35] \cup [54,71]. 
\end{split}
\end{equation} 
From (\ref{lbl1}), we have
$$\psi(f\arrowvert_{x_{1,2}=0})=\left(\zeta_6^2\zeta_6^2\zeta_6^2\zeta_6^2\zeta_6^4\zeta_6^0\zeta_6^2\zeta_6^0\zeta_6^4\zeta_6^2\zeta_6^2\zeta_6^2\zeta_6^2\zeta_6^4\zeta_6^0
\zeta_6^2\zeta_6^0\zeta_6^4\mbf{0}_{18}\zeta_6^2\zeta_6^2\zeta_6^2\zeta_6^2\zeta_6^4\zeta_6^0\zeta_6^2\zeta_6^0\zeta_6^4\zeta_6^5\zeta_6^5\zeta_6^5\zeta_6^5\zeta_6^1\zeta_6^3\zeta_6^5
\zeta_6^3\zeta_6^1\mbf{0}_{18}\right).$$
\begin{figure}[H]
	\includegraphics[height=4cm,width=8cm]{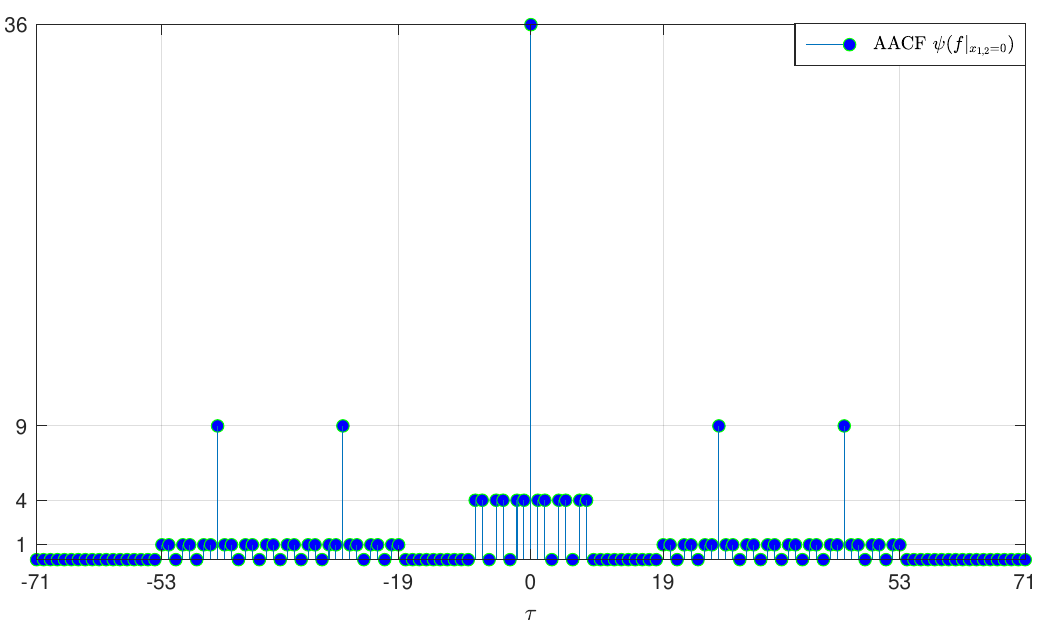}
	\caption{Correlation Plotting of $\psi(f\arrowvert_{x_{1,2}=0})$}
	\label{ccexan}
\end{figure}
Now following (\ref{tures}), let us find $\mal{T}_0$, i.e., we have to find those $\tau$ for which $\mal{A}_\tau(0)\neq \emptyset$.  From (\ref{corr_restric_c}), for some $\tau$, the set $\mal{A}_\tau(0)$ is non-empty if there exist at least one pair $(x,x+\tau)$ in $\mal{N}_0\times \mal{N}_0$ for some values of $x$ in $\mal{N}_0$, 
which says $\tau\in [0,17]\cup [19,53]$. Similarly as above, one can obtain $\psi(f\arrowvert_{x_{1,2}=1})$ and $\mal{T}_1$. 

\ccn 
\end{example}
\section{Proposed $q$-ary Functions and CCCs}\label{contsec}
\begin{figure}
	\centering
	\begin{tikzpicture}[line width=.6pt]
	\filldraw 
	(0, 0) circle [radius=5pt]
	(2, 0) circle [radius=5pt]
	(2.5,0) circle [radius=0.5pt]
	(3,0) circle [radius=0.5pt]
	(3.5,0) circle [radius=0.5pt]
	(4, 0) circle [radius=5pt]
	(6, 0) circle [radius=5pt]
	
	
	(6.5, -1.5) circle [radius=0.5pt]
	(7, -1.5) circle [radius=0.5pt]
	(7.5, -1.5) circle [radius=0.5pt]
	
	(6.5, -3) circle [radius=0.5pt]
	(7, -3) circle [radius=0.5pt]
	(7.5, -3) circle [radius=0.5pt]

	(1, -3) circle [radius=5pt]
	(3, -3) circle [radius=5pt]
	(3.5,-3) circle [radius=0.5pt]
	(4,-3) circle [radius=0.5pt]
	(4.5,-3) circle [radius=0.5pt]
	(5, -3) circle [radius=5pt];

	\draw[densely dotted] (3, -3.3) ellipse [x radius=5 pt, y radius=10pt];
	\draw[densely dotted] (3, -3.4) ellipse [x radius=6 pt, y radius=13pt];
	\draw[densely dotted] (3, -3.5) ellipse [x radius=8 pt, y radius=15pt];
	
	\draw[densely dotted] (1, -3.3) ellipse [x radius=5 pt, y radius=10pt];
	\draw[densely dotted] (1, -3.4) ellipse [x radius=6 pt, y radius=13pt];
	\draw[densely dotted] (1, -3.5) ellipse [x radius=8 pt, y radius=15pt];
	\draw[densely dotted] (5, -3.3) ellipse [x radius=5 pt, y radius=10pt];
	\draw[densely dotted] (5, -3.4) ellipse [x radius=6 pt, y radius=13pt];
	\draw[densely dotted] (5, -3.5) ellipse [x radius=8 pt, y radius=15pt];

	\draw    
	(0, 1) node {$x_{1,\pi_1(1)}$} 
	(1, .3) node {$\frac{q}{p_1}$}
	(2, 1) node {$x_{1,\pi_1(2)}$} 
	(4, 1) node {$x_{1,\pi_1(m_1-n_1-1)}$}
	(5.3, 0.3) node {$\frac{q}{p_1}$}
	(6.5, 1) node {$x_{1,\pi_1(m_1-n_1)}$}
	(1,-4.5) node {$x_{1,j_{1,1}}$}
	(3,-4.5) node {$x_{1,j_{1,2}}$}
	(5,-4.5) node {$x_{1,j_{1,n_1}}$};
	
	\draw(0,0) -- (2,0); \draw (4,0) -- (6,0);
	\draw[loosely dotted](2,0)--(4,0);
	\draw[loosely dotted](3,-3)--(5,-3);
	\draw[loosely dotted](5,-3)--(9,-3);
	\draw[loosely dotted](11,-3)--(13,-3);
	\draw[loosely dotted](10,0)--(12,0);
	\draw(1, -3) -- (0,0);
	\draw(1, -3) -- (2,0);
	\draw[loosely dotted](1, -3) -- (2.5,0);
	\draw[loosely dotted](1, -3) -- (3,0);
	\draw[loosely dotted](1, -3) -- (3.5,0);
	\draw(1, -3) -- (4,0);
	\draw(1, -3) -- (6,0);
	\draw(3,-3)--(0,0);
	\draw(3,-3)--(2,0);
	\draw[loosely dotted](3,-3)--(2.5,0);
	\draw[loosely dotted](3,-3)--(3,0);
	\draw[loosely dotted](3,-3)--(3.5,0);
	\draw(3,-3)--(4,0);
	\draw[loosely dotted](3, -3.4) -- (3,-3.5);
	\draw(1,-3)--(3,-3);
	\draw[densely dotted](1,-3) -- (1.92,-2.3);
	\draw[densely dotted](1,-3) -- (2,-2.7);
	\draw[densely dotted](1,-3) -- (1.92,-3.4);
	\draw[densely dotted](1,-3) -- (2.5,-3.2);
	\draw[densely dotted](1,-3) -- (2.5,-2.8);
	
	\draw[densely dotted](3,-3) -- (4,-2.6);
	\draw[densely dotted](3,-3) -- (3.92,-3.4);
	\draw[densely dotted](3,-3) -- (4.5,-3.2);
	\draw[densely dotted](3,-3) -- (4.5,-2.8);
	\draw(5,-3)--(6,0);
	\draw(3,-3)--(6,0);
	\draw(5,-3)--(0,0);
	\draw(5,-3)--(2,0);
	\draw(5,-3)--(4,0);
	\draw(5,-3)--(6,0);
	
	\draw[densely dotted](5,-3) -- (6,-2.7);
	\draw[densely dotted](5,-3) -- (5.92,-3.4);
	\draw[densely dotted](5,-3) -- (6.5,-3.2);
	\draw[densely dotted](5,-3) -- (6.5,-2.8);  
	
	\draw[densely dotted](9,-3) -- (8,-2.7);
	\draw[densely dotted](9,-3) -- (8.08,-3.4);
	\draw[densely dotted](9,-3) -- (7.5,-3.2);
	\draw[densely dotted](9,-3) -- (7.5,-2.8);  
	
	\draw[densely dotted](11,-3) -- (10,-2.7);
	\draw[densely dotted](11,-3) -- (10.08,-3.4);
	\draw[densely dotted](11,-3) -- (9.5,-3.2);
	\draw[densely dotted](11,-3) -- (9.5,-2.8); 

	\draw[densely dotted](13,-3) -- (12,-2.7);
    \draw[densely dotted](13,-3) -- (12.08,-3.4);
    \draw[densely dotted](13,-3) -- (11.5,-3.2);
    \draw[densely dotted](13,-3) -- (11.5,-2.8); 

	\draw(9,-3) -- (11,-3);

	\draw[loosely dotted](5,-3)--(2.5,0);
	\draw[loosely dotted](5,-3)--(3,0);
	\draw[loosely dotted](5,-3)--(3.5,0);

	\filldraw (8, 0) circle [radius=5pt]
	(10, 0) circle [radius=5pt]
	(10.5,0) circle [radius=0.5pt]
	(11,0) circle [radius=0.5pt]
	(11.5,0) circle [radius=0.5pt]
	(12, 0) circle [radius=5pt]
	(14, 0) circle [radius=5pt]
	(9, -3) circle [radius=5pt]
	(11, -3) circle [radius=5pt]
	(11.5,-3) circle [radius=0.5pt]
	(12,-3) circle [radius=0.5pt]
	(12.5,-3) circle [radius=0.5pt]
	(13, -3) circle [radius=5pt];

	\draw[densely dotted] (11, -3.3) ellipse [x radius=5 pt, y radius=10pt];
	\draw[densely dotted] (11, -3.4) ellipse [x radius=6 pt, y radius=13pt];
	\draw[densely dotted] (11, -3.5) ellipse [x radius=8 pt, y radius=15pt];
	
	\draw[densely dotted] (9, -3.3) ellipse [x radius=5 pt, y radius=10pt];
	\draw[densely dotted] (9, -3.4) ellipse [x radius=6 pt, y radius=13pt];
	\draw[densely dotted] (9, -3.5) ellipse [x radius=8 pt, y radius=15pt];
	
	\draw[densely dotted] (13, -3.3) ellipse [x radius=5 pt, y radius=10pt];
	\draw[densely dotted] (13, -3.4) ellipse [x radius=6 pt, y radius=13pt];
	\draw[densely dotted] (13, -3.5) ellipse [x radius=8 pt, y radius=15pt];
	
	\draw    
	(8.2, 1) node {$x_{k,\pi_k(1)}$} 
	(9, 0.5) node {$\frac{q}{p_k}$}
	(10, 1) node {$x_{k,\pi_k(2)}$} 
	(12, 1) node {$x_{k,\pi_k(m_k-n_k-1)}$}
	(13, 0.5) node {$\frac{q}{p_k}$}
	(14.5, 1) node {$x_{k,\pi_k(m_k-n_k)}$}
	(9,-4.5) node {$x_{k,j_{k,1}}$}
	(11,-4.5) node {$x_{k,j_{k,2}}$}
	(13,-4.5) node {$x_{k,j_{k,n_k}}$};

	
	\draw(8,0) -- (10,0); 
	\draw (12,0) -- (14,0);
	\draw(9, -3) -- (8,0);
	\draw(9, -3) -- (10,0);
	\draw[loosely dotted](9, -3) -- (10.5,0);
	\draw[loosely dotted](9, -3) -- (11,0);
	\draw[loosely dotted](9, -3) -- (11.5,0);
	\draw(9, -3) -- (12,0);
	\draw(9, -3) -- (14,0);

	\draw(11, -3) -- (8,0);
	\draw(11, -3) -- (10,0);
	\draw(11, -3) -- (12,0);
	\draw(11, -3) -- (14,0);
	\draw(13,-3)--(14,0);
	\draw(13,-3)--(12,0);
	\draw[loosely dotted](13,-3)--(11.5,0);
	\draw[loosely dotted](13,-3)--(11,0);
	\draw[loosely dotted](13,-3)--(10.5,0);
\draw[loosely dotted](5,-3)--(8,0);	
	\draw[loosely dotted](9,-3)--(6,0);
	\draw[densely dotted](11, -3.4) -- (11,-3.5);	
	\end{tikzpicture}
	\caption{Graph of proposed $q$-ary Functions of Hamming Degree $2$}
	\label{fig_somvf}
\end{figure} 
Let $f:\mal{V}_L \rightarrow \mathbb{Z}_q $ be a $q$-ary function of Hamming degree $r$ 
such that $G(f\arrowvert_{\mathbf{x}=\mathbf{c}})$ is a collection of $k$ Hamiltonian paths $P_1,P_2,\hdots,P_k$, for $\mathbf{c} \in \mal{V}_{L'}$,
where $P_i=G(f_i)$ and
\begin{equation}\label{func1}
\begin{split}
f_i=\frac{q}{p_i}\left(\sum_{\alpha=1}^{m_i-n_i-1} x_{i,\pi_i^{c}(\alpha)}x_{i,\pi_i^{c}(\alpha+1)}\right)+g_i(x_{i,\pi_i^{c}(1)},x_{i,\pi_i^{c}(2)},\hdots,x_{i,\pi_i^{c}(m_i-n_i)}),
\end{split}
\end{equation} 
where $c\in \bZ_{L'}$ is the decimal representation of $\mbf{c}$, $\pi_i^{c}$ is a one-to-one mapping from $\{1,2,\hdots,m_i-n_i\}$ to the set $\{1,2,\hdots,m_i\}\setminus J_i$, and $g_i$ is a linear function over the variables $x_{i,\pi_i^{c}(1)},x_{i,\pi_i^{c}(2)},\hdots $, and $x_{i,\pi_i^{c}(m_i-n_i)}$, where $i=1,2,\hdots,k$. Another way we can say $f\arrowvert_{\mathbf{x}_J=\mathbf{c}}=f_1+f_2+\hdots+f_k$.
As an example, Figure \ref{fig_somvf} represents the class of $q$-ary functions of Hamming degree two that satisfies the afore-mentioned graphical properties that after deletion of the variables $x_{i,\alpha_1}$, where  $\alpha_1=j_{i,\alpha_2}$, $\alpha_2=1,2,\hdots,n_i$, and $i=1,2,\hdots,k$, which says after restricting $\mbf{x}_J$ at $\mbf{c}\in \calV_{L'}$, the graph results to a collection of $k$ Hamiltonian paths represented by the quadratic parts in the expression of $f_i$ for $i=1,2,\hdots,k$ in (\ref{func1}). 
For a fixed choice of $\mbf{x}_J$, the class of Hamming degree two $q$-ary functions that can be represented by the graph appears in Figure \ref{fig_somvf} can be expressed as follows:
\begin{equation}\label{ankita009}
	\begin{split}
\mathfrak{C}_{J,2}=\left\{\sum_{i=1}^k \frac{q}{p_i}\left(\sum_{\alpha=1}^{m_i-n_i-1} x_{i,\pi_i(\alpha)}x_{i,\pi_i(\alpha+1)}\right)  +\sum_{\mbf{e}\in \chi} c_{\mbf{e}} x^{\mbf{e}}: c_{\mbf{e}} \in \mathbb{Z}_q, \pi_i\in \mathfrak{P}_{m_i,n_i}, i=1,2,\hdots,k \right\}, 
\end{split}
\end{equation}
where $\mathfrak{P}_{m_i,n_i}$ is the set of all those one-to-one mappings, from $\{1,2,\hdots,m_i-n_i\}$ to the set $\{1,2,\hdots,m_i\}\setminus J_i$,
for which the quadratic term in the expression of $f_i$ given in (\ref{func1}) represenst distinct functions  and it says, $|\mathfrak{P}_{m_i,n_i}|=\frac{(m_i-n_i)!}{2}$. The set $\chi$ in (\ref{ankita009}) can be expressed as follows:
 \[\chi =\chi_1\cup \chi_2 \cup \chi_3,\]
where
\begin{equation}\nonumber
\begin{split}
\chi_1&=\left\{ \mbf{e}\in \calV_L : 0\leq \wt(\mbf{e}) =\wt(\mbf{e}_J)\leq 2 \right\},\\
\chi_2&=\left\{ \mbf{e}\in \calV_L :  \wt(\mbf{e})=2, \wt(\mbf{e}_J)=1,e_{i,j}\in\{0,1\}\forall j\in \{1,2,\hdots,m_i\}\setminus J_i, i=1,2,\hdots,k \right\},\\
\chi_3&=\left\{ \mbf{e}\in \calV_L :  \wt(\mbf{e})=1, \wt(\mbf{e}_J)=0,e_{i,j}\in\{0,1\}\forall j\in \{1,2,\hdots,m_i\}\setminus J_i, i=1,2,\hdots,k \right\}.
\end{split}
\end{equation}
The cardinality of $\chi$ is $|\chi_1|+|\chi_2|+|\chi_3|$, where
\[|\chi_1|=1+\sum_{i=1}^k \left( n_i(p_i-1)+ \binom{n_i}{2}  (p_i-1)^2 \right)+\sum_{i<j}n_i n_j (p_i-1)(p_j-1),  \]
\[|\chi_2|=(m-n)\sum_{i=1}^k n_i(p_i-1), \]
and
\[|\chi_3|=m-n.\]
Then we obtain the cardinality of $\mathfrak{C}_{J,2}$ as  \[|\mathfrak{C}_{J,2}|=\left(\prod_{i=1}^k \mathfrak{P}_{m_i,n_i}\right)q^{|\chi|} =\left(\prod_{i=1}^k \frac{(m_i-n_i)!}{2}\right) q^{|\chi|}. \]
As Figure \ref{fig_somvf} represents a class of functions of Hamming degree $2$, in (\ref{ankita009}), 
$\pi_i^c=\pi_i$ for all $c$. Besides $\mathfrak{C}_{J,2}\subset \mho_{L,2}$, and the graph of any function in (\ref{ankita009}) satisfies the graphical properties as introduced in Figure \ref{fig_somvf}. 
Let us go through the following example to illustrate the Figure \ref{fig_somvf} and the aforementioned matters as discussed in this section.
\begin{example}
Let us consider $k=2$, $m_1=4$, $m_2=3$, $p_1=2$, $p_2=3$, $q=6$, and $n_1=n_2=1$. Then $\mathbf{x}_J=(\mathbf{x_1}_{J_1},\mathbf{x_2}_{J_2})$, where we assume $\mathbf{x_1}_{J_1}=(x_{1,1})$ and $\mathbf{x_2}_{J_2}=(x_{2,1})$.
Following (\ref{func1}), we obtain
\begin{equation}\begin{split}\label{exchih21}
\mathfrak{C}_{J,2}=&\left\{3\left(x_{1,\pi_1(1)}x_{1,\pi_1(2)}+x_{1,\pi_1(2)}x_{1,\pi_1(3)}\right)\right. \\ &\left. + 2x_{2,\pi_2(1)}x_{2,\pi_2(2)}+\sum_{\mbf{e}\in \chi} c_{\mbf{e}} x^{\mbf{e}}: c_{\mbf{e}} \in \mathbb{Z}_6, \pi_i\in \mathfrak{P}_{m_i,n_i}, i=1,2\right\}. 
\end{split}
\end{equation}
Here 
\begin{equation}\label{exchih22}
\begin{split}
\mathfrak{P}_{4,1}&=\left\{(\pi_1(1),\pi_1(2),\pi_1(3)):\pi_1:\{1,2,3\}\rightarrow \{2,3,4\}~ \textnormal{is an one-to-one mapping} \right\}\\
&=\left\{(2,3,4),(2,4,3),(3,2,4)\right\},\\
\mathfrak{P}_{3,1}&=\left\{(\pi_2(1),\pi_2(2)):\pi_2:\{1,2\}\rightarrow \{2,3\}~ \textnormal{is an one-to-one mapping}\right\}\\
&=\left\{(2,3)\right\}.
\end{split}
\end{equation}
\begin{table}[]
	\centering
	\caption{List of Elements in $\chi_1$ and $\chi_3$}\label{chih21}
	\begin{tabular}{|llllllll|}
		\hline
		\multicolumn{8}{|l|}{~~~~~~~~~~~~~~~~~~~~~~~~~~$\chi_1$}                                                                                                                                                                                   \\ \hline
		\multicolumn{1}{|l|}{$x_{1,1}$} & \multicolumn{1}{l|}{$x_{1,2}$} & \multicolumn{1}{l|}{$x_{1,3}$} & \multicolumn{1}{l|}{$x_{1,4}$} & \multicolumn{1}{l|}{$x_{2,1}$} & \multicolumn{1}{l|}{$x_{2,2}$} &\multicolumn{1}{l|}{$x_{2,3}$}& $\mbf{x}^{\mbf{e}}$  \\ \hline
		\multicolumn{1}{|l|}{}          & \multicolumn{1}{l|}{}          & \multicolumn{1}{l|}{}          & \multicolumn{1}{l|}{}          & \multicolumn{1}{l|}{}          & \multicolumn{1}{l|}{}          &      \multicolumn{1}{l|}{}   &  $1$  \\ \hline
		\multicolumn{1}{|l|}{}          & \multicolumn{1}{l|}{}          & \multicolumn{1}{l|}{}          & \multicolumn{1}{l|}{}          & \multicolumn{1}{l|}{$1$}       & \multicolumn{1}{l|}{}          &     \multicolumn{1}{l|}{$$}  & $x_{2,1}$    \\ \hline
		\multicolumn{1}{|l|}{}          & \multicolumn{1}{l|}{}          & \multicolumn{1}{l|}{}          & \multicolumn{1}{l|}{}          & \multicolumn{1}{l|}{$2$}       & \multicolumn{1}{l|}{}          & \multicolumn{1}{l|}{}& $x_{2,1}^2$          \\ \hline
		\multicolumn{1}{|l|}{$1$}       & \multicolumn{1}{l|}{}          & \multicolumn{1}{l|}{}          & \multicolumn{1}{l|}{}          & \multicolumn{1}{l|}{}          & \multicolumn{1}{l|}{}          & \multicolumn{1}{l|}{}& $x_{1,1}$          \\ \hline
		\multicolumn{1}{|l|}{$1$}       & \multicolumn{1}{l|}{}          & \multicolumn{1}{l|}{}          & \multicolumn{1}{l|}{}          & \multicolumn{1}{l|}{$1$}       & \multicolumn{1}{l|}{}          & \multicolumn{1}{l|}{} & $x_{1,1}x_{2,1}$          \\ \hline
		\multicolumn{1}{|l|}{$1$}       & \multicolumn{1}{l|}{}          & \multicolumn{1}{l|}{}          & \multicolumn{1}{l|}{}          &
		 \multicolumn{1}{l|}{$2$}       & \multicolumn{1}{l|}{}          &    \multicolumn{1}{l|}{} & $x_{1,1}x_{2,1}^2$       \\ \hline
		 \multicolumn{8}{|l|}{~~~~~~~~~~~~~~~~~~~~~~~~~~$\chi_3$} \\\hline
		 \multicolumn{1}{|l|}{}          & \multicolumn{1}{l|}{$1$}          & \multicolumn{1}{l|}{}          & \multicolumn{1}{l|}{}          & \multicolumn{1}{l|}{}          & \multicolumn{1}{l|}{}          &\multicolumn{1}{l|}{}& $x_{1,2}$           \\ \hline
		 \multicolumn{1}{|l|}{}          & \multicolumn{1}{l|}{}          & \multicolumn{1}{l|}{$1$}          & \multicolumn{1}{l|}{}          & \multicolumn{1}{l|}{}          & \multicolumn{1}{l|}{}          &       \multicolumn{1}{l|}{}& $x_{1,3}$    \\ \hline
		 
		 \multicolumn{1}{|l|}{}          & \multicolumn{1}{l|}{}          & \multicolumn{1}{l|}{}          & \multicolumn{1}{l|}{1}          & \multicolumn{1}{l|}{}          & \multicolumn{1}{l|}{}          &      \multicolumn{1}{l|}{} & $x_{1,4}$     \\ \hline
		 \multicolumn{1}{|l|}{}          & \multicolumn{1}{l|}{}          & \multicolumn{1}{l|}{}          & \multicolumn{1}{l|}{}          & \multicolumn{1}{l|}{}          & \multicolumn{1}{l|}{}          &  \multicolumn{1}{l|}{$1$}& $x_{2,3}$      \\ \hline
		 \multicolumn{1}{|l|}{}          & \multicolumn{1}{l|}{}          & \multicolumn{1}{l|}{}          & \multicolumn{1}{l|}{}          & \multicolumn{1}{l|}{}          & \multicolumn{1}{l|}{$1$}          &    \multicolumn{1}{l|}{}& $x_{2,2}$       \\ \hline
	\end{tabular}
\end{table}

\begin{table}[]
	\centering
		\caption{List of Elements in $\chi_2$}\label{chih22}
	\begin{tabular}{|l|l|l|l|l|l|l|l|}
				\hline
			\multicolumn{8}{|l|}{~~~~~~~~~~~~~~~~~~~~~~~~~~$\chi_2$} 
			\\\hline                                                                                                                                            
		$x_{1,1}$ & $x_{1,2}$ & $x_{1,3}$ & $x_{1,4}$ & $x_{2,1}$ & $x_{2,2}$ & $x_{2,3}$&$\mbf{x}^{\mbf{e}}$ \\ \hline
		$1$       & $1$       &           &           &           &           &   & $x_{1,1}x_{1,2}$        \\ \hline
		$1$       &           & $1$       &           &           &           &   & $x_{1,1}x_{1,3}$        \\ \hline
		$1$       &           &           & $1$       &           &           &    &$x_{1,1}x_{1,4}$       \\ \hline
		$1$       &           &           &           &           &           & $1$ & $x_{1,1}x_{2,3}$      \\ \hline
		$1$       &           &           &           &           & $1$       &  & $x_{1,1}x_{2,2}$         \\ \hline
		& $1$       &           &           & $1$       &           &         & $x_{1,2}x_{2,1}$  \\ \hline
		&           & $1$       &           & $1$       &           &        & $x_{1,3}x_{2,1}$   \\ \hline
		&           &           & $1$       & $1$       &           &   & $x_{1,4}x_{2,1}$        \\ \hline
		&           &           &           & $1$       &           & $1$  & $x_{2,1}x_{2,3}$     \\ \hline
		&           &           &           & $1$       & $1$       &      & $x_{2,1}x_{2,2}$     \\ \hline
		& $1$       &           &           & $2$       &           &      & $x_{1,2}x_{2,1}^2$     \\ \hline
		&           & $1$       &           & $2$       &           &      & $x_{1,3}x_{2,1}^2$     \\ \hline
		&           &           & $1$       & $2$       &           &      & $x_{1,4}x_{2,1}^2$     \\ \hline
		&           &           &           & $2$       &           & $1$  & $x_{2,1}^2 x_{2,3}$     \\ \hline
		&           &           &           & $2$       & $1$       &       & $x_{2,1}^2 x_{2,2}$    \\ \hline
	\end{tabular}
\end{table}

\begin{figure}
	\centering
	\begin{tikzpicture}[line width=.6pt]
	\filldraw 
	(0, 0) circle [radius=5pt]
	(1, 0) circle [radius=5pt]
	(2,0) circle [radius=5pt]
	(1,-1.5) circle [radius=5pt]
	(4, 0) circle [radius=5pt]
	(5,-1.5) circle [radius=5pt]
	(6, 0) circle [radius=5pt];
	\draw[densely dotted](1,-1.5) -- (0,0);
	\draw[densely dotted](1,-1.5) -- (1,0);
	\draw[densely dotted](1,-1.5) -- (2,0);
	\draw[densely dotted](1,-1.5) -- (4,0);
	\draw[densely dotted](1,-1.5) -- (6,0);
	
	\draw[densely dotted](5,-1.5) -- (0,0);
	\draw[densely dotted](5,-1.5) -- (1,0);
	\draw[densely dotted](5,-1.5) -- (2,0);
	\draw[densely dotted](5,-1.5) -- (4,0);
	\draw[densely dotted](5,-1.5) -- (6,0);
	\draw[](1,-1.5) -- (5,-1.5);
	
	\draw (0,0) -- (1,0);
	\draw (1,0) -- (2,0);
	\draw (4,0) -- (6,0);

		\draw    
		(-.5, .6) node {$x_{1,\pi_1(1)}$}
		(1, .6) node {$x_{1,\pi_1(2)}$}
		(2.4, .6) node {$x_{1,\pi_1(3)}$}
		(4, .6) node {$x_{2,\pi_2(1)}$}
		(6, .6) node {$x_{2,\pi_2(2)}$}
		(.5, .2) node {$3$}
		(1.5, .2) node {$3$}
		(5, .2) node {$2$}
		(0.4, -1.5) node {$x_{1,1}$}
		(5.6, -1.5) node {$x_{2,1}$}
		;
	
	
%
%
%
%
\draw[densely dotted] (5, -1.7) ellipse [x radius=6 pt, y radius=8pt];

\end{tikzpicture}
	\caption{Graph of Proposed $6$-ary Functions of Hamming Degree $2$ with $x_{1,1}$ and $x_{2,1}$ as the Restricted Variables}
	\label{fignewsomvf}
\end{figure}	
The set of elements $\mbf{e}\in \chi$ and their corresponding monomials $\mbf{x}^{\mbf{e}}$ are listed in Table \ref{chih21} and Table \ref{chih22}, and $|\chi|=26$. From (\ref{exchih21}), (\ref{exchih22}), Table \ref{chih21} and Table \ref{chih22}, we have 
\begin{equation}
|\mathfrak{C}_{J,2}|=|\mathfrak{P}_{4,1}|\cdot|\mathfrak{P}_{3,1}|\cdot 6^{26}=3\cdot  6^{26}.
\end{equation}
We can also verify that all the $3\cdot 6^{26}$ functions $f:\mal{V}_{432}\rightarrow \mathbb{Z}_6$ in (\ref{exchih21}) satisfy the graph as presented in Figure \ref{fignewsomvf}, where $\mal{V}_{432}$ is presnted in Table \ref{3aryvec432} of Appendix \ref{Appendix:B}.

\end{example}
Let $f:\mal{V}_L\rightarrow \mathbb{Z}_q$ be a $q$-ary functions of Hamming degree at most $r$, and satisfies all the graphical and functional properties as introduced in the beginning of Section \ref{contsec} and in (\ref{func1}), respectively. \ccr In the case $r=2$, $f\in \mathfrak{C}_{J,2} $.
Now using $f$, let us define the following set of $q$-ary functions:\ccn
\begin{equation}\label{mainset1}
\begin{split}
C_t=\left\{f+\sum_{i=1}^k\frac{q}{p_i}\left((\mathbf{d}_i\cdot \mathbf{x}_{J_i}+d_i x_{i,\pi_i^{c}(m_i-n_i)})+(\mathbf{t}_i\cdot \mathbf{x}_{J_i}+t_i x_{i,\pi_i^{c}(1)})\right): \mathbf{d}_i\in \mathbf{Z}_{p_i}^{n_i},d_i\in\mathbf{Z}_{p_i},\right. \\ \left. i=1,2,\hdots,k\right\},
\end{split}
\end{equation}
where $\mathbf{t}_i\in \mathbf{Z}_{p_i}^{n_i}$, $t_i\in \mathbf{Z}_{p_i}$, and $(\mathbf{t}_i,t_i)$ is the vectorial form of $t$, and
\begin{equation}\label{verep1}
t=\sum_{i=1}^{k-1}\left((\mathbf{t}_i,t_i)\cdot (p_i^{n_i},p_i^{n_i-1},\hdots,1)\right)\prod_{j=1}^{k-i} p_{k-j+1}^{n_{k-j+1}+1} +(\mathbf{t}_k,t_k)\cdot (p_k^{n_k},p_k^{m_k-2},\hdots,1).
\end{equation}
 From (\ref{mainset1}), for each of $t$ in $\{0,1,\hdots,(\prod_{i=1}^k p_i^{n_i+1})-1\}$, it is clear that $\psi(C_t)$ forms a complex-valued code over $\mbb{Z}_q$, containing $K$
 sequences each having length of $L$. Let us construct the following set of codes 
 \begin{equation}\label{ccc1}
 \mal{C}=\{\psi(C_t):0\leq t<\prod_{i=1}^k p_i^{n_i+1}\}.
 \end{equation}
Now let us construct the following set using the restricted $q$-ary function $f\arrowvert_{\mathbf{x}=\mathbf{c}}$ as
\begin{equation}\label{ankit2}
	\begin{split}
	\mal{S}_u^{\mbf{c}}=\left\{\sum_{i=1}^k f_i+\frac{q}{p_i}\left(d_i x_{i,\pi^{c}_i(m_i-n_i)}+t_i x_{i,\pi_i^{c}(1)}\right): d_i\in\mathbf{Z}_{p_i}, i=1,2,\hdots,k\right\},
	\end{split}
\end{equation}
where $u=\sum_{i=1}^{k-1}t_i\prod_{j=1}^{k-i} p_{k-j+1}^{n_{k-j+1}+1} +t_k$ is the decimal representation of $(t_1,t_2,\hdots,t_k)$. 
Then we define the following sets of SNC complex-valued codes:
\begin{equation}
	\mathcal{S}_\mbf{c}=\left\{ \psi(\mal{S}_u^{\mbf{c}}):0\leq u<\prod_{i=1}^k p_i\right\}.
\end{equation}
In the next section, we shall discuss the correlation properties of the code sets $\mathcal{S}_\mbf{c}$ and $\mal{C}$.
\begin{theorem}\label{teorem1}
	For any $\mbf{c}\in \calV_{L'}$, the code set $\mal{S}_\mbf{c}$ forms $(\prod_{i=1}^k p_i,L)$-CCC with $\mal{N}_\mbf{c}$ as support.
\end{theorem}
\begin{IEEEproof}
Let us define $h:\calV_L^\mbf{c}\rightarrow \mathbb{Z}_q $ be a function given by
$$h=\left(f+\sum_{i=1}^k \frac{q}{p_i}\left(d_i x_{i,\pi_i^{c}(m_i-n_i)}+t_i x_{i,\pi_i^{c}(1)}\right)\right)\Big\arrowvert_{\mbf{x}_J=\mbf{c}}.$$
Then 
\begin{equation}\label{srest1}
\psi(h)=
\begin{cases}
\zeta_q^{h_x}, & x\in \mal{N}_\mbf{c},\\
0,& x\in \mbf{Z}_L \setminus \mal{N}_\mbf{c}.
\end{cases}
\end{equation}
As $x\in\mal{N}_\mbf{c}$, 
$$h_x=h(\mathbf{x})=\sum_{i=1}^k \left(f_i(\mbf{x}_i)+\frac{q}{p_i}\left(d_i x_{i,\pi_i^{c}(m_i-n_i)}+t_i x_{i,\pi_i^{c}(1)}\right)\right).$$
Then
\begin{equation}\label{srest2}
\psi(h)=
\begin{cases}
\prod_{i=1}^k \zeta_q^{\left(f_i(\mbf{x}_i)+\frac{q}{p_i}\left(d_i x_{i,\pi_i^{c}(m_i-n_i)}+t_i x_{i,\pi_i^{c}(1)}\right)\right) }, & x\in \mal{N}_\mbf{c},\\
0,& x\in \mbf{Z}_L \setminus \mal{N}_\mbf{c}.
\end{cases}
\end{equation}
Similarly as $h$, for $h':\calV_L^\mbf{c}\rightarrow \mathbb{Z}_q $, given by
$$h'=\left(f+\sum_{i=1}^k\frac{q}{p_i}\left(d_i x_{i,\pi_i^{c}(m_i-n_i)}+t_i' x_{i,\pi_i^{c}(1)}\right)\right)\Big\arrowvert_{\mbf{x}_J=\mbf{c}},$$
\begin{equation}\label{srest2}
\psi(h')=
\begin{cases}
\zeta_q^{h_x'}, & x\in \mal{N}_\mbf{c},\\
0,& x\in \mbf{Z}_L \setminus \mal{N}_\mbf{c}.
\end{cases}
\end{equation}
Let us assume $\tau\geq 0$ and $y\in \mal{N}_\mbf{c}$. For $\mbf{y}=(\mbf{y}_1,\mbf{y}_2,\hdots,\mbf{y}_k)\in \calV_L^\mbf{c}$, as the vector representation of $y$, and $y=x+\tau$, similarly as (\ref{srest2}), $\psi(h')$ can also be expressed as 
\begin{equation}\label{srest3}
\psi(h')=
\begin{cases}
\prod_{i=1}^k \zeta_q^{\left(f_i(\mbf{y}_i)+\frac{q}{p_i}\left(d_i y_{i,\pi_i^{c}(m_i-n_i)}+t_i' y_{i,\pi_i^{c}(1)}\right)\right) }, & y\in \mal{N}_\mbf{c},\\
0,& y\in \mbf{Z}_L \setminus \mal{N}_\mbf{c}.
\end{cases}
\end{equation}
Let 
\begin{equation}\label{kta1}
F_{d_i,t_i}(\mbf{x}_i)=f_i(\mbf{x}_i)+\frac{q}{p_i}\left(d_i x_{i,\pi_i^{c}(m_i-n_i)}+t_i x_{i,\pi_i^{c}(1)}\right),  
\end{equation}
and
\begin{equation}\label{kit2}
F_{d_i,t_i'}(\mbf{y}_i)=f_i(\mbf{y}_i)+\frac{q}{p_i}\left(d_i y_{i,\pi_i^{c}(m_i-n_i)}+t_i' y_{i,\pi_i^{c}(1)}\right).
\end{equation} 
Then from (\ref{srest1}), and (\ref{srest2}), for $u\neq u'$, where $u'=\sum_{i=1}^{k-1}t_i'\prod_{j=1}^{k-i} p_{k-j+1}^{n_{k-j+1}+1} +t_k'$, the ACCF between $\psi(\mal{S}_u^{\mbf{c}})$ and $\psi(\mal{S}_{u'}^{\mbf{c}})$ at $\tau\in\mal{T}_\mbf{c}$ can be expressed as 
\begin{equation}\label{snckit1}
\begin{split}
\Theta(\psi(\mal{S}_u^{\mbf{c}}),\psi(\mal{S}_{u'}^{\mbf{c}}))(\tau)=\sum_{\mbf{d}\in \calV_{L''}} 
&\tta\left(\psi\left(\left(f+\sum_{i=1}^k\frac{q}{p_i}\left(d_i x_{i,\pi_i^{c}(m_i-n_i)}+t_i x_{i,\pi_i^{c}(1)}\right)\right)\Big\arrowvert_{\mbf{x}_J=\mbf{c}}\right),\right.\\&~~~~~ \left.\psi\left(\left(f+\sum_{i=1}^k\frac{q}{p_i}\left(d_i x_{i,\pi_i^{c}(m_i-n_i)}+t_i' x_{i,\pi_i^{c}(1)}\right)\right)\Big\arrowvert_{\mbf{x}_J=\mbf{c}}\right)\right)(\tau)\\
& =\sum_{\mbf{d}\in \calV_{L''} } \sum_{(x,y)\in \mal{A}_\tau(\mathbf{c})} \zeta_q^{h_x-h_y'}\\
&= \sum_{(x,y)\in \mal{A}_\tau(\mathbf{c})} \prod_{i=1}^k \sum_{d_i\in \mbf{Z}_{p_i}}\zeta_q^{F_{d_i,t_i}(\mbf{x}_i)-F_{d_i,t_i'}(\mbf{y}_i)},
\end{split}
\end{equation}
where $\mbf{d}=(d_1,d_2,\hdots,d_k)$, and $\calV_{L''}=\mbf{Z}_{p_1}\times \mbf{Z}_{p_2}\times\cdots\times \mbf{Z}_{p_k}$, where $L''=\prod_{i=1}^k p_i$.	
As $\mbf{x}=(\mbf{x}_1,\mbf{x}_2,\hdots,\mbf{x}_k)$ and $\mbf{y}=(\mbf{y}_1,\mbf{y}_2,\hdots,\mbf{y}_k)$ are in $\calV_L^\mbf{c}$, from (\ref{func1}),
(\ref{kta1}), and (\ref{snckit1}), we have
\begin{equation}\label{apnk1}
	\begin{split}
\dis\sum_{d_i\in\mbf{Z}_{p_i}}& \zeta_q^{F_{d_i,t_i}(\mbf{x}_i)-F_{d_i,t_i'}(\mbf{y}_i)}\\
&=  \zeta_q^{f_i(\mbf{x}_i)-f_i(\mbf{y}_i)}\zeta_{p_i}^{\left(t_i x_{i,\pi_i^{c}(1)}-t_i'y_{i,\pi_i^{c}(1)}\right)}\dis\sum_{d_i\in\mbf{Z}_{p_i}}\zeta_{p_i}^{d_i\left(x_{i,\pi_i^{c}(m_i-n_i)}-y_{i,\pi_i^{c}(m_i-n_i)}\right)}
	\end{split}
\end{equation}
For some $\tau\in \mal{T}_\mbf{c}$, there is possiblity that there exist  $\{i_1,i_2,\hdots,i_v\}\subset \{1,2,\hdots,k\}$, where $0\leq v\leq k$, the following condition holds:
$\mbf{x}_{i} = \mbf{y}_{i}$ for $i\in  \{i_1,i_2,\hdots,i_v\}$. Then $\mbf{x}_{i} \neq  \mbf{y}_{i}$ for $i\in  \{1,2,\hdots,k\}\setminus \{i_1,i_2,\hdots,i_v\}$.
\setcounter{case}{0}
\begin{case}[{$\mbf{x}_{i} = \mbf{y}_{i}$},~{$i\in  \{i_1,i_2,\hdots,i_v\}$}]
In this case, from (\ref{apnk1}), we have 
\begin{equation}\label{apnk10}
\begin{split}
\dis\sum_{d_i\in\mbf{Z}_{p_i}}& \zeta_q^{F_{d_i,t_i}(\mbf{x}_i)-F_{d_i,t_i'}(\mbf{y}_i)}\\
&=  p_i \zeta_{p_i}^{\left(t_i -t_i'\right)x_{i,\pi_i^{c}(1)}}
\end{split}
\end{equation}
As $(x,y)\in \mal{A}_\tau(\mbf{c})$, $\tau=y-x$, we have 
\begin{equation}\label{nket1}
\tau =\sum_{\substack{{i=1}\\{i\notin \{i_1,i_2,\hdots,i_v\}} }}^k (y_i-x_i)\prod_{j=1}^{k-i} L_{k-j+1},
\end{equation}
where $\prod_{j=1}^{k-i} L_{k-j+1}=1$ if $k\notin \{i_1,i_2,\hdots,i_v\}$. In (\ref{nket1}),
$\tau$ is constant for all possible values of $x_i=y_i\in \mbf{Z}_{L_i}$, which says $\mal{A}_\tau(\mbf{c})$ contains all those $(x,y)\in\mal{N}_\mbf{c}\times \mal{N}_\mbf{c}$, for which the vector representations $\mbf{x}$ and $\mbf{y}$ in $\calV_L^\mbf{c}$ of $x$ and $y$, respectively, follows  $\mbf{x}_i=\mbf{y}_i$.
Then from (\ref{apnk10}), we have 
\begin{equation}\label{apnk21}
\begin{split}
\sum_{\substack{{(x,y)\in\mal{A}_\tau(\mbf{c})}\\{x_i=y_i}\\{i\in \{i_1,i_2,\hdots,i_v\}} }}&\prod_{i\in\{i_1,i_2,\hdots,i_v\}}\dis\sum_{d_i\in\mbf{Z}_{p_i}} \zeta_q^{F_{d_i,t_i}(\mbf{x}_i)-F_{d_i,t_i'}(\mbf{y}_i)}\\
&=\begin{cases} 
\dis\prod_{i\in\{i_1,i_2,\hdots,i_v\}} p_i^{m_i-n_i+1},&t_i=t_i',~ i=1,2,\hdots,v,\\
0,& \textnormal{otherwise.} 
\end{cases}
\end{split}
\end{equation}
\end{case}
\begin{case}[{$\mbf{x}_{i} \neq \mbf{y}_{i}$},~{$i\notin  \{i_1,i_2,\hdots,i_v\}$}]
From the expression in (\ref{apnk1}), this case can be defined into the following sub-cases:
\begin{scase}[$x_{i,\pi_i^{c}(m_i-n_i)}\neq y_{i,\pi_i^{c}(m_i-n_i)} $]
Since $x_{i,\pi_i^{c}(m_i-n_i)}\neq y_{i,\pi_i^{c}(m_i-n_i)} $,  $$\dis\sum_{d_i\in\mbf{Z}_{p_i}}\zeta_{p_i}^{d_i\left(x_{i,\pi_i^{c}(m_i-n_i)}-y_{i,\pi_i^{c}(m_i-n_i)}\right)}=0.$$
Then again from (\ref{apnk1}),
$$\dis\sum_{d_i\in\mbf{Z}_{p_i}} \zeta_q^{F_{d_i,t_i}(\mbf{x}_i)-F_{d_i,t_i'}(\mbf{y}_i)}=0.$$	
\end{scase}
\begin{scase}[$x_{i,\pi_i^{c}(m_i-n_i)}= y_{i,\pi_i^{c}(m_i-n_i)} $]
\begin{equation}\label{kintu1}
{F_{d_i,t_i}(\mbf{x}_i)-{F_{d_i,t_i'}}(\mbf{y}_i)}={f_i(\mbf{x}_i)-f_i(\mbf{y}_i)} + {\left(t_i x_{i,\pi_i^{c}(1)}-t_i'y_{i,\pi_i^{c}(1)}\right)}.
\end{equation}
Let $(x^\kappa,y^\kappa)\in \mal{N}_\mbf{c}\times \mal{N}_\mbf{c}$ such that $\mbf{x}^\kappa=(\mbf{x}_1^\kappa,\mbf{x}_2^\kappa,\hdots,\mbf{x}_k^\kappa)$ and $\mbf{y}^\kappa=(\mbf{y}_1^\kappa,\mbf{y}_2^\kappa,\hdots,\mbf{y}_k^\kappa)$ are the vector representation of $x^\kappa$ and $y^\kappa$ in $\calV_L^\mbf{c}$, and given by
\begin{equation}
\begin{split}
\mbf{x}_i^\kappa &=\left( x_{i,1},\hdots,x_{i,\pi_i^{c}(1)},\hdots, \left(\left(x_{i,\pi_i^{c}(l+1)}-\kappa\right)\!\!\!\!\!\!\mod p_i\right),\hdots,x_{i,\pi_i^{c}(m_i-n_i)},\hdots,x_{i,m_i} \right),\\
\mbf{y}_i^\kappa &=\left( y_{i,1},\hdots,y_{i,\pi_i^{c}(1)},\hdots, \left(\left( y_{i,\pi_i^{c}(l+1)}-\kappa\right)\!\!\!\!\!\!\mod p_i\right),\hdots,y_{i,\pi_i^{c}(m_i-n_i)},\hdots,y_{i,m_i} \right),
\end{split}
\end{equation}
where $\kappa=1,2,\hdots,p_i-1$, $1\leq l<m_i-n_i$, and $\mbf{x}_i^\kappa$ and $\mbf{y}_i^\kappa$ differs from $\mbf{x}_i$ and $\mbf{y_i}$ only at $\pi_i^{c}(l)$th position, respectively. We also assume $\pi_i^{c}(l)$ is the smallest integer such that 
$x_{i,\pi_i^{c}(l)} \neq y_{i,\pi_i^{c}(l)}$, and $x_{i,\pi_i^{c}(j)} = y_{i,\pi_i^{c}(j)}$, where $l<j\leq m_i-n_i$. Then $y^\kappa=x^\kappa+\tau$ and $(x^\kappa,y^\kappa)\in \mal{A}_\tau(\mbf{c})$, for $\kappa=1,2,\hdots,p_i-1$. 
Now 
\begin{equation}\label{kintu2}
 \begin{split}
f_i(\mbf{x}_i)-f_i(\mbf{x}_i^\kappa)
&=\frac{q}{p_i}\left( (x_{i,\pi_i^{c}(l)}x_{i,\pi_i^{c}(l+1)}+x_{i,\pi_i^{c}(l+1)}x_{i,\pi_i^{c}(l+2)})-
(x_{i,\pi_i^{c}(l)}(x_{i,\pi_i^{c}(l+1)}-\kappa+\lambda_ip_i)\right. \\&\left.-(x_{i,\pi_i^{c}(l+1)}-\kappa+\lambda_ip_i)x_{i,\pi_i^{c}(l+2)}) \right)+\lambda'_i(x_{i,\pi_i^{c}(l+1)}-x_{i,\pi_i^{c}(l+1)}+\kappa-\lambda_ip_i) \\
&=\frac{q}{p_i}(\kappa-\lambda_i p_i)(x_{i,\pi_i^{c}(l)}+x_{i,\pi_i^{c}(l+2)})+\lambda_i' (\kappa-\lambda_ip_i),
 \end{split}
\end{equation}
where $\lambda_i=0$ when $x_{i,\pi_i^{c}(l+1)}-k\geq 0 $, and $1$, else, and $\lambda_i'\in \mbf{Z}_q$, $i=1,2,\hdots,k$. Similarly, 
\begin{equation}\label{kintu3}
\begin{split}
f_i(\mbf{y}_i)-f_i(\mbf{y}_i^\kappa)
&=\frac{q}{p_i}(\kappa-\lambda_i p_i)(y_{i,\pi_i^{c}(l)}+y_{i,\pi_i^{c}(l+2)})+\lambda_i' (\kappa-\lambda_ip_i),
\end{split}
\end{equation}
From (\ref{kintu1}), (\ref{kintu2}), and (\ref{kintu3}), we have
\begin{equation}
\begin{split}
{F_{d_i,t_i}(\mbf{x}_i)-{F_{d_i,t_i'}}(\mbf{y}_i)}-\left({F_{d_i,t_i}(\mbf{x}_i^\kappa)-{F_{d_i,t_i'}}(\mbf{y}_i^\kappa)}\right)\\={f_i(\mbf{x}_i)-f_i(\mbf{x}_i^\kappa)}-\left({f_i(\mbf{y}_i)-{f_i(\mbf{y}_i^\kappa)}}\right)\\
=\frac{q}{p_i}(\kappa-\lambda_i p_i)(x_{i,\pi_i^{c}(l)}-y_{i,\pi_i^{c}(l)}).
\end{split}
\end{equation}
Then 
\begin{equation}
	\begin{split}
	\sum_{\kappa=1}^{p_i-1}\om^{-(F_{d_i,t_i}(\mbf{x}_i)-{F_{d_i,t_i'}}(\mbf{y}_i))}   \om^{{F_{d_i,t_i}(\mbf{x}_i^\kappa)-{F_{d_i,t_i'}}(\mbf{y}_i^\kappa)}}=\sum_{\kappa=1}^{p_i-1} \zeta_{p_i}^{\kappa (x_{i,\pi_i^{c}(l)}-y_{i,\pi_i^{c}(l)})}=-1,
	\end{split}
\end{equation}
which says 
\begin{equation}
\begin{split}
\om^{F_{d_i,t_i}(\mbf{x}_i)-{F_{d_i,t_i'}}(\mbf{y}_i)}+\sum_{\kappa=1}^{p_i-1}  \om^{{F_{d_i,t_i}(\mbf{x}_i^\kappa)-{F_{d_i,t_i'}}(\mbf{y}_i^\kappa)}}=0,
\end{split}
\end{equation}
Then
\begin{equation}\label{apnk21}
\begin{split}
\sum_{\substack{{(x,y)\in\mal{A}_\tau(\mbf{c})}\\{x_i\neq y_i}\\{i\notin \{i_1,i_2,\hdots,i_v\}} }}\prod_{i\notin\{i_1,i_2,\hdots,i_v\}}\dis\sum_{d_i\in\mbf{Z}_{p_i}} \zeta_q^{F_{d_i,t_i}(\mbf{x}_i)-F_{d_i,t_i'}(\mbf{y}_i)}=0.
\end{split}
\end{equation}
\end{scase}
Using all the above cases in (\ref{snckit1}), we have
\begin{equation}
\Theta(\psi(\mal{S}_u^{\mbf{c}}),\psi(\mal{S}_{u'}^{\mbf{c}}))(\tau)
=\begin{cases}
 \dis\prod_{i=1}^k p_i^{m_i-n_i+1},& \tau=0, u=u',\\
 0,& \textnormal{otherwise.}
\end{cases}
\end{equation}
\end{case}
Therefore the code set $\mathcal{S}_\mbf{c}$ forms $(\prod_{i=1}^k p_i,L)$-SNC-CCC and from
(\ref{srest1}), it is clear that the set $\mal{N}_\mbf{c}$ is support for $\mathcal{S}_\mbf{c}$. 
%
\end{IEEEproof}
As $\mbf{c}\in \calV_{L'}$, for a $q$-ary function $f$, there exist $\prod_{i=1}^k p_i^{n_i}$ number of SNC-CCCs $\mal{S}^\mbf{c}$.	
In the below example, we illustrate Theorem \ref{teorem1} with the help of the $6$-ary function $f$ as given in Example \ref{exmo1}.
\begin{example}\label{ex2folded1}
Let $f:\mal{V}_{72}\rightarrow \mathbb{Z}_6$ be the same $6$-ary function as defined in (\ref{ankby1}). It can be expressed as below:	
$$f(x_{1,1},x_{1,2},x_{1,3},x_{2,1},x_{2,2})=2x_{1,1}x_{1,2}+4x_{1,2}x_{1,3}+x_{1,2}x_{2,1}+x_{1,2}x_{2,2}+3x_{1,1}x_{1,3}+2x_{2,1}x_{2,2}+x_{1,2}+2.$$
From Figure \ref{fig_somvf1}, for $c\in\{0,1\}$, $G(f\arrowvert_{x_{1,2}=c})$ is a collection of two Hamiltonian paths identified by the monomials $x_{1,1}x_{1,3}$ and $x_{2,1}x_{2,2}$. 
Then $n_1=1$, $n_2=0$, and $J=\{J_1,J_2\}=\{J_{1,2}\}=\{2\}$, where $J_2=\emptyset$. Hence 
$\mbf{x}_J=({\mathbf{x}_1}_{J_1},{\mathbf{x}_2}_{J_2})=(x_{1,2})$. Then 
$$\calV_{72}^0=\left\{(x_{1,1},x_{1,2},x_{1,3},x_{2,1},x_{2,2})\in \mal{V}_{72}:x_{1,2}=0 \right\},$$
and 
\begin{equation}
\begin{split}
\mal{N}_0 & =\{x\in \mbf{Z}_{72}:x=(x_{1,1},0,x_{1,3})\cdot (4,2,1)+(x_{2,1},x_{2,1})\cdot (3,1), x_{1,1}, x_{1,3}\in \mbf{Z}_2,~x_{2,1},x_{2,2}\in \mbf{Z}_3\}\\
&=\{0,1,\hdots,17\} \cup \{36,37,\hdots,53\}. 
\end{split}
\end{equation}  
Then $f\arrowvert_{x_{1,2}=0}=3x_{1,1}x_{1,3}+2x_{2,1}x_{2,2}+2$ and 
$f\arrowvert_{x_{1,2}=1}=3x_{1,1}x_{1,3}+2x_{2,1}x_{2,2}+2x_{1,1}+4x_{1,3}+x_{2,1}+x_{2,2}+3$, which says, $\pi_1(1)=1$, $\pi_1(2)=3$, $\pi_2(1)=1$, and $\pi_2(2)=2$.  
Then following (\ref{ankit2}), we derive the following set $6$-ary functions:
\begin{equation}\label{ankita1}
\begin{split}
\mal{S}_u^{0}=\left\{3x_{1,1}x_{1,3}+2x_{2,1}x_{2,2}+2+3\left(d_1 x_{1,3}+t_1 x_{1,1}\right)+2(d_2x_{2,2}+t_2x_{2,1}): d_1\in\mathbf{Z}_{2},d_2\in\mbf{Z}_3\right\},
\end{split}
\end{equation}
where $t_1\in \mbf{Z}_2$, and $t_2\in \mbf{Z}_3$, and $u=3t_1+t_2$. We present the set of SNC codes $\mal{S}_0=\{\psi(\mal{S}_u^0):u=0,1,\hdots,5\}$ in Table \ref{ankut2}. 
  The AACFs and ACCFs of the codes in $\mal{S}_0$ follow the correlation plot as given in Figure \ref{futus1}, which says $\mal{S}_0$ forms $(5,72)$-SNC-CCC. Similarly, we can obtain $\mal{S}_0$ and the corresponding SNC code set $\mal{S}_1$, alo and verify that it forms $(5,72)$-SNC-CCC.
   Hence the presented example verifies the proposed result in Theorem \ref{teorem1}. 
\begin{figure}[H]
	\centering
	\begin{tikzpicture}[line width=.6pt]
	\filldraw 
	(0, 0) circle [radius=5pt]
	(2, 0) circle [radius=5pt]
	(1, -1.5) circle [radius=5pt]
	(0, -3) circle [radius=5pt]
	(2, -3) circle [radius=5pt]
	(4.5,0) circle [radius=5pt]
    (4.5,-3) circle [radius=5pt]
	(6.5,0) circle [radius=5pt]
	(6.5,-3) circle [radius=5pt]
	;
	
\draw[->] (1.5, -1.5)--(4.05,-1.5) ;
		
\draw(0, .5) node {$x_{1,1}$}; 
\draw(1, 0.3) node {$3$};
\draw(2, 0.5) node {$x_{1,3}$};
\draw(-0.2,-1.5) node {$x_{1,2}$};
\draw(0,-3.5) node {$x_{2,1}$};
\draw(2,-3.5) node {$x_{2,2}$};
\draw(1,-3.3) node {$2$};
\draw(4.5, 0.5) node {$x_{1,1}$};
\draw(5.5, 0.3) node {$3$};
\draw(6.5, 0.5) node {$x_{1,3}$};
\draw(4.5,-3.5) node {$x_{2,1}$};
\draw(5.5,-3.3) node {$2$};
\draw(6.5,-3.5) node {$x_{2,2}$};
\draw(2.7,-1.2) node{$x_{1,2}=0$ or $1$};
	
	\draw(0,0) -- (2,0); 
	\draw(1,-1.5) -- (0,-3);
	\draw(1,-1.5) -- (2,-3);
	\draw(1,-1.5) -- (0,0);
	\draw(1,-1.5) -- (2,0);
	\draw(0,-3) -- (2,-3);
	\draw(4.5,0) -- (6.5,0);
	\draw(4.5,-3) -- (6.5,-3);
	\end{tikzpicture}
	\caption{Proposed $q$-ary Functions}
	\label{fig_somvf1}
\end{figure} 
	\begin{figure}[H]
\includegraphics[width=10cm]{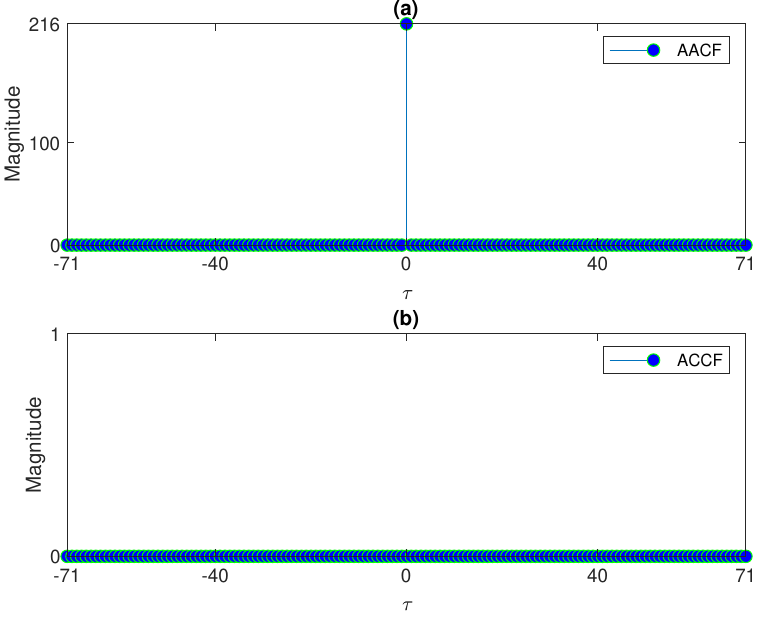}
\caption{Correlation Plotting for the codes in $\mathcal{S}_0$}
\label{futus1}
\end{figure}
\end{example}

In the following theorem, we shall prove that the set of codes $\mal{C}$ forms $(K,L)$-CCC.  
\begin{theorem}\label{teorem2}
	The code set $\mal{C}$ forms $(K,L)$-CCC over the alphabet $\mathbb{Z}_q$.
\end{theorem} 
	\begin{IEEEproof}
From (\ref{mainset1}), $C_t$ can be expressed as 
	\begin{equation}
	\begin{split}
	C_t = \left\{f+\sum_{i=1}^k\frac{q}{p_i}\left((\mathbf{d}_i+\mathbf{t}_i)\cdot \mathbf{x}_{J_i}+(d_i x_{i,\pi_i^{c}(m_i-n_i)}+t_i x_{i,\pi_i^{c}(1)})\right): \mathbf{d}_i\in \mathbf{Z}_{p_i}^{n_i},d_i\in\mathbf{Z}_{p_i},i=1,2,\hdots,k\right\}.
	\end{split}
	\end{equation}
Let us assume $$\mathcal{F}=f+\sum_{i=1}^k\frac{q}{p_i}\left((\mathbf{d}_i+\mathbf{t}_i)\cdot \mathbf{x}_{J_i}+(d_i x_{i,\pi_i^{c}(m_i-n_i)}+t_i x_{i,\pi_i^{c}(1)})\right),$$ and $$\mathcal{F}'=f+\sum_{i=1}^k\frac{q}{p_i}\left((\mathbf{d}_i+\mathbf{t}'_i)\cdot \mathbf{x}_{J_i}+(d_i x_{i,\pi_i^{c'}(m_i-n_i)}+t_i' x_{i,\pi_i^{c'}(1)})\right).$$
 Let $\mathbf{c}=(\mathbf{c}_1,\mathbf{c}_2,\hdots,\mathbf{c}_k)$ and $\mathbf{c}'=(\mathbf{c}_1',\mathbf{c}_2',\hdots,\mathbf{c}_k')$, where $\mathbf{c}_i,\mathbf{c}_i'\in\mathbf{Z}_{p_i}^{n_i}$, $i=1,2,\hdots,k$, which says, $\mathbf{c},\mathbf{c}'\in \calV_{L'}$.
Now 
\begin{equation}\label{ankita1}
\begin{split}
	\psi(\mal{F}\arrowvert_{\mbf{x}_J=\mbf{c}})=\psi\left(\left(f+\sum_{i=1}^k\frac{q}{p_i}\left(d_i x_{i,\pi_i^{c}(m_i-n_i)}+t_i x_{i,\pi_i^{c}(1)}\right)\right)\Big\arrowvert_{\mbf{x}_J=\mbf{c}}\right)\prod_{i=1}^k\zeta_{p_i}^{(\mathbf{d}_i+\mathbf{t}_i)\cdot \mathbf{c}_i}
	\end{split}
\end{equation} 
Similarly, 
\begin{equation}
\begin{split}
\psi(\mal{F}'\arrowvert_{\mbf{x}_J=\mbf{c}'})=\psi\left(\left(f+\sum_{i=1}^k\frac{q}{p_i}\left(d_i x_{i,\pi_i^{c'}(m_i-n_i)}+t_i' x_{i,\pi_i^{c'}(1)}\right)\right)\Big\arrowvert_{\mbf{x}_J=\mbf{c}'}\right)\prod_{i=1}^k\zeta_{p_i}^{(\mathbf{d}_i+\mathbf{t}_i')\cdot \mathbf{c}_i'}
\end{split}
\end{equation} 
Now 
\begin{equation}
\begin{split}
\tta&\left(\psi(\mal{F}\arrowvert_{\mbf{x}_J=\mbf{c}}),\psi(\mal{F}'\arrowvert_{\mbf{x}_J=\mbf{c}'})\right)(\tau)\\&=\tta\left(\psi\left(\left(f+\sum_{i=1}^k\frac{q}{p_i}\left(d_i x_{i,\pi_i^{c}(m_i-n_i)}+t_i x_{i,\pi_i^{c}(1)}\right)\right)\Big\arrowvert_{\mbf{x}_J=\mbf{c}}\right),\right.\\&\left.\psi\left(\left(f+\sum_{i=1}^k\frac{q}{p_i}\left(d_i x_{i,\pi_i^{c'}(m_i-n_i)}+t_i' x_{i,\pi_i^{c'}(1)}\right)\right)\Big\arrowvert_{\mbf{x}_J=\mbf{c}'}\right)\right)(\tau)\prod_{i=1}^k\zeta_{p_i}^{\left(\mathbf{d}_i\cdot(\mathbf{c}_i-\mathbf{c}_i')\right)+\left((\mathbf{t}_i\cdot\mathbf{c}_i)-(\mathbf{t}_i'\cdot\mathbf{c}_i')\right)}
\end{split}
\end{equation}
%

For a non-negative integer $\tau$,
\begin{equation}\label{core1}
\begin{split}
\tta&(\psi(C_t),\psi(C_{t'}))(\tau)\\&=\sum_{\mbf{D}\in  \calV_{L'},\mbf{d}\in \calV_{L''} }\tta\left(\sum_{\mbf{c}\in \mal{D}'}\psi(\mal{F}\arrowvert_{\mbf{x}_J=\mbf{c}}),\sum_{\mbf{c}'\in \mal{D}'}\psi(\mal{F}'\arrowvert_{\mbf{x}_J=\mbf{c}'})\right)(\tau)\\
&=\sum_{\mbf{D}\in \calV_{L'},\mbf{d}\in \calV_{L''} \ccn}\sum_{\mbf{c},\mbf{c}'\in\calV_{L'}} 
\tta\left(\psi\left(\left(f+\sum_{i=1}^k\frac{q}{p_i}\left(d_i x_{i,\pi_i^{c}(m_i-n_i)}+t_i x_{i,\pi_i^{c}(1)}\right)\right)\Big\arrowvert_{\mbf{x}_J=\mbf{c}}\right),\right.\\&\left.\psi\left(\left(f+\sum_{i=1}^k\frac{q}{p_i}\left(d_i x_{i,\pi_i^{c'}(m_i-n_i)}+t_i' x_{i,\pi_i^{c'}(1)}\right)\right)\Big\arrowvert_{\mbf{x}_J=\mbf{c}'}\right)\right)(\tau)\prod_{j=1}^k\zeta_{p_j}^{\left(\mathbf{d}_j\cdot(\mathbf{c}_j-\mathbf{c}_j')\right)+\left((\mathbf{t}_j\cdot\mathbf{c}_j)-(\mathbf{t}_j'\cdot\mathbf{c}_j')\right)},
\end{split}
\end{equation}	
where $\mbf{D}=(\mbf{d}_1,\mbf{d}_2,\hdots,\mbf{d}_k)$, $\mbf{d}=(d_1,d_2,\hdots,d_k)$, and $\mal{D}''=\mbf{Z}_{p_1}\times \mbf{Z}_{p_2}\times\cdots\times \mbf{Z}_{p_k}$. 
From (\ref{core1}), we have 
\begin{equation}\label{core2}
\begin{split}
\tta&(\psi(C_t),\psi(C_{t'}))(\tau)\\&=\sum_{\mbf{d}\in \calV_{L''} }\sum_{\mbf{c},\mbf{c}'\in\calV_{L'}} 
\tta\left(\psi\left(\left(f+\sum_{i=1}^k\frac{q}{p_i}\left(d_i x_{i,\pi_i^{c}(m_i-n_i)}+t_i x_{i,\pi_i^{c}(1)}\right)\right)\Big\arrowvert_{\mbf{x}_J=\mbf{c}}\right),\right.\\&\left.\psi\left(\left(f+\sum_{i=1}^k\frac{q}{p_i}\left(d_i x_{i,\pi_i^{c'}(m_i-n_i)}+t_i' x_{i,\pi_i^{c'}(1)}\right)\right)\Big\arrowvert_{\mbf{x}_J=\mbf{c}'}\right)\right)(\tau)\sum_{\mbf{D}\in \calV_{L'}}\prod_{j=1}^k\zeta_{p_j}^{\left(\mathbf{d}_j\cdot(\mathbf{c}_j-\mathbf{c}_j')\right)+\left((\mathbf{t}_j\cdot\mathbf{c}_j)-(\mathbf{t}_j'\cdot\mathbf{c}_j')\right)}\\
&=|\calV_{L'}|\sum_{\mbf{d}\in \calV_{L''} }\sum_{\mbf{c}\in\calV_{L'}} 
\tta\left(\psi\left(\left(f+\sum_{i=1}^k\frac{q}{p_i}\left(d_i x_{i,\pi_i^{c}(m_i-n_i)}+t_i x_{i,\pi_i^{c}(1)}\right)\right)\Big\arrowvert_{\mbf{x}_J=\mbf{c}}\right),\right.\\&\left.\psi\left(\left(f+\sum_{i=1}^k\frac{q}{p_i}\left(d_i x_{i,\pi_i^{c}(m_i-n_i)}+t_i' x_{i,\pi_i^{c}(1)}\right)\right)\Big\arrowvert_{\mbf{x}_J=\mbf{c}}\right)\right)(\tau) \prod_{j=1}^k\zeta_{p_j}^{(\mathbf{t}_j-\mathbf{t}_j')\cdot\mathbf{c}_j}\\
&=|\calV_{L'}|\dis\sum_{\mbf{c}\in\calV_{L'}} \Theta(\psi(\mal{S}_u^{\mbf{c}},\mal{S}_{u'}^{\mbf{c}}))(\tau)
\prod_{j=1}^k\zeta_{p_j}^{(\mathbf{t}_j-\mathbf{t}_j')\cdot\mathbf{c}_j}.
\end{split}
\end{equation}
In Theorem 1, we have
\begin{equation}\label{core3}
\begin{split}
\Theta(\psi(\mal{S}_u^{\mbf{c}}),\psi(\mal{S}_{u'}^{\mbf{c}}))(\tau)=\sum_{\mbf{d}\in \calV_{L''} } 
&\tta\left(\psi\left(\left(f+\sum_{i=1}^k\frac{q}{p_i}\left(d_i x_{i,\pi_i^{c}(m_i-n_i)}+t_i x_{i,\pi_i^{c}(1)}\right)\right)\Big\arrowvert_{\mbf{x}_J=\mbf{c}}\right),\right.\\&~~~~~ \left.\psi\left(\left(f+\sum_{i=1}^k\frac{q}{p_i}\left(d_i x_{i,\pi_i^{c}(m_i-n_i)}+t_i' x_{i,\pi_i^{c}(1)}\right)\right)\Big\arrowvert_{\mbf{x}_J=\mbf{c}}\right)\right)(\tau)\\
&=\begin{cases}
\dis\prod_{i=1}^k p_i^{m_i-n_i+1},& \tau=0, u=u',\\
0,& \textnormal{otherwise.}
\end{cases}
\end{split}
\end{equation}
From (\ref{core2}) and (\ref{core3}), we have
\begin{equation}\label{corefi}
\begin{split}
\tta(\psi(C_t),\psi(C_{t'}))(\tau)
&=\begin{cases}
|\calV_{L'}|^2 \prod_{i=1}^k	p_i^{m_i-n_i+1}, & \mbf{t}_i=\mbf{t}_i',~ t_i=t_i', ~i=1,2,\hdots,k,\\
0,& \text{otherwise,}
\end{cases}\\
&=\begin{cases}
 \prod_{i=1}^k	p_i^{m_i+n_i+1}, & t=t',\\
0,& \text{otherwise},
\end{cases}
\end{split}
\end{equation}
where $u=u'\implies t_i=t_i'$, $i=1,2,\hdots,k$. 
Therefore the code set $\mal{C}$ forms $(K,L)$-CCC.
\ccb
%
%
%
%
%
%
	\end{IEEEproof}	
	\begin{example}
	Let us consider the same function as defined in Example \ref{ex2folded1} to illustrate Theorem \ref{teorem2}.
	Then following (\ref{mainset1}), we derive the following set of $6$-ary functions.
	\begin{equation}
	\begin{split}
	C_t&=\left\{f+\sum_{i=1}^2\frac{q}{p_i}\left((\mathbf{d}_i\cdot \mathbf{x}_{J_i}+d_i x_{i,\pi_i^{c}(m_i-n_i)})+(\mathbf{t}_i\cdot \mathbf{x}_{J_i}+t_i x_{i,\pi_i^{c}(1)})\right): \mathbf{d}_i\in \mathbf{Z}_{p_i}^{n_i},d_i\in\mathbf{Z}_{p_i},\right.\\&\left. i=1,2\right\}\\
	&=\left\{f+3\left((d_{1,1}x_{1,2}+d_1 x_{1,3})+3(t_{1,1}x_{1,2}+t_1 x_{1,1})+2(d_2x_{2,2}+t_2x_{2,1})\right):d_{1,1},d_1\in\mathbf{Z}_{2},d_2\in\mbf{Z}_3\right\},
	\end{split}
	\end{equation}
	where $\mbf{t}_1=t_{1,1},~t_1\in\mbf{Z}_2$, $\mbf{d}_1=d_{1,1}$, and as $n_2=0$, $\mbf{d}_2,~\mbf{t}_2\in\emptyset$, $t_2\in \mbf{Z}_3$, and $t=3(2t_{1,1}+t_1)+t_2$. The $\mbf{Z}_6$-valued codes $C_0,C_1,\hdots,C_{11}$ are presented in Table \ref{ankita200}, Table \ref{ankita201}, Table \ref{ankita202}, and Table \ref{ankita203}. \ccn
	The AACFs and ACCFs of the codes in $\mal{C}=\{\psi(C_t):0\leq t< 12\}$, follow the correlation plot as given in Figure \ref{ankut111}. Then $\mal{C}$ forms $(12,72)$-CCC over $\mbf{Z}_6$ and it follows the proposed Theorem \ref{teorem2}.
	\begin{figure}[H]
	\includegraphics[width=10cm]{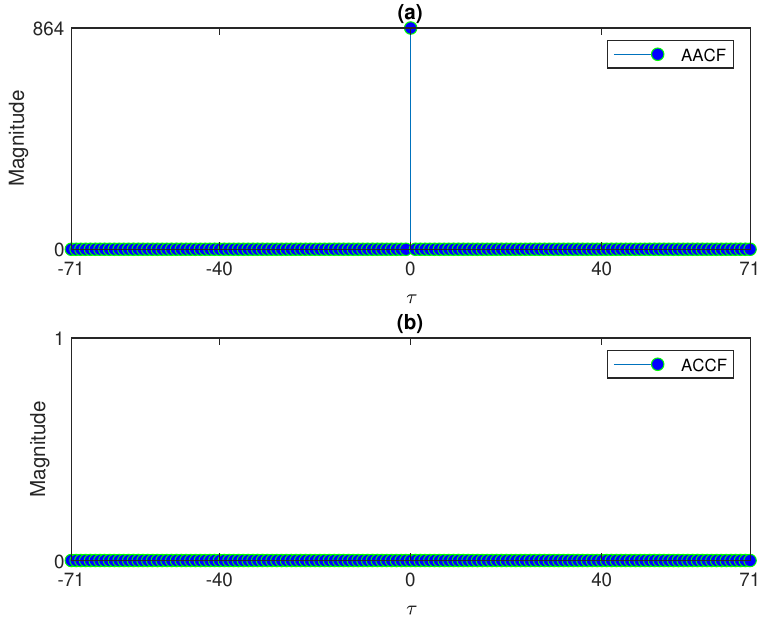}
	\caption{Correlation Plotting for the codes in $\mathcal{C}$}
	\label{ankut111}
	\end{figure}
	\end{example}
\begin{corollary}
	The relationship between the SNC CC $\psi(\mal{S}_u^{\mbf{c}})$ and the traditional CC $\psi(C_t)$ can be deduced from (\ref{ankit2}) and (\ref{ankita1}) in the following form:
	\begin{equation}
	\psi(C_t)=\left\{\sum_{\mbf{c}\in \mal{D}} \psi(\mal{S}_u^\mbf{c})\prod_{i=1}^k \zeta_{p_i}^{(\mbf{d}_i+\mbf{t}_i)\cdot \mbf{c}_i}:\mbf{d}_i\in \mbf{Z}_{p_i}^{n_i}~i=1,2,\hdots,k\right\}.
	\end{equation} 
\end{corollary}
The above equation establishes a connection between the proposed SNC-CCC introduced in Theorem \ref{teorem1} and the traditional CCC presented in Theorem \ref{teorem2}.

In Theorem \ref{teorem2}, a variety of outcomes are obtained, some of which encompass existing constructions. The following remark delves into these aspects in depth:

In the following remarks, we compare our proposed results with existing related results.
	\begin{remark}[Comparison with \cite{Davis1999, pater2000, rati}]
		In Theorem \ref{teorem2}, the variable $K$ signifies the count of CCs within the set of CCCs, including their constituent sequences, while $L$ represents the sequence length. Additionally, $\mathbb{Z}_q$ denotes the alphabet set. We can revisit the expressions for $K$, $L$, and $q$ as follows:
		\[
		K=\prod_{i=1}^k p_i^{n_i+1},~L=\prod_{i=1}^kp_i^{m_i},~q=\prod_{i=1}^k p_i.
		\]
		 For $k=1$ and $p_1=2$, the parameters can be expressed as:
		\[
		K=2^{n_1+1},~L=2^{m_1},~q=2,
		\] 
		underscoring that the findings reported in \cite{Davis1999, pater2000, rati} emerge as special cases of Theorem \ref{teorem2}, where \cite{Davis1999, pater2000, rati} introduce the construction of GCPs, CCs, and CCCs with the parameters as mentioned above. It is to be noted here that our proposed construction also works if we choose $q$ to be a positive integer such that $p_i|q$, $i=1,2,\hdots,k$. As we are interested in small alphabet, throughout the paper we fix $q$ to its minimum value that is $\prod_{i=1}^k p_i$. 
	\end{remark} 
\begin{remark}
In \cite{wang2020new}, Wang \emph{et~al} proposed a PU matrix based construction of CCCs, followed by a systematic approach to
extract $p$-ary functions, for prime $p$, as the generators for CCCs of non-power-of-two lengths.
The construction method in \cite{wang2020new} needs to perform the following steps {(unlike our proposed one)}
to generate CCCs:
\begin{itemize}
\item It needs to follow a generic framework for constructing desired PU matrices.
\item A series of operations need to be performed to extract multivariable functions from the constructed PU matrices.
\end{itemize}
In \cite[Th. 4]{wangong}, Wang \emph{et~al} proposed a construction of $(p,p,p^m)$-CCC using multivariable functions which are obtained through the
	following steps:
	\begin{itemize}
	\item First, construct a Butson-type Hadamard (BH) matrix of order $p^n$ with the help of $p$-ary sequences with two-level autocorrelation of period $p^n-1$.
	\item Then, construct $\delta$-quadratic terms with the help of the constructed BH matrix of order $p^n$.
	\item Finally, with the help of $\delta$-quadratic terms and suitable permutation polynomials, multivariable functions are obtained to
	generate $(p,p,p^m)$-CCC over the alphabet $\mathbb{Z}_p$.
	\end{itemize}
	Therefore, the afore-mentioned constructions may not be suitable for rapid hardware generations of
	CCCs. Besides, the proposed construction generates more flexible parameters as compared to \cite{wang2020new} and  \cite{wangong}.

\end{remark}
\begin{remark}
In the context of the current constructions, the importance of defining the domain and co-domain of a function becomes apparent when considering the alphabet size of the resultant sequences. This study characterizes the domain as $\calV_L \subset \mathbb{Z}_q^m$ and the co-domain as $\mathbb{Z}_q$. This configuration generates sequences of length $L$ over the alphabet $\mathbb{Z}_q$, which is determined by the prime factors of $L$. Notably, when co-primes values for $m_i \neq m_j$ occur for some specific indices $i \neq j \in \{1, 2, \ldots, k\}$, the proposed construction leads to the creation of CCCs featuring the minimal alphabet $q=\prod_{i=1}^k p_i$. This outcome deviates from existing constructions, particularly in cases where $k \geq 2$. For instance, in a recent work \cite{shen2023}, the authors adopt a domain $\mathbb{Z}_q^m$, where $m \geq 2$, and a co-domain of $\mathbb{Z}_q$, yielding $(q^{n+1}, q^m)$-CCCs with an alphabet size of $q$, where $0\leq n\leq m-1$. When dealing with scenarios where $L = \prod_{i=1}^k p_i^{m_i}$ and some of  $m_1,m_2,\hdots,m_k$ are co-primes to each-other, \cite{shen2023} can also generate a sequence length of $L$ by setting $m = 1$ and $q = L$. However, this yields both the alphabet size and set size as $L$, which is relatively larger than our outcomes. In practical applications, CCCs characterized by modest set and alphabet sizes tend to outperform those with larger dimensions, owing to considerations like the Peak-to-Mean Envelope Power Ratio (PMEPR) of constituent sequences \cite{pater2000, Davis1999}. This is where our proposed construction showcases its strengths, achieving superior performance when scrutinized through vital parameters such as alphabet and set size. 
\end{remark}
In Table \ref{cmptle}, we also compare our proposed construction with the existing constructions of CCCs \cite{rati,liumc,chencommlett,shen2023}.
\begin{table*}[h!]\small
	\centering
	\caption{Comparison with \cite{rati,liumc,chencommlett,shen2023}}\label{cmptle}
	\resizebox{\textwidth}{!}{
		\begin{tabular}{ |c|c|c|c|c|c| }
				\hline
				Sequence Class&Type &Set size ($K$) & Length ($L$)&Alphabet size&Constraints \\\hline\hline
				\cite{rati}&Traditional&$2^{k+1}$&$2^m$&$q$& $m\geq 1$, $k\geq 0$, $q$ is even \\\hline
				 \cite{liumc}&Traditional&$2^{k+1}$&$2^m$&$q$& $m\geq 1$, $k\geq 0$,$q$ is even\\\hline
				 \cite{chencommlett}&Traditional&$2^{k+1}$& $2^m$&$q$& $m\geq 1$, $k\geq 0$, $q$ is even\\\hline
			 \cite[Thm. 2]{shen2023}        &Traditional& $q^{k+1}$ & $q^m$&$q$ &$m\geq 1$, $k\geq 0$, $q~(\geq 2)$ is integer \\\hline
				\cite[Thm. 3]{shen2023}&SNC        & $q^{k+1}$ & $U(q^m-1)+V(q-1)+1$ &$q+1$& $m\geq 1$, $\ccr k\geq 0 \ccn$, $U,V\geq 1$, $q~(\geq 2)$ is integer\\\hline
				
			 \textit{Theorem 1}&SNC         &$\prod_{i=1}^kp_i$ & $\prod_{i=1}^kp_i^{m_i}$&$q+1$& $q=\prod_{i=1}^k p_i$, $m_i\geq 2$\\\hline
			 \textit{Theorem 2}&Traditional& $\prod_{i=1}^k p_i^{n_i+1}$ & $\prod_{i=1}^kp_i^{m_i}$& $q$&$q=\prod_{i=1}^k p_i$, $m_i\geq 2$, $n_i\geq 0$, $p_i$ is a prime number \\\hline
	\end{tabular}}
\end{table*}

\section{Conclusion}
In this paper, we have firstly proposed $q$-ary function based construction of SNC-CCCs of length $\prod_{i=1}^k p_i^{m_i}$, set sizes $\prod_{i=1}^k p_i$, and alphabet size is $q+1$, where $q=\prod_{i=1}^k p_i$. Then the idea of SNC-CCCs has been extended to 
traditional CCCs of lengths and alphabet size same as of SNC-CCCs, set size  $\prod_{i=1}^k p_i^{n_i+1}$, and the alphabet size is $q$. The proposed construction is able maintain a low aplphabet size for a given set of other parameters as compared to the existing works of CCCs, specially, when $k\geq 2$ and some of $m_1,m_2,\hdots,m_k$ are co-prime to each other. A detailed connection has also been established between the proposed $q$-ary functions and the undirected graphs.  
\appendices
\section{Table \ref{ankut2}, \ref{ankita200}, \ref{ankita201}, \ref{ankita202}, \ref{ankita203}}
\begin{table}[H]
	\caption{$(5,72)$-SNC-CCC from the $6$-ary function of (\ref{ankby1})}\label{ankut2}
	\resizebox{\columnwidth}{!}{	\begin{tabular}{|l|llllll|}
			\hline
			$\psi(\mal{S}_0^0)$  & \multicolumn{6}{l|}{\begin{tabular}[c]{@{}l@{}}$\zeta_6^2\zeta_6^2\zeta_6^2\zeta_6^2\zeta_6^4\zeta_6^0\zeta_6^2\zeta_6^0\zeta_6^4\zeta_6^2\zeta_6^2\zeta_6^2\zeta_6^2\zeta_6^4\zeta_6^0
					\zeta_6^2\zeta_6^0\zeta_6^4\mbf{0}_{18}\zeta_6^2\zeta_6^2\zeta_6^2\zeta_6^2\zeta_6^4\zeta_6^0\zeta_6^2\zeta_6^0\zeta_6^4\zeta_6^5\zeta_6^5\zeta_6^5\zeta_6^5\zeta_6^1\zeta_6^3\zeta_6^5
					\zeta_6^3\zeta_6^1\mbf{0}_{18}$\\
					$\zeta_6^2\zeta_6^4\zeta_6^0\zeta_6^2 \zeta_6^0\zeta_6^4\zeta_6^2\zeta_6^2\zeta_6^2\zeta_6^2\zeta_6^4\zeta_6^0\zeta_6^2\zeta_6^0\zeta_6^4\zeta_6^2\zeta_6^2\zeta_6^2\mbf{0}_{18}\zeta_6^2\zeta_6^4\zeta_6^0\zeta_6^2\zeta_6^0\zeta_6^4\zeta_6^2\zeta_6^2\zeta_6^2\zeta_6^5\zeta_6^1\zeta_6^3\zeta_6^5\zeta_6^3\zeta_6^1\zeta_6^5\zeta_6^5\zeta_6^5\mbf{0}_{18}$\\
					$\zeta_6^2\zeta_6^0\zeta_6^4\zeta_6^2\zeta_6^2\zeta_6^2\zeta_6^2\zeta_6^4\zeta_6^0\zeta_6^2\zeta_6^0\zeta_6^4\zeta_6^2\zeta_6^2\zeta_6^2\zeta_6^2\zeta_6^4\zeta_6^0\mbf{0}_{18} \zeta_6^2\zeta_6^0\zeta_6^4\zeta_6^2\zeta_6^2\zeta_6^2\zeta_6^2\zeta_6^4\zeta_6^0\zeta_6^5\zeta_6^3\zeta_6^1\zeta_6^5\zeta_6^5\zeta_6^5\zeta_6^5\zeta_6^1\zeta_6^3\mbf{0}_{18}$\\ $\zeta_6^2\zeta_6^2\zeta_6^2\zeta_6^2\zeta_6^4\zeta_6^0\zeta_6^2\zeta_6^0\zeta_6^4\zeta_6^5\zeta_6^5\zeta_6^5\zeta_6^5\zeta_6^1\zeta_6^3\zeta_6^5\zeta_6^3\zeta_6^1
					\mbf{0}_{18}\zeta_6^2\zeta_6^2\zeta_6^2\zeta_6^2\zeta_6^4\zeta_6^0\zeta_6^2\zeta_6^0\zeta_6^4\zeta_6^2\zeta_6^2\zeta_6^2\zeta_6^2\zeta_6^4\zeta_6^0\zeta_6^2\zeta_6^0\zeta_6^4\mbf{0}_{18}$\\ $\zeta_6^2\zeta_6^4\zeta_6^0\zeta_6^2\zeta_6^0\zeta_6^4\zeta_6^2\zeta_6^2\zeta_6^2\zeta_6^5\zeta_6^1\zeta_6^3\zeta_6^5\zeta_6^3\zeta_6^1\zeta_6^5\zeta_6^5\zeta_6^5\mbf{0}_{18}\zeta_6^2\zeta_6^4\zeta_6^0\zeta_6^2\zeta_6^0\zeta_6^4\zeta_6^2\zeta_6^2\zeta_6^2\zeta_6^2\zeta_6^4\zeta_6^0\zeta_6^2\zeta_6^0\zeta_6^4\zeta_6^2\zeta_6^2\zeta_6^2\mbf{0}_{18}$\\ $\zeta_6^2\zeta_6^0\zeta_6^4\zeta_6^2\zeta_6^2\zeta_6^2\zeta_6^2\zeta_6^4\zeta_6^0\zeta_6^5\zeta_6^3\zeta_6^1\zeta_6^5\zeta_6^5\zeta_6^5\zeta_6^5\zeta_6^1\zeta_6^3\mbf{0}_{18}\zeta_6^2\zeta_6^0\zeta_6^4\zeta_6^2\zeta_6^2\zeta_6^2\zeta_6^2\zeta_6^4\zeta_6^0\zeta_6^2\zeta_6^0\zeta_6^4\zeta_6^2\zeta_6^2\zeta_6^2\zeta_6^2\zeta_6^4\zeta_6^0\mbf{0}_{18}$\end{tabular}}\\
			\hline
			$\psi(\mal{S}_1^0)$ &                  \multicolumn{6}{l|}{\begin{tabular}[c]{@{}l@{}}
					$\zeta_6^2\zeta_6^2\zeta_6^2\zeta_6^4\zeta_6^0\zeta_6^2\zeta_6^0\zeta_6^4\zeta_6^2\zeta_6^2\zeta_6^2\zeta_6^2\zeta_6^4\zeta_6^0\zeta_6^2\zeta_6^0\zeta_6^4\zeta_6^2\mbf{0}_{18}\zeta_6^2\zeta_6^2\zeta_6^2\zeta_6^4\zeta_6^0\zeta_6^2\zeta_6^0\zeta_6^4\zeta_6^2\zeta_6^5\zeta_6^5\zeta_6^5\zeta_6^1\zeta_6^3\zeta_6^5\zeta_6^3\zeta_6^1\zeta_6^5\mbf{0}_{18}$\\ $\zeta_6^2\zeta_6^4\zeta_6^0\zeta_6^4\zeta_6^2\zeta_6^0\zeta_6^0\zeta_6^0\zeta_6^0\zeta_6^2\zeta_6^4\zeta_6^0\zeta_6^4\zeta_6^2\zeta_6^0\zeta_6^0\zeta_6^0\zeta_6^0\mbf{0}_{18}\zeta_6^2\zeta_6^4\zeta_6^0\zeta_6^4\zeta_6^2\zeta_6^0\zeta_6^0\zeta_6^0\zeta_6^0\zeta_6^5\zeta_6^1\zeta_6^3\zeta_6^1\zeta_6^5\zeta_6^3\zeta_6^3\zeta_6^3\zeta_6^3\mbf{0}_{18}$\\ $\zeta_6^2\zeta_6^0\zeta_6^4\zeta_6^4\zeta_6^4\zeta_6^4\zeta_6^0\zeta_6^2\zeta_6^4\zeta_6^2\zeta_6^0\zeta_6^4\zeta_6^4\zeta_6^4\zeta_6^4\zeta_6^0\zeta_6^2\zeta_6^4\mbf{0}_{18}\zeta_6^2\zeta_6^0\zeta_6^4\zeta_6^4\zeta_6^4\zeta_6^4\zeta_6^0\zeta_6^2\zeta_6^4\zeta_6^5\zeta_6^3\zeta_6^1\zeta_6^1\zeta_6^1\zeta_6^1\zeta_6^3\zeta_6^5\zeta_6^1\mbf{0}_{18}$\\ $\zeta_6^2\zeta_6^2\zeta_6^2\zeta_6^4\zeta_6^0\zeta_6^2\zeta_6^0\zeta_6^4\zeta_6^2\zeta_6^5\zeta_6^5\zeta_6^5\zeta_6^1\zeta_6^3\zeta_6^5\zeta_6^3\zeta_6^1\zeta_6^5\mbf{0}_{18}\zeta_6^2\zeta_6^2\zeta_6^2\zeta_6^4\zeta_6^0\zeta_6^2\zeta_6^0\zeta_6^4\zeta_6^2\zeta_6^2\zeta_6^2\zeta_6^2\zeta_6^4\zeta_6^0\zeta_6^2\zeta_6^0\zeta_6^4\zeta_6^2\mbf{0}_{18}$\\ $\zeta_6^2\zeta_6^4\zeta_6^0\zeta_6^4\zeta_6^2\zeta_6^0\zeta_6^0\zeta_6^0\zeta_6^0\zeta_6^5\zeta_6^1\zeta_6^3\zeta_6^1\zeta_6^5\zeta_6^3\zeta_6^3\zeta_6^3\zeta_6^3\mbf{0}_{18}\zeta_6^2\zeta_6^4\zeta_6^0\zeta_6^4\zeta_6^2\zeta_6^0\zeta_6^0\zeta_6^0\zeta_6^0\zeta_6^2\zeta_6^4\zeta_6^0\zeta_6^4\zeta_6^2\zeta_6^0\zeta_6^0\zeta_6^0\zeta_6^0\mbf{0}_{18}$\\ $\zeta_6^2\zeta_6^0\zeta_6^4\zeta_6^4\zeta_6^4\zeta_6^4\zeta_6^0\zeta_6^2\zeta_6^4\zeta_6^5\zeta_6^3\zeta_6^1\zeta_6^1\zeta_6^1\zeta_6^1\zeta_6^3\zeta_6^5\zeta_6^1\mbf{0}_{18}\zeta_6^2\zeta_6^0\zeta_6^4\zeta_6^4\zeta_6^4\zeta_6^4\zeta_6^0\zeta_6^2\zeta_6^4\zeta_6^2\zeta_6^0\zeta_6^4\zeta_6^4\zeta_6^4\zeta_6^4\zeta_6^0\zeta_6^2\zeta_6^4\mbf{0}_{18}$\end{tabular}}                  \\ \hline
			$\psi(\mal{S}_2^0)$ & \multicolumn{6}{l|}{\begin{tabular}[c]{@{}l@{}}
					$\zeta_6^2\zeta_6^2\zeta_6^2\zeta_6^0\zeta_6^2\zeta_6^4\zeta_6^4\zeta_6^2\zeta_6^0\zeta_6^2\zeta_6^2\zeta_6^2\zeta_6^0\zeta_6^2\zeta_6^4\zeta_6^4\zeta_6^2\zeta_6^0\mbf{0}_{18}\zeta_6^2\zeta_6^2\zeta_6^2\zeta_6^0\zeta_6^2\zeta_6^4\zeta_6^4\zeta_6^2\zeta_6^0\zeta_6^5\zeta_6^5\zeta_6^5\zeta_6^3\zeta_6^5\zeta_6^1\zeta_6^1\zeta_6^5\zeta_6^3\mbf{0}_{18}$\\ $\zeta_6^2\zeta_6^4\zeta_6^0\zeta_6^0\zeta_6^4\zeta_6^2\zeta_6^4\zeta_6^4\zeta_6^4\zeta_6^2\zeta_6^4\zeta_6^0\zeta_6^0\zeta_6^4\zeta_6^2\zeta_6^4\zeta_6^4\zeta_6^4\mbf{0}_{18}\zeta_6^2\zeta_6^4\zeta_6^0\zeta_6^0\zeta_6^4\zeta_6^2\zeta_6^4\zeta_6^4\zeta_6^4\zeta_6^5\zeta_6^1\zeta_6^3\zeta_6^3\zeta_6^1\zeta_6^5\zeta_6^1\zeta_6^1\zeta_6^1\mbf{0}_{18}$\\ $\zeta_6^2\zeta_6^0\zeta_6^4\zeta_6^0\zeta_6^0\zeta_6^0\zeta_6^4\zeta_6^0\zeta_6^2\zeta_6^2\zeta_6^0\zeta_6^4\zeta_6^0\zeta_6^0\zeta_6^0\zeta_6^4\zeta_6^0\zeta_6^2\mbf{0}_{18}\zeta_6^2\zeta_6^0\zeta_6^4\zeta_6^0\zeta_6^0\zeta_6^0\zeta_6^4\zeta_6^0\zeta_6^2\zeta_6^5\zeta_6^3\zeta_6^1\zeta_6^3\zeta_6^3\zeta_6^3\zeta_6^1\zeta_6^3\zeta_6^5\mbf{0}_{18}$\\ $\zeta_6^2\zeta_6^2\zeta_6^2\zeta_6^0\zeta_6^2\zeta_6^4\zeta_6^4\zeta_6^2\zeta_6^0\zeta_6^5\zeta_6^5\zeta_6^5\zeta_6^3\zeta_6^5\zeta_6^1\zeta_6^1\zeta_6^5\zeta_6^3\mbf{0}_{18}\zeta_6^2\zeta_6^2\zeta_6^2\zeta_6^0\zeta_6^2\zeta_6^4\zeta_6^4\zeta_6^2\zeta_6^0\zeta_6^2\zeta_6^2\zeta_6^2\zeta_6^0\zeta_6^2\zeta_6^4\zeta_6^4\zeta_6^2\zeta_6^0\mbf{0}_{18}$\\ $\zeta_6^2\zeta_6^4\zeta_6^0\zeta_6^0\zeta_6^4\zeta_6^2\zeta_6^4\zeta_6^4\zeta_6^4\zeta_6^5\zeta_6^1\zeta_6^3\zeta_6^3\zeta_6^1\zeta_6^5\zeta_6^1\zeta_6^1\zeta_6^1\mbf{0}_{18}\zeta_6^2\zeta_6^4\zeta_6^0\zeta_6^0\zeta_6^4\zeta_6^2\zeta_6^4\zeta_6^4\zeta_6^4\zeta_6^2\zeta_6^4\zeta_6^0\zeta_6^0\zeta_6^4\zeta_6^2\zeta_6^4\zeta_6^4\zeta_6^4\mbf{0}_{18}$\\ $\zeta_6^2\zeta_6^0\zeta_6^4\zeta_6^0\zeta_6^0\zeta_6^0\zeta_6^4\zeta_6^0\zeta_6^2\zeta_6^5\zeta_6^3\zeta_6^1\zeta_6^3\zeta_6^3\zeta_6^3\zeta_6^1\zeta_6^3\zeta_6^5\mbf{0}_{18}\zeta_6^2\zeta_6^0\zeta_6^4\zeta_6^0\zeta_6^0\zeta_6^0\zeta_6^4\zeta_6^0\zeta_6^2\zeta_6^2\zeta_6^0\zeta_6^4\zeta_6^0\zeta_6^0\zeta_6^0\zeta_6^4\zeta_6^0\zeta_6^2\mbf{0}_{18}$
			\end{tabular}}\\ \hline
			$\psi(\mal{S}_3^0)$& \multicolumn{6}{l|}{\begin{tabular}[c]{@{}l@{}}
					$\zeta_6^2\zeta_6^2\zeta_6^2\zeta_6^2\zeta_6^4\zeta_6^0\zeta_6^2\zeta_6^0\zeta_6^4\zeta_6^2\zeta_6^2\zeta_6^2\zeta_6^2\zeta_6^4\zeta_6^0\zeta_6^2\zeta_6^0\zeta_6^4\mbf{0}_{18}\zeta_6^5\zeta_6^5\zeta_6^5\zeta_6^5\zeta_6^1\zeta_6^3\zeta_6^5\zeta_6^3\zeta_6^1\zeta_6^2\zeta_6^2\zeta_6^2\zeta_6^2\zeta_6^4\zeta_6^0\zeta_6^2\zeta_6^0\zeta_6^4\mbf{0}_{18}$\\ $\zeta_6^2\zeta_6^4\zeta_6^0\zeta_6^2\zeta_6^0\zeta_6^4\zeta_6^2\zeta_6^2\zeta_6^2\zeta_6^2\zeta_6^4\zeta_6^0\zeta_6^2\zeta_6^0\zeta_6^4\zeta_6^2\zeta_6^2\zeta_6^2\mbf{0}_{18}\zeta_6^5\zeta_6^1\zeta_6^3\zeta_6^5\zeta_6^3\zeta_6^1\zeta_6^5\zeta_6^5\zeta_6^5\zeta_6^2\zeta_6^4\zeta_6^0\zeta_6^2\zeta_6^0\zeta_6^4\zeta_6^2\zeta_6^2\zeta_6^2\mbf{0}_{18}$\\ $\zeta_6^2\zeta_6^0\zeta_6^4\zeta_6^2\zeta_6^2\zeta_6^2\zeta_6^2\zeta_6^4\zeta_6^0\zeta_6^2\zeta_6^0\zeta_6^4\zeta_6^2\zeta_6^2\zeta_6^2\zeta_6^2\zeta_6^4\zeta_6^0\mbf{0}_{18}\zeta_6^5\zeta_6^3\zeta_6^1\zeta_6^5\zeta_6^5\zeta_6^5\zeta_6^5\zeta_6^1\zeta_6^3\zeta_6^2\zeta_6^0\zeta_6^4\zeta_6^2\zeta_6^2\zeta_6^2\zeta_6^2\zeta_6^4\zeta_6^0\mbf{0}_{18}$\\ $\zeta_6^2\zeta_6^2\zeta_6^2\zeta_6^2\zeta_6^4\zeta_6^0\zeta_6^2\zeta_6^0\zeta_6^4\zeta_6^5\zeta_6^5\zeta_6^5\zeta_6^5\zeta_6^1\zeta_6^3\zeta_6^5\zeta_6^3\zeta_6^1\mbf{0}_{18}\zeta_6^5\zeta_6^5\zeta_6^5\zeta_6^5\zeta_6^1\zeta_6^3\zeta_6^5\zeta_6^3\zeta_6^1\zeta_6^5\zeta_6^5\zeta_6^5\zeta_6^5\zeta_6^1\zeta_6^3\zeta_6^5\zeta_6^3\zeta_6^1\mbf{0}_{18}$\\ $\zeta_6^2\zeta_6^4\zeta_6^0\zeta_6^2\zeta_6^0\zeta_6^4\zeta_6^2\zeta_6^2\zeta_6^2\zeta_6^5\zeta_6^1\zeta_6^3\zeta_6^5\zeta_6^3\zeta_6^1\zeta_6^5\zeta_6^5\zeta_6^5\mbf{0}_{18}\zeta_6^5\zeta_6^1\zeta_6^3\zeta_6^5\zeta_6^3\zeta_6^1\zeta_6^5\zeta_6^5\zeta_6^5\zeta_6^5\zeta_6^1\zeta_6^3\zeta_6^5\zeta_6^3\zeta_6^1\zeta_6^5\zeta_6^5\zeta_6^5\mbf{0}_{18}$\\ $\zeta_6^2\zeta_6^0\zeta_6^4\zeta_6^2\zeta_6^2\zeta_6^2\zeta_6^2\zeta_6^4\zeta_6^0\zeta_6^5\zeta_6^3\zeta_6^1\zeta_6^5\zeta_6^5\zeta_6^5\zeta_6^5\zeta_6^1\zeta_6^3\mbf{0}_{18}\zeta_6^5\zeta_6^3\zeta_6^1\zeta_6^5\zeta_6^5\zeta_6^5\zeta_6^5\zeta_6^1\zeta_6^3\zeta_6^5\zeta_6^3\zeta_6^1\zeta_6^5\zeta_6^5\zeta_6^5\zeta_6^5\zeta_6^1\zeta_6^3\mbf{0}_{18}$\end{tabular}}\\ \hline
			$\psi(\mal{S}_4^0)$                  & \multicolumn{6}{l|}{\begin{tabular}[c]{@{}l@{}}
					$\zeta_6^2\zeta_6^2\zeta_6^2\zeta_6^4\zeta_6^0\zeta_6^2\zeta_6^0\zeta_6^4\zeta_6^2\zeta_6^2\zeta_6^2\zeta_6^2\zeta_6^4\zeta_6^0
					\zeta_6^2\zeta_6^0\zeta_6^4\zeta_6^2\mbf{0}_{18}\zeta_6^5\zeta_6^5\zeta_6^5\zeta_6^1\zeta_6^3\zeta_6^5\zeta_6^3\zeta_6^1\zeta_6^5\zeta_6^2\zeta_6^2\zeta_6^2\zeta_6^4\zeta_6^0\zeta_6^2\zeta_6^0\zeta_6^4\zeta_6^2\mbf{0}_{18}$\\ $\zeta_6^2\zeta_6^4\zeta_6^0\zeta_6^4\zeta_6^2\zeta_6^0\zeta_6^0\zeta_6^0\zeta_6^0\zeta_6^2\zeta_6^4\zeta_6^0\zeta_6^4\zeta_6^2\zeta_6^0\zeta_6^0\zeta_6^0\zeta_6^0\mbf{0}_{18}\zeta_6^5\zeta_6^1\zeta_6^3\zeta_6^1\zeta_6^5\zeta_6^3\zeta_6^3\zeta_6^3\zeta_6^3\zeta_6^2\zeta_6^4\zeta_6^0\zeta_6^4\zeta_6^2\zeta_6^0\zeta_6^0\zeta_6^0\zeta_6^0\mbf{0}_{18}$\\ 
					$\zeta_6^2\zeta_6^0\zeta_6^4\zeta_6^4\zeta_6^4\zeta_6^4\zeta_6^0\zeta_6^2\zeta_6^4\zeta_6^2\zeta_6^0\zeta_6^4\zeta_6^4\zeta_6^4\zeta_6^4\zeta_6^0\zeta_6^2\zeta_6^4\mbf{0}_{18}\zeta_6^5\zeta_6^3\zeta_6^1\zeta_6^1\zeta_6^1\zeta_6^1\zeta_6^3\zeta_6^5\zeta_6^1\zeta_6^2\zeta_6^0\zeta_6^4\zeta_6^4\zeta_6^4\zeta_6^4\zeta_6^0\zeta_6^2\zeta_6^4\mbf{0}_{18}$\\ 
					$\zeta_6^2\zeta_6^2\zeta_6^2\zeta_6^4\zeta_6^0\zeta_6^2\zeta_6^0\zeta_6^4\zeta_6^2\zeta_6^5\zeta_6^5\zeta_6^5\zeta_6^1\zeta_6^3\zeta_6^5\zeta_6^3\zeta_6^1\zeta_6^5\mbf{0}_{18}\zeta_6^5\zeta_6^5\zeta_6^5\zeta_6^1\zeta_6^3\zeta_6^5\zeta_6^3\zeta_6^1\zeta_6^5\zeta_6^5\zeta_6^5\zeta_6^5\zeta_6^1\zeta_6^3\zeta_6^5\zeta_6^3\zeta_6^1\zeta_6^5\mbf{0}_{18}$\\ 
					$\zeta_6^2\zeta_6^4\zeta_6^0\zeta_6^4\zeta_6^2\zeta_6^0\zeta_6^0\zeta_6^0\zeta_6^0\zeta_6^5\zeta_6^1\zeta_6^3\zeta_6^1\zeta_6^5\zeta_6^3\zeta_6^3\zeta_6^3\zeta_6^3\mbf{0}_{18}\zeta_6^5\zeta_6^1\zeta_6^3\zeta_6^1\zeta_6^5\zeta_6^3\zeta_6^3\zeta_6^3\zeta_6^3\zeta_6^5\zeta_6^1\zeta_6^3\zeta_6^1\zeta_6^5\zeta_6^3\zeta_6^3\zeta_6^3\zeta_6^3\mbf{0}_{18}$\\ $\zeta_6^2\zeta_6^0\zeta_6^4\zeta_6^4\zeta_6^4\zeta_6^4\zeta_6^0\zeta_6^2\zeta_6^4\zeta_6^5\zeta_6^3\zeta_6^1\zeta_6^1\zeta_6^1\zeta_6^1\zeta_6^3\zeta_6^5\zeta_6^1\mbf{0}_{18}\zeta_6^5\zeta_6^3\zeta_6^1\zeta_6^1\zeta_6^1\zeta_6^1\zeta_6^3\zeta_6^5\zeta_6^1\zeta_6^5\zeta_6^3\zeta_6^1\zeta_6^1\zeta_6^1\zeta_6^1\zeta_6^3\zeta_6^5\zeta_6^1\mbf{0}_{18}$\end{tabular}}                  \\ \hline
			$\psi(\mal{S}_5^0)$                  & \multicolumn{6}{l|}{\begin{tabular}[c]{@{}l@{}}
					$\zeta_6^2\zeta_6^2\zeta_6^2\zeta_6^0\zeta_6^2\zeta_6^4\zeta_6^4\zeta_6^2\zeta_6^0\zeta_6^2\zeta_6^2\zeta_6^2\zeta_6^0\zeta_6^2\zeta_6^4\zeta_6^4\zeta_6^2\zeta_6^0\mbf{0}_{18}\zeta_6^5\zeta_6^5\zeta_6^5\zeta_6^3\zeta_6^5\zeta_6^1\zeta_6^1\zeta_6^5\zeta_6^3\zeta_6^2\zeta_6^2\zeta_6^2\zeta_6^0\zeta_6^2\zeta_6^4\zeta_6^4\zeta_6^2\zeta_6^0\mbf{0}_{18}$\\ $\zeta_6^2\zeta_6^4\zeta_6^0\zeta_6^0\zeta_6^4\zeta_6^2\zeta_6^4\zeta_6^4\zeta_6^4\zeta_6^2\zeta_6^4\zeta_6^0\zeta_6^0\zeta_6^4\zeta_6^2\zeta_6^4\zeta_6^4\zeta_6^4\mbf{0}_{18}\zeta_6^5\zeta_6^1\zeta_6^3\zeta_6^3\zeta_6^1\zeta_6^5\zeta_6^1\zeta_6^1\zeta_6^1\zeta_6^2\zeta_6^4\zeta_6^0\zeta_6^0\zeta_6^4\zeta_6^2\zeta_6^4\zeta_6^4\zeta_6^4\mbf{0}_{18}$\\ $\zeta_6^2\zeta_6^0\zeta_6^4\zeta_6^0\zeta_6^0\zeta_6^0\zeta_6^4\zeta_6^0\zeta_6^2\zeta_6^2\zeta_6^0\zeta_6^4\zeta_6^0\zeta_6^0\zeta_6^0\zeta_6^4\zeta_6^0\zeta_6^2\mbf{0}_{18}\zeta_6^5\zeta_6^3\zeta_6^1\zeta_6^3\zeta_6^3\zeta_6^3\zeta_6^1\zeta_6^3\zeta_6^5\zeta_6^2\zeta_6^0\zeta_6^4\zeta_6^0\zeta_6^0\zeta_6^0\zeta_6^4\zeta_6^0\zeta_6^2\mbf{0}_{18}$\\ $\zeta_6^2\zeta_6^2\zeta_6^2\zeta_6^0\zeta_6^2\zeta_6^4\zeta_6^4\zeta_6^2\zeta_6^0\zeta_6^5\zeta_6^5\zeta_6^5\zeta_6^3\zeta_6^5\zeta_6^1\zeta_6^1\zeta_6^5\zeta_6^3\mbf{0}_{18}\zeta_6^5\zeta_6^5\zeta_6^5\zeta_6^3\zeta_6^5\zeta_6^1\zeta_6^1\zeta_6^5\zeta_6^3\zeta_6^5\zeta_6^5\zeta_6^5\zeta_6^3\zeta_6^5\zeta_6^1\zeta_6^1\zeta_6^5\zeta_6^3\mbf{0}_{18}$\\ $\zeta_6^2\zeta_6^4\zeta_6^0\zeta_6^0\zeta_6^4\zeta_6^2\zeta_6^4\zeta_6^4\zeta_6^4\zeta_6^5\zeta_6^1\zeta_6^3\zeta_6^3\zeta_6^1\zeta_6^5\zeta_6^1\zeta_6^1\zeta_6^1\mbf{0}_{18}\zeta_6^5\zeta_6^1\zeta_6^3\zeta_6^3\zeta_6^1\zeta_6^5\zeta_6^1\zeta_6^1\zeta_6^1\zeta_6^5\zeta_6^1\zeta_6^3\zeta_6^3\zeta_6^1\zeta_6^5\zeta_6^1\zeta_6^1\zeta_6^1\mbf{0}_{18}$\\ $\zeta_6^2\zeta_6^0\zeta_6^4\zeta_6^0\zeta_6^0\zeta_6^0\zeta_6^4\zeta_6^0\zeta_6^2\zeta_6^5\zeta_6^3\zeta_6^1\zeta_6^3\zeta_6^3\zeta_6^3\zeta_6^1\zeta_6^3\zeta_6^5\mbf{0}_{18}\zeta_6^5\zeta_6^3\zeta_6^1\zeta_6^3\zeta_6^3\zeta_6^3\zeta_6^1\zeta_6^3\zeta_6^5\zeta_6^5\zeta_6^3\zeta_6^1\zeta_6^3\zeta_6^3\zeta_6^3\zeta_6^1\zeta_6^3\zeta_6^5\mbf{0}_{18}$\end{tabular}}                  \\ \hline
	\end{tabular}}
\end{table}
\begin{table}[H]
	\caption{Codes $C_0$, $C_1$, and $C_2$ over $\mbf{Z}_6$}\label{ankita200}
	\resizebox{\columnwidth}{!}{	\begin{tabular}{|l|llllll|}
			\hline
			$C_0$  & \multicolumn{6}{l|}{\begin{tabular}[c]{@{}l@{}}
					$2	2	2	2	4	0	2	0	4	2	2	2	2	4	0	2	0	4	3	4	5	4	1	4	5	4	3	1	2	3	2	5	2	3	2	1	2	2	2	2	4	0	2	0	4	5	5	5	5	1	3	5	3	1	5	0	1	0	3	0	1	0	5	0	1	2	1	4	1	2	1	0
					$\\
					$2	4	0	2	0	4	2	2	2	2	4	0	2	0	4	2	2	2	3	0	3	4	3	2	5	0	1	1	4	1	2	1	0	3	4	5	2	4	0	2	0	4	2	2	2	5	1	3	5	3	1	5	5	5	5	2	5	0	5	4	1	2	3	0	3	0	1	0	5	2	3	4
					$\\	
					$2	0	4	2	2	2	2	4	0	2	0	4	2	2	2	2	4	0	3	2	1	4	5	0	5	2	5	1	0	5	2	3	4	3	0	3	2	0	4	2	2	2	2	4	0	5	3	1	5	5	5	5	1	3	5	4	3	0	1	2	1	4	1	0	5	4	1	2	3	2	5	2
					$\\	
					$2	2	2	2	4	0	2	0	4	5	5	5	5	1	3	5	3	1	3	4	5	4	1	4	5	4	3	4	5	0	5	2	5	0	5	4	2	2	2	2	4	0	2	0	4	2	2	2	2	4	0	2	0	4	5	0	1	0	3	0	1	0	5	3	4	5	4	1	4	5	4	3
					$\\
					$2	4	0	2	0	4	2	2	2	5	1	3	5	3	1	5	5	5	3	0	3	4	3	2	5	0	1	4	1	4	5	4	3	0	1	2	2	4	0	2	0	4	2	2	2	2	4	0	2	0	4	2	2	2	5	2	5	0	5	4	1	2	3	3	0	3	4	3	2	5	0	1
					$\\
					$2	0	4	2	2	2	2	4	0	5	3	1	5	5	5	5	1	3	3	2	1	4	5	0	5	2	5	4	3	2	5	0	1	0	3	0	2	0	4	2	2	2	2	4	0	2	0	4	2	2	2	2	4	0	5	4	3	0	1	2	1	4	1	3	2	1	4	5	0	5	2	5
					$\\
					$2	2	2	2	4	0	2	0	4	2	2	2	2	4	0	2	0	4	0	1	2	1	4	1	2	1	0	4	5	0	5	2	5	0	5	4	2	2	2	2	4	0	2	0	4	5	5	5	5	1	3	5	3	1	2	3	4	3	0	3	4	3	2	3	4	5	4	1	4	5	4	3
					$\\
					$2	4	0	2	0	4	2	2	2	2	4	0	2	0	4	2	2	2	0	3	0	1	0	5	2	3	4	4	1	4	5	4	3	0	1	2	2	4	0	2	0	4	2	2	2	5	1	3	5	3	1	5	5	5	2	5	2	3	2	1	4	5	0	3	0	3	4	3	2	5	0	1
					$\\
					$2	0	4	2	2	2	2	4	0	2	0	4	2	2	2	2	4	0	0	5	4	1	2	3	2	5	2	4	3	2	5	0	1	0	3	0	2	0	4	2	2	2	2	4	0	5	3	1	5	5	5	5	1	3	2	1	0	3	4	5	4	1	4	3	2	1	4	5	0	5	2	5
					$\\
					$2	2	2	2	4	0	2	0	4	5	5	5	5	1	3	5	3	1	0	1	2	1	4	1	2	1	0	1	2	3	2	5	2	3	2	1	2	2	2	2	4	0	2	0	4	2	2	2	2	4	0	2	0	4	2	3	4	3	0	3	4	3	2	0	1	2	1	4	1	2	1	0
					$\\
					$2	4	0	2	0	4	2	2	2	5	1	3	5	3	1	5	5	5	0	3	0	1	0	5	2	3	4	1	4	1	2	1	0	3	4	5	2	4	0	2	0	4	2	2	2	2	4	0	2	0	4	2	2	2	2	5	2	3	2	1	4	5	0	0	3	0	1	0	5	2	3	4
					$\\
					$2	0	4	2	2	2	2	4	0	5	3	1	5	5	5	5	1	3	0	5	4	1	2	3	2	5	2	1	0	5	2	3	4	3	0	3	2	0	4	2	2	2	2	4	0	2	0	4	2	2	2	2	4	0	2	1	0	3	4	5	4	1	4	0	5	4	1	2	3	2	5	2
					$
			\end{tabular}}\\ \hline
			$C_1$  & \multicolumn{6}{l|}{\begin{tabular}[c]{@{}l@{}}
					$2	2	2	4	0	2	0	4	2	2	2	2	4	0	2	0	4	2	3	4	5	0	3	0	3	2	1	1	2	3	4	1	4	1	0	5	2	2	2	4	0	2	0	4	2	5	5	5	1	3	5	3	1	5	5	0	1	2	5	2	5	4	3	0	1	2	3	0	3	0	5	4
					$\\
					$2	4	0	4	2	0	0	0	0	2	4	0	4	2	0	0	0	0	3	0	3	0	5	4	3	4	5	1	4	1	4	3	2	1	2	3	2	4	0	4	2	0	0	0	0	5	1	3	1	5	3	3	3	3	5	2	5	2	1	0	5	0	1	0	3	0	3	2	1	0	1	2
					$\\	
					$2	0	4	4	4	4	0	2	4	2	0	4	4	4	4	0	2	4	3	2	1	0	1	2	3	0	3	1	0	5	4	5	0	1	4	1	2	0	4	4	4	4	0	2	4	5	3	1	1	1	1	3	5	1	5	4	3	2	3	4	5	2	5	0	5	4	3	4	5	0	3	0
					$\\
					$2	2	2	4	0	2	0	4	2	5	5	5	1	3	5	3	1	5	3	4	5	0	3	0	3	2	1	4	5	0	1	4	1	4	3	2	2	2	2	4	0	2	0	4	2	2	2	2	4	0	2	0	4	2	5	0	1	2	5	2	5	4	3	3	4	5	0	3	0	3	2	1
					$\\
					$2	4	0	4	2	0	0	0	0	5	1	3	1	5	3	3	3	3	3	0	3	0	5	4	3	4	5	4	1	4	1	0	5	4	5	0	2	4	0	4	2	0	0	0	0	2	4	0	4	2	0	0	0	0	5	2	5	2	1	0	5	0	1	3	0	3	0	5	4	3	4	5
					$\\
					$2	0	4	4	4	4	0	2	4	5	3	1	1	1	1	3	5	1	3	2	1	0	1	2	3	0	3	4	3	2	1	2	3	4	1	4	2	0	4	4	4	4	0	2	4	2	0	4	4	4	4	0	2	4	5	4	3	2	3	4	5	2	5	3	2	1	0	1	2	3	0	3
					$\\
					$2	2	2	4	0	2	0	4	2	2	2	2	4	0	2	0	4	2	0	1	2	3	0	3	0	5	4	4	5	0	1	4	1	4	3	2	2	2	2	4	0	2	0	4	2	5	5	5	1	3	5	3	1	5	2	3	4	5	2	5	2	1	0	3	4	5	0	3	0	3	2	1
					$\\
					$2	4	0	4	2	0	0	0	0	2	4	0	4	2	0	0	0	0	0	3	0	3	2	1	0	1	2	4	1	4	1	0	5	4	5	0	2	4	0	4	2	0	0	0	0	5	1	3	1	5	3	3	3	3	2	5	2	5	4	3	2	3	4	3	0	3	0	5	4	3	4	5
					$\\
					$2	0	4	4	4	4	0	2	4	2	0	4	4	4	4	0	2	4	0	5	4	3	4	5	0	3	0	4	3	2	1	2	3	4	1	4	2	0	4	4	4	4	0	2	4	5	3	1	1	1	1	3	5	1	2	1	0	5	0	1	2	5	2	3	2	1	0	1	2	3	0	3
					$\\
					$2	2	2	4	0	2	0	4	2	5	5	5	1	3	5	3	1	5	0	1	2	3	0	3	0	5	4	1	2	3	4	1	4	1	0	5	2	2	2	4	0	2	0	4	2	2	2	2	4	0	2	0	4	2	2	3	4	5	2	5	2	1	0	0	1	2	3	0	3	0	5	4
					$\\
					$2	4	0	4	2	0	0	0	0	5	1	3	1	5	3	3	3	3	0	3	0	3	2	1	0	1	2	1	4	1	4	3	2	1	2	3	2	4	0	4	2	0	0	0	0	2	4	0	4	2	0	0	0	0	2	5	2	5	4	3	2	3	4	0	3	0	3	2	1	0	1	2
					$\\
					$2	0	4	4	4	4	0	2	4	5	3	1	1	1	1	3	5	1	0	5	4	3	4	5	0	3	0	1	0	5	4	5	0	1	4	1	2	0	4	4	4	4	0	2	4	2	0	4	4	4	4	0	2	4	2	1	0	5	0	1	2	5	2	0	5	4	3	4	5	0	3	0
					$
			\end{tabular}}\\ \hline	
			$C_2$  & \multicolumn{6}{l|}{\begin{tabular}[c]{@{}l@{}}
					$222024420222024420345252105123030543222024420555351153501414321012525432$\\
					$240042444240042444303210123141054501240042444513315111525432345030543450$\\
					$204000402204000402321234141105012525204000402531333135543450303054501414$\\
					$222024420555351153345252105450303210222024420222024420501414321345252105$\\
					$240042444513315111303210123414321234240042444240042444525432345303210123$\\
					$204000402531333135321234141432345252204000402204000402543450303321234141$\\
					$222024420222024420012525432450303210222024420555351153234141054345252105$\\
					$240042444240042444030543450414321234240042444513315111252105012303210123$\\
					$204000402204000402054501414432345252204000402531333135210123030321234141$\\
					$222024420555351153012525432123030543222024420222024420234141054012525432$\\
					$240042444513315111030543450141054501240042444240042444252105012030543450$\\
					$204000402531333135054501414105012525204000402204000402210123030054501414$\\		
			\end{tabular}}\\ \hline	
	\end{tabular}}
\end{table}	
\begin{table}[H]
	\caption{Codes $C_3$, $C_4$, and $C_5$ over $\mbf{Z}_6$}\label{ankita201}
	\resizebox{\columnwidth}{!}{	\begin{tabular}{|l|llllll|}
			\hline
			$C_3$  & \multicolumn{6}{l|}{\begin{tabular}[c]{@{}l@{}}
					$222240204222240204345414543123252321555513531222240204234303432345414543$\\
					$240204222240204222303432501141210345513531555240204222252321450303432501$\\
					$204222240204222240321450525105234303531555513204222240210345414321450525$\\
					$222240204555513531345414543450525054555513531555513531234303432012141210$\\
					$240204222513531555303432501414543012513531555513531555252321450030105234$\\
					$204222240531555513321450525432501030531555513531555513210345414054123252$\\
					$222240204222240204012141210450525054555513531222240204501030105012141210$\\
					$240204222240204222030105234414543012513531555240204222525054123030105234$\\
					$204222240204222240054123252432501030531555513204222240543012141054123252$\\
					$222240204555513531012141210123252321555513531555513531501030105345414543$\\
					$240204222513531555030105234141210345513531555513531555525054123303432501$\\
					$204222240531555513054123252105234303531555513531555513543012141321450525$\\
			\end{tabular}}\\ \hline
			$C_4$  & \multicolumn{6}{l|}{\begin{tabular}[c]{@{}l@{}}		
					$222402042222402042345030321123414105555135315222402042234525210345030321$\\
					$240420000240420000303054345141432123513153333240420000252543234303054345$\\
					$204444024204444024321012303105450141531111351204444024210501252321012303$\\
					$222402042555135315345030321450141432555135315555135315234525210012303054$\\
					$240420000513153333303054345414105450513153333513153333252543234030321012$\\
					$204444024531111351321012303432123414531111351531111351210501252054345030$\\
					$222402042222402042012303054450141432555135315222402042501252543012303054$\\
					$240420000240420000030321012414105450513153333240420000525210501030321012$\\
					$204444024204444024054345030432123414531111351204444024543234525054345030$\\
					$222402042555135315012303054123414105555135315555135315501252543345030321$\\
					$240420000513153333030321012141432123513153333513153333525210501303054345$\\
					$204444024531111351054345030105450141531111351531111351543234525321012303$\\		
			\end{tabular}}\\ \hline		
			
			$C_5$  & \multicolumn{6}{l|}{\begin{tabular}[c]{@{}l@{}}	
					$222024420222024420345252105123030543555351153222024420234141054345252105$\\
					$240042444240042444303210123141054501513315111240042444252105012303210123$\\
					$204000402204000402321234141105012525531333135204000402210123030321234141$\\
					$222024420555351153345252105450303210555351153555351153234141054012525432$\\
					$240042444513315111303210123414321234513315111513315111252105012030543450$\\
					$204000402531333135321234141432345252531333135531333135210123030054501414$\\
					$222024420222024420012525432450303210555351153222024420501414321012525432$\\
					$240042444240042444030543450414321234513315111240042444525432345030543450$\\
					$204000402204000402054501414432345252531333135204000402543450303054501414$\\
					$222024420555351153012525432123030543555351153555351153501414321345252105$\\
					$240042444513315111030543450141054501513315111513315111525432345303210123$\\
					$204000402531333135054501414105012525531333135531333135543450303321234141$\\	
			\end{tabular}}\\ \hline		
	\end{tabular}}	
\end{table}	
\begin{table}[H]
	\caption{Codes $C_6$, $C_7$, and $C_8$ over $\mbf{Z}_6$}\label{ankita202}
	\resizebox{\columnwidth}{!}{\begin{tabular}{|l|llllll|}
			\hline
			$C_6$  & \multicolumn{6}{l|}{\begin{tabular}[c]{@{}l@{}}
					$222240204222240204012141210450525054222240204555513531234303432345414543$\\
					$240204222240204222030105234414543012240204222513531555252321450303432501$\\
					$204222240204222240054123252432501030204222240531555513210345414321450525$\\
					$222240204555513531012141210123252321222240204222240204234303432012141210$\\
					$240204222513531555030105234141210345240204222240204222252321450030105234$\\
					$204222240531555513054123252105234303204222240204222240210345414054123252$\\
					$222240204222240204345414543123252321222240204555513531501030105012141210$\\
					$240204222240204222303432501141210345240204222513531555525054123030105234$\\
					$204222240204222240321450525105234303204222240531555513543012141054123252$\\
					$222240204555513531345414543450525054222240204222240204501030105345414543$\\
					$240204222513531555303432501414543012240204222240204222525054123303432501$\\
					$204222240531555513321450525432501030204222240204222240543012141321450525$\\
			\end{tabular}}\\ \hline
			$C_7$  & \multicolumn{6}{l|}{\begin{tabular}[c]{@{}l@{}}
					$222402042222402042012303054450141432222402042555135315234525210345030321$\\
					$240420000240420000030321012414105450240420000513153333252543234303054345$\\
					$204444024204444024054345030432123414204444024531111351210501252321012303$\\
					$222402042555135315012303054123414105222402042222402042234525210012303054$\\
					$240420000513153333030321012141432123240420000240420000252543234030321012$\\
					$204444024531111351054345030105450141204444024204444024210501252054345030$\\
					$222402042222402042345030321123414105222402042555135315501252543012303054$\\
					$240420000240420000303054345141432123240420000513153333525210501030321012$\\
					$204444024204444024321012303105450141204444024531111351543234525054345030$\\
					$222402042555135315345030321450141432222402042222402042501252543345030321$\\
					$240420000513153333303054345414105450240420000240420000525210501303054345$\\
					$204444024531111351321012303432123414204444024204444024543234525321012303$\\
			\end{tabular}}\\ \hline
			$C_8$  & \multicolumn{6}{l|}{\begin{tabular}[c]{@{}l@{}}
					$222024420222024420012525432450303210222024420555351153234141054345252105$\\
					$240042444240042444030543450414321234240042444513315111252105012303210123$\\
					$204000402204000402054501414432345252204000402531333135210123030321234141$\\
					$222024420555351153012525432123030543222024420222024420234141054012525432$\\
					$240042444513315111030543450141054501240042444240042444252105012030543450$\\
					$204000402531333135054501414105012525204000402204000402210123030054501414$\\
					$222024420222024420345252105123030543222024420555351153501414321012525432$\\
					$240042444240042444303210123141054501240042444513315111525432345030543450$\\
					$204000402204000402321234141105012525204000402531333135543450303054501414$\\
					$222024420555351153345252105450303210222024420222024420501414321345252105$\\
					$240042444513315111303210123414321234240042444240042444525432345303210123$\\
					$204000402531333135321234141432345252204000402204000402543450303321234141$\\
			\end{tabular}}\\ \hline
			
	\end{tabular}}	
\end{table}

\begin{table}[H]
	\caption{Codes $C_9$, $C_{10}$, and $C_{11}$ over $\mbf{Z}_6$}\label{ankita203}
	\resizebox{\columnwidth}{!}{\begin{tabular}{|l|llllll|}
			\hline
			$C_9$  & \multicolumn{6}{l|}{\begin{tabular}[c]{@{}l@{}}
					$222240204222240204012141210450525054555513531222240204501030105012141210$\\
					$240204222240204222030105234414543012513531555240204222525054123030105234$\\
					$204222240204222240054123252432501030531555513204222240543012141054123252$\\
					$222240204555513531012141210123252321555513531555513531501030105345414543$\\
					$240204222513531555030105234141210345513531555513531555525054123303432501$\\
					$204222240531555513054123252105234303531555513531555513543012141321450525$\\
					$222240204222240204345414543123252321555513531222240204234303432345414543$\\
					$240204222240204222303432501141210345513531555240204222252321450303432501$\\
					$204222240204222240321450525105234303531555513204222240210345414321450525$\\
					$222240204555513531345414543450525054555513531555513531234303432012141210$\\
					$240204222513531555303432501414543012513531555513531555252321450030105234$\\
					$204222240531555513321450525432501030531555513531555513210345414054123252$\\			
			\end{tabular}}\\ \hline
			$C_{10}$  & \multicolumn{6}{l|}{\begin{tabular}[c]{@{}l@{}}
					$222402042222402042012303054450141432555135315222402042501252543012303054$\\
					$240420000240420000030321012414105450513153333240420000525210501030321012$\\
					$204444024204444024054345030432123414531111351204444024543234525054345030$\\
					$222402042555135315012303054123414105555135315555135315501252543345030321$\\
					$240420000513153333030321012141432123513153333513153333525210501303054345$\\
					$204444024531111351054345030105450141531111351531111351543234525321012303$\\
					$222402042222402042345030321123414105555135315222402042234525210345030321$\\
					$240420000240420000303054345141432123513153333240420000252543234303054345$\\
					$204444024204444024321012303105450141531111351204444024210501252321012303$\\
					$222402042555135315345030321450141432555135315555135315234525210012303054$\\
					$240420000513153333303054345414105450513153333513153333252543234030321012$\\
					$204444024531111351321012303432123414531111351531111351210501252054345030$\\
			\end{tabular}}\\ \hline
			$C_{11}$  & \multicolumn{6}{l|}{\begin{tabular}[c]{@{}l@{}}
					$222024420222024420012525432450303210555351153222024420501414321012525432$\\
					$240042444240042444030543450414321234513315111240042444525432345030543450$\\
					$204000402204000402054501414432345252531333135204000402543450303054501414$\\
					$222024420555351153012525432123030543555351153555351153501414321345252105$\\
					$240042444513315111030543450141054501513315111513315111525432345303210123$\\
					$204000402531333135054501414105012525531333135531333135543450303321234141$\\
					$222024420222024420345252105123030543555351153222024420234141054345252105$\\
					$240042444240042444303210123141054501513315111240042444252105012303210123$\\
					$204000402204000402321234141105012525531333135204000402210123030321234141$\\
					$222024420555351153345252105450303210555351153555351153234141054012525432$\\
					$240042444513315111303210123414321234513315111513315111252105012030543450$\\
					$204000402531333135321234141432345252531333135531333135210123030054501414$\\		
			\end{tabular}}\\ \hline			
	\end{tabular}}	
\end{table}	
\section{Set of Elements in $\mal{V}_{432}$}\label{Appendix:B}		
\begin{table}[ht]
\caption{The Set $\mal{V}_{432}$ for $p_1=2$, $p_2=3$, $m_1=4$, and $m_2=3$ }\label{3aryvec432}
\resizebox{\columnwidth}{!}{\begin{tabular}{|l|l|l|l|l|l|l|l|l|l|l|l|}
	\hline
	\begin{tabular}[c]{@{}l@{}}$0 0 0 0 0 0 0$\\ $0 0 0 0 0 0 1$\\ $0 0 0 0 0 0 2$\\ $0 0 0 0 0 1 0$\\ $0 0 0 0 0 1 1$\\ $0 0 0 0 0 1 2$\\ $0 0 0 0 0 2 0$\\ $0 0 0 0 0 2 1$\\ $0 0 0 0 0 2 2$\\ $0 0 0 0 1 0 0$\\ $0 0 0 0 1 0 1$\\ $0 0 0 0 1 0 2$\\ $0 0 0 0 1 1 0$\\ $0 0 0 0 1 1 1$\\ $0 0 0 0 1 1 2$\\ $0 0 0 0 1 2 0$\\ $0 0 0 0 1 2 1$\\ $0 0 0 0 1 2 2$\\ $0 0 0 0 2 0 0$\\ $0 0 0 0 2 0 1$\\ $0 0 0 0 2 0 2$\\ $0 0 0 0 2 1 0$\\ $0 0 0 0 2 1 1$\\ $0 0 0 0 2 1 2$\\ $0 0 0 0 2 2 0$\\ $0 0 0 0 2 2 1$\\ $0 0 0 0 2 2 2$\\ $0 0 0 1 0 0 0$\\ $0 0 0 1 0 0 1$\\ $0 0 0 1 0 0 2$\\ $0 0 0 1 0 1 0$\\ $0 0 0 1 0 1 1$\\ $0 0 0 1 0 1 2$\\ $0 0 0 1 0 2 0$\\ $0 0 0 1 0 2 1$\\ $0 0 0 1 0 2 2$\\ $0 0 0 1 1 0 0$\\ $0 0 0 1 1 0 1$\\ $0 0 0 1 1 0 2$\end{tabular} & \begin{tabular}[c]{@{}l@{}}
	$0 0 0 1 1 1 0$\\ $0 0 0 1 1 1 1$\\ $0 0 0 1 1 1 2$\\ $0 0 0 1 1 2 0$\\ $0 0 0 1 1 2 1$\\ $0 0 0 1 1 2 2$\\ $0 0 0 1 2 0 0$\\ $0 0 0 1 2 0 1$\\ $0 0 0 1 2 0 2$\\ $0 0 0 1 2 1 0$\\ $0 0 0 1 2 1 1$\\ $0 0 0 1 2 1 2$\\ $0 0 0 1 2 2 0$\\ $0 0 0 1 2 2 1$\\ $0 0 0 1 2 2 2$\\ $0 0 1 0 0 0 0$\\ $0 0 1 0 0 0 1$\\ $0 0 1 0 0 0 2$\\ $0 0 1 0 0 1 0$\\ $0 0 1 0 0 1 1$\\ $0 0 1 0 0 1 2$\\ $0 0 1 0 0 2 0$\\ $0 0 1 0 0 2 1$\\ $0 0 1 0 0 2 2$\\ $0 0 1 0 1 0 0$\\ $0 0 1 0 1 0 1$\\ $0 0 1 0 1 0 2$\\ $0 0 1 0 1 1 0$\\ $0 0 1 0 1 1 1$\\ $0 0 1 0 1 1 2$\\ $0 0 1 0 1 2 0$\\ $0 0 1 0 1 2 1$\\ $0 0 1 0 1 2 2$\\ $0 0 1 0 2 0 0$\\ $0 0 1 0 2 0 1$\\ $0 0 1 0 2 0 2$\\ $0 0 1 0 2 1 0$\\ $0 0 1 0 2 1 1$\\ $0 0 1 0 2 1 2$\end{tabular} & \begin{tabular}[c]{@{}l@{}}$0 0 1 0 2 2 0$\\ $0 0 1 0 2 2 1$\\ $0 0 1 0 2 2 2$\\ $0 0 1 1 0 0 0$\\ $0 0 1 1 0 0 1$\\ $0 0 1 1 0 0 2$\\ $0 0 1 1 0 1 0$\\ $0 0 1 1 0 1 1$\\ $0 0 1 1 0 1 2$\\ $0 0 1 1 0 2 0$\\ $0 0 1 1 0 2 1$\\ $0 0 1 1 0 2 2$\\ $0 0 1 1 1 0 0$\\ $0 0 1 1 1 0 1$\\ $0 0 1 1 1 0 2$\\ $0 0 1 1 1 1 0$\\ $0 0 1 1 1 1 1$\\ $0 0 1 1 1 1 2$\\ $0 0 1 1 1 2 0$\\ $0 0 1 1 1 2 1$\\ $0 0 1 1 1 2 2$\\ $0 0 1 1 2 0 0$\\ $0 0 1 1 2 0 1$\\ $0 0 1 1 2 0 2$\\ $0 0 1 1 2 1 0$\\ $0 0 1 1 2 1 1$\\ $0 0 1 1 2 1 2$\\ $0 0 1 1 2 2 0$\\ $0 0 1 1 2 2 1$\\ $0 0 1 1 2 2 2$\\ $1 0 0 0 0 0 0$\\ $1 0 0 0 0 0 1$\\ $1 0 0 0 0 0 2$\\ $1 0 0 0 0 1 0$\\ $1 0 0 0 0 1 1$\\ $1 0 0 0 0 1 2$\\ $1 0 0 0 0 2 0$\\ $1 0 0 0 0 2 1$\\ $1 0 0 0 0 2 2$\end{tabular} & \begin{tabular}[c]{@{}l@{}}
	$1 0 0 0 1 0 0$\\ $1 0 0 0 1 0 1$\\ $1 0 0 0 1 0 2$\\ $1 0 0 0 1 1 0$\\ $1 0 0 0 1 1 1$\\ $1 0 0 0 1 1 2$\\ $1 0 0 0 1 2 0$\\ $1 0 0 0 1 2 1$\\ $1 0 0 0 1 2 2$\\ $1 0 0 0 2 0 0$\\ $1 0 0 0 2 0 1$\\ $1 0 0 0 2 0 2$\\ $1 0 0 0 2 1 0$\\ $1 0 0 0 2 1 1$\\ $1 0 0 0 2 1 2$\\ $1 0 0 0 2 2 0$\\ $1 0 0 0 2 2 1$\\ $1 0 0 0 2 2 2$\\ $1 0 0 1 0 0 0$\\ $1 0 0 1 0 0 1$\\ $1 0 0 1 0 0 2$\\ $1 0 0 1 0 1 0$\\ $1 0 0 1 0 1 1$\\ $1 0 0 1 0 1 2$\\ $1 0 0 1 0 2 0$\\ $1 0 0 1 0 2 1$\\ $1 0 0 1 0 2 2$\\ $1 0 0 1 1 0 0$\\ $1 0 0 1 1 0 1$\\ $1 0 0 1 1 0 2$\\ $1 0 0 1 1 1 0$\\ $1 0 0 1 1 1 1$\\ $1 0 0 1 1 1 2$\\ $1 0 0 1 1 2 0$\\ $1 0 0 1 1 2 1$\\ $1 0 0 1 1 2 2$\\ $1 0 0 1 2 0 0$\\ $1 0 0 1 2 0 1$\\ $1 0 0 1 2 0 2$\end{tabular} & \begin{tabular}[c]{@{}l@{}}$1 0 0 1 2 1 0$\\ $1 0 0 1 2 1 1$\\ $1 0 0 1 2 1 2$\\ $1 0 0 1 2 2 0$\\ $1 0 0 1 2 2 1$\\ $1 0 0 1 2 2 2$\\ $1 0 1 0 0 0 0$\\ $1 0 1 0 0 0 1$\\$ 1 0 1 0 0 0 2$\\$ 1 0 1 0 0 1 0$\\ $1 0 1 0 0 1 1$\\$ 1 0 1 0 0 1 2$\\$ 1 0 1 0 0 2 0$\\$ 1 0 1 0 0 2 1$\\$ 1 0 1 0 0 2 2$\\$ 1 0 1 0 1 0 0$\\$ 1 0 1 0 1 0 1$\\$ 1 0 1 0 1 0 2$\\$ 1 0 1 0 1 1 0$\\$ 1 0 1 0 1 1 1$\\$ 1 0 1 0 1 1 2$\\$ 1 0 1 0 1 2 0$\\$ 1 0 1 0 1 2 1$\\$ 1 0 1 0 1 2 2$\\$ 1 0 1 0 2 0 0$\\$ 1 0 1 0 2 0 1$\\$ 1 0 1 0 2 0 2$\\$ 1 0 1 0 2 1 0$\\$ 1 0 1 0 2 1 1$\\$ 1 0 1 0 2 1 2$\\$ 1 0 1 0 2 2 0$\\$ 1 0 1 0 2 2 1$\\$ 1 0 1 0 2 2 2$\\$ 1 0 1 1 0 0 0$\\$ 1 0 1 1 0 0 1$\\$ 1 0 1 1 0 0 2$\\$ 1 0 1 1 0 1 0$\\$ 1 0 1 1 0 1 1$\\$ 1 0 1 1 0 1 2$
	\end{tabular} & \begin{tabular}[c]{@{}l@{}}$1 0 1 1 0 2 0$\\ $1 0 1 1 0 2 1$\\ $1 0 1 1 0 2 2$\\ $1 0 1 1 1 0 0$\\ $1 0 1 1 1 0 1$\\ $1 0 1 1 1 0 2$\\ $1 0 1 1 1 1 0$\\ $1 0 1 1 1 1 1$\\ $1 0 1 1 1 1 2$\\ $1 0 1 1 1 2 0$\\ $1 0 1 1 1 2 1$\\ $1 0 1 1 1 2 2$\\ $1 0 1 1 2 0 0$\\ $1 0 1 1 2 0 1$\\ $1 0 1 1 2 0 2$\\ $1 0 1 1 2 1 0$\\ $1 0 1 1 2 1 1$\\ $1 0 1 1 2 1 2$\\ $1 0 1 1 2 2 0$\\ $1 0 1 1 2 2 1$\\ $1 0 1 1 2 2 2$\\ $0 1 0 0 0 0 0$\\ $0 1 0 0 0 0 1$\\ $0 1 0 0 0 0 2$\\ $0 1 0 0 0 1 0$\\ $0 1 0 0 0 1 1$\\ $0 1 0 0 0 1 2$\\ $0 1 0 0 0 2 0$\\ $0 1 0 0 0 2 1$\\ $0 1 0 0 0 2 2$\\ $0 1 0 0 1 0 0$\\ $0 1 0 0 1 0 1$\\ $0 1 0 0 1 0 2$\\ $0 1 0 0 1 1 0$\\ $0 1 0 0 1 1 1$\\ $0 1 0 0 1 1 2$\\ $0 1 0 0 1 2 0$\\ $0 1 0 0 1 2 1$\\ $0 1 0 0 1 2 2$\end{tabular} & 
	\begin{tabular}[c]{@{}l@{}}
	$0 1 0 0 2 0 0$\\ $0 1 0 0 2 0 1$\\$ 0 1 0 0 2 0 2$\\$ 0 1 0 0 2 1 0$\\$ 0 1 0 0 2 1 1$\\ 
	$0 1 0 0 2 1 2$\\ $0 1 0 0 2 2 0$\\$0 1 0 0 2 2 1$\\$ 0 1 0 0 2 2 2$\\$ 0 1 0 1 0 0 0$\\ 
	$0 1 0 1 0 0 1$\\ $0 1 0 1 0 0 2$\\$0 1 0 1 0 1 0$\\$ 0 1 0 1 0 1 1$\\$ 0 1 0 1 0 1 2$\\ 
	$0 1 0 1 0 2 0$\\ $0 1 0 1 0 2 1$\\$ 0 1 0 1 0 2 2$\\$ 0 1 0 1 1 0 0$\\$ 0 1 0 1 1 0 1$\\ 
	$0 1 0 1 1 0 2$\\ $0 1 0 1 1 1 0$\\$ 0 1 0 1 1 1 1$\\$ 0 1 0 1 1 1 2$\\$ 0 1 0 1 1 2 0$\\ 
	$0 1 0 1 1 2 1$\\ $0 1 0 1 1 2 2$\\$ 0 1 0 1 2 0 0$\\$ 0 1 0 1 2 0 1$\\$ 0 1 0 1 2 0 2$\\ 
	$0 1 0 1 2 1 0$\\ $0 1 0 1 2 1 1$\\$ 0 1 0 1 2 1 2$\\$ 0 1 0 1 2 2 0$\\$ 0 1 0 1 2 2 1$\\$0 1 0 1 2 2 2$\\ $0 1 1 0 0 0 0$\\$ 0 1 1 0 0 0 1$\\$ 0 1 1 0 0 0 2$
	\end{tabular} & 
	\begin{tabular}[c]{@{}l@{}}
	$	0 1 1 0 0 1 0$\\$ 0 1 1 0 0 1 1$\\$ 0 1 1 0 0 1 2$\\$ 0 1 1 0 0 2 0$\\$ 0 1 1 0 0 2 1$\\$ 0 1 1 0 0 2 2$\\$ 0 1 1 0 1 0 0$\\$ 0 1 1 0 1 0 1$\\$ 0 1 1 0 1 0 2$\\$ 0 1 1 0 1 1 0$\\$ 0 1 1 0 1 1 1$\\$ 0 1 1 0 1 1 2$\\$ 0 1 1 0 1 2 0$\\$ 0 1 1 0 1 2 1$\\$ 0 1 1 0 1 2 2$\\$ 0 1 1 0 2 0 0$\\$ 0 1 1 0 2 0 1$\\$ 0 1 1 0 2 0 2$\\$ 0 1 1 0 2 1 0$\\$ 0 1 1 0 2 1 1$\\$ 0 1 1 0 2 1 2$\\$ 0 1 1 0 2 2 0$\\$ 0 1 1 0 2 2 1$\\$ 0 1 1 0 2 2 2$\\$ 0 1 1 1 0 0 0$\\$ 0 1 1 1 0 0 1$\\$ 0 1 1 1 0 0 2$\\$ 0 1 1 1 0 1 0$\\$ 0 1 1 1 0 1 1$\\$ 0 1 1 1 0 1 2$\\$ 0 1 1 1 0 2 0$\\$ 0 1 1 1 0 2 1$\\$ 0 1 1 1 0 2 2$\\$ 0 1 1 1 1 0 0$\\$ 0 1 1 1 1 0 1$\\$ 0 1 1 1 1 0 2$\\$ 0 1 1 1 1 1 0$\\$ 0 1 1 1 1 1 1$\\$ 0 1 1 1 1 1 2$
	\end{tabular}
	& \begin{tabular}[c]{@{}l@{}}
	$0 1 1 1 1 2 0$\\$ 0 1 1 1 1 2 1$\\$ 0 1 1 1 1 2 2$\\$ 0 1 1 1 2 0 0$\\$ 0 1 1 1 2 0 1$\\$ 0 1 1 1 2 0 2$\\$ 0 1 1 1 2 1 0$\\$ 0 1 1 1 2 1 1$\\$ 0 1 1 1 2 1 2$\\$ 0 1 1 1 2 2 0$\\$ 0 1 1 1 2 2 1$\\$ 0 1 1 1 2 2 2$\\$ 1 1 0 0 0 0 0$\\ $1 1 0 0 0 0 1$\\$ 1 1 0 0 0 0 2$\\$ 1 1 0 0 0 1 0$\\$ 1 1 0 0 0 1 1$\\$ 1 1 0 0 0 1 2$\\$ 1 1 0 0 0 2 0$\\$ 1 1 0 0 0 2 1$\\$ 1 1 0 0 0 2 2$\\$ 1 1 0 0 1 0 0$\\$ 1 1 0 0 1 0 1$\\$ 1 1 0 0 1 0 2$\\$ 1 1 0 0 1 1 0$\\$ 1 1 0 0 1 1 1$\\ $1 1 0 0 1 1 2$\\$ 1 1 0 0 1 2 0$\\$ 1 1 0 0 1 2 1$\\$ 1 1 0 0 1 2 2$\\$ 1 1 0 0 2 0 0$\\$ 1 1 0 0 2 0 1$\\$ 1 1 0 0 2 0 2$\\$ 1 1 0 0 2 1 0$\\$ 1 1 0 0 2 1 1$\\$ 1 1 0 0 2 1 2$\\$ 1 1 0 0 2 2 0$\\$ 1 1 0 0 2 2 1$\\$ 1 1 0 0 2 2 2$\\$	1 1 0 1 0 0 0$\\
	\end{tabular} & 
	\begin{tabular}[c]{@{}l@{}}
	$ 1 1 0 1 0 0 1$\\$ 1 1 0 1 0 0 2$\\$ 1 1 0 1 0 1 0$\\$ 1 1 0 1 0 1 1$\\$ 1 1 0 1 0 1 2$\\$ 1 1 0 1 0 2 0$\\$ 1 1 0 1 0 2 1$\\$ 1 1 0 1 0 2 2$\\$ 1 1 0 1 1 0 0$\\$ 1 1 0 1 1 0 1$\\$ 1 1 0 1 1 0 2$\\$ 1 1 0 1 1 1 0$\\$ 1 1 0 1 1 1 1$\\$ 1 1 0 1 1 1 2$\\$1 1 0 1 1 2 0$\\$ 1 1 0 1 1 2 1$\\$ 1 1 0 1 1 2 2$\\$ 1 1 0 1 2 0 0$\\$ 1 1 0 1 2 0 1$\\$ 1 1 0 1 2 0 2$\\$ 1 1 0 1 2 1 0$\\$ 1 1 0 1 2 1 1$\\$ 1 1 0 1 2 1 2$\\$ 1 1 0 1 2 2 0$\\$ 1 1 0 1 2 2 1$\\$ 1 1 0 1 2 2 2$\\$ 1 1 1 0 0 0 0$\\$ 1 1 1 0 0 0 1$\\$ 1 1 1 0 0 0 2$\\$ 1 1 1 0 0 1 0$\\$ 1 1 1 0 0 1 1$\\$ 1 1 1 0 0 1 2$\\$ 1 1 1 0 0 2 0$\\$ 1 1 1 0 0 2 1$\\$ 1 1 1 0 0 2 2$\\$ 1 1 1 0 1 0 0$\\$ 1 1 1 0 1 0 1$\\$ 1 1 1 0 1 0 2$\\$1 1 1 0 1 1 0$\\$ 1 1 1 0 1 1 1$
	\end{tabular} & 
	\begin{tabular}[c]{@{}l@{}}
	$ 1 1 1 0 1 1 2$\\$ 1 1 1 0 1 2 0$\\$ 1 1 1 0 1 2 1$\\$ 1 1 1 0 1 2 2$\\$ 1 1 1 0 2 0 0$\\$ 1 1 1 0 2 0 1$\\$ 1 1 1 0 2 0 2$\\$ 1 1 1 0 2 1 0$\\$ 1 1 1 0 2 1 1$\\$ 1 1 1 0 2 1 2$\\$ 1 1 1 0 2 2 0$\\$ 1 1 1 0 2 2 1$\\$ 1 1 1 0 2 2 2$\\$ 1 1 1 1 0 0 0$\\$ 1 1 1 1 0 0 1$\\ $1 1 1 1 0 0 2$\\$ 1 1 1 1 0 1 0$\\$ 1 1 1 1 0 1 1$\\$ 1 1 1 1 0 1 2$\\$ 1 1 1 1 0 2 0$\\$ 1 1 1 1 0 2 1$\\$ 1 1 1 1 0 2 2$\\$ 1 1 1 1 1 0 0$\\$ 1 1 1 1 1 0 1$\\$ 1 1 1 1 1 0 2$\\$ 1 1 1 1 1 1 0$\\$ 1 1 1 1 1 1 1$\\$ 1 1 1 1 1 1 2$\\$ 1 1 1 1 1 2 0$\\$ 1 1 1 1 1 2 1$\\$ 1 1 1 1 1 2 2$\\$ 1 1 1 1 2 0 0$\\$ 1 1 1 1 2 0 1$\\$ 1 1 1 1 2 0 2$\\$ 1 1 1 1 2 1 0$\\$ 1 1 1 1 2 1 1$\\$ 1 1 1 1 2 1 2$\\ $1 1 1 1 2 2 0$\\ $1 1 1 1 2 2 1 $\\ $1 1 1 1 2 2 2$
	\end{tabular} 
	\\ \hline
	\end{tabular}}
\end{table}
\bibliographystyle{IEEEtran}
\bibliography{CCC_SNC_CCC}

\begin{thebibliography}{10}
\providecommand{\url}[1]{#1}
\csname url@samestyle\endcsname
\providecommand{\newblock}{\relax}
\providecommand{\bibinfo}[2]{#2}
\providecommand{\BIBentrySTDinterwordspacing}{\spaceskip=0pt\relax}
\providecommand{\BIBentryALTinterwordstretchfactor}{4}
\providecommand{\BIBentryALTinterwordspacing}{\spaceskip=\fontdimen2\font plus
\BIBentryALTinterwordstretchfactor\fontdimen3\font minus
  \fontdimen4\font\relax}
\providecommand{\BIBforeignlanguage}[2]{{%
\expandafter\ifx\csname l@#1\endcsname\relax
\typeout{** WARNING: IEEEtran.bst: No hyphenation pattern has been}%
\typeout{** loaded for the language `#1'. Using the pattern for}%
\typeout{** the default language instead.}%
\else
\language=\csname l@#1\endcsname
\fi
#2}}
\providecommand{\BIBdecl}{\relax}
\BIBdecl

\bibitem{chinchong}
C.-C. Tseng and C.~Liu, ``Complementary sets of sequences,'' \emph{IEEE Trans.
  Inf. Theory}, vol.~18, no.~5, pp. 644--652, Sept. 1972.

\bibitem{pater2000}
K.~G. Paterson, ``Generalized {Reed-Muller} codes and power control in {OFDM}
  modulation,'' \emph{IEEE Trans. Inf. Theory}, vol.~46, no.~1, pp. 104--120,
  Jan. 2000.

\bibitem{liug}
Z.~Liu, Y.~Li, and Y.~L. Guan, ``New constructions of general {QAM} {Golay}
  complementary sequences,'' \emph{IEEE Trans. Inf. Theory}, vol.~59, no.~11,
  pp. 7684--7692, Nov. 2013.

\bibitem{Thesis_1949golay}
M.~J.~E. Golay, ``Multislit spectroscopy,'' \emph{Journal of the Optical
  Society of America}, vol.~39, no.~6, pp. 437--444, June 1949.

\bibitem{rati}
A.~Rathinakumar and A.~K. Chaturvedi, ``Complete mutually orthogonal {Golay}
  complementary sets from {Reed-Muller} codes,'' \emph{IEEE Trans. Inf.
  Theory}, vol.~54, no.~3, pp. 1339--1346, Mar. 2008.

\bibitem{hator}
N.~{Suehiro} and M.~{Hatori}, ``{N}-shift cross-orthogonal sequences,''
  \emph{IEEE Trans. Inf. Theory}, vol.~34, no.~1, pp. 143--146, 1988.

\bibitem{Davis1999}
J.~A. Davis and J.~Jedwab, ``Peak-to-mean power control in {OFDM}, {Golay}
  complementary sequences, and {Reed-Muller} codes,'' \emph{IEEE Trans. Inf.
  Theory}, vol.~45, no.~7, pp. 2397--2417, Nov. 1999.

\bibitem{chen2007next}
H.-H. Chen, \emph{The Next Generation {CDMA} Technologies}.\hskip 1em plus
  0.5em minus 0.4em\relax Wiley, 2007.

\bibitem{liumc}
Z.~Liu, Y.~L. Guan, and U.~Parampalli, ``New complete complementary codes for
  peak-to-mean power control in multi-carrier {CDMA},'' \emph{IEEE Trans.
  Commun.}, vol.~62, no.~3, pp. 1105--1113, Mar. 2014.

\bibitem{Wang2007}
S.~Wang and A.~Abdi, ``{MIMO ISI channel estimation using uncorrelated Golay
  complementary sets of polyphase sequences},'' \emph{IEEE Trans. Veh.
  Technol.}, vol.~56, no.~5, pp. 3024--3039, Sept. 2007.

\bibitem{Pezeshiki2008}
A.~Pezeshiki, A.~R. Calderbank, W.~Moran, and S.~D. Howard, ``{Doppler
  resilient Golay complementary waveforms},'' \emph{IEEE Trans. Inf. Theory},
  vol.~54, no.~9, pp. 4254--4266, Sept. 2008.

\bibitem{Tang2014}
J.~Tang, N.~Zhang, Z.~Ma, and B.~Tang, ``{Construction of Doppler resilient
  complete complementary code in MIMO radar},'' \emph{IEEE Trans. Signal
  Process.}, vol.~62, no.~18, pp. 4704--4712, Sept. 2014.

\bibitem{Thesis_2014Kojima}
T.~Kojima, T.~Tachikawa, A.~Oizumi, Y.~Yamaguchi, and U.~Parampalli, ``A
  disaster prevention broadcasting based on data hiding scheme using complete
  complementary codes,'' in \emph{Proc. International Symposium on Information
  Theory and its Applications (ISITA)}, Oct. 2014, pp. 45--49.

\bibitem{psktcom}
P.~{Sarkar}, S.~{Majhi}, and Z.~{Liu}, ``Optimal {$Z$} -complementary code set
  from generalized {Reed-Muller} codes,'' \emph{IEEE Trans. Commun.}, vol.~67,
  no.~3, pp. 1783--1796, Mar. 2019.

\bibitem{pa_pbf}
P.~Sarkar, S.~Majhi, and Z.~Liu, ``Pseudo-boolean functions for optimal
  z-complementary code sets with flexible lengths,'' \emph{IEEE Signal Process.
  Lett.}, vol.~28, pp. 1350--1354, 2021.

\bibitem{avikr_qccs}
A.~R. Adhikary, Y.~Feng, Z.~Zhou, and P.~Fan, ``Asymptotically optimal and
  near-optimal aperiodic quasi-complementary sequence sets based on florentine
  rectangles,'' \emph{IEEE Trans. Commun.}, vol.~70, no.~3, pp. 1475--1485,
  Mar. 2022.

\bibitem{zhu_qccs}
Z.~Zhou, F.~Liu, A.~R. Adhikary, and P.~Fan, ``A generalized construction of
  multiple complete complementary codes and asymptotically optimal aperiodic
  quasi-complementary sequence sets,'' \emph{IEEE Trans. Commun.}, vol.~68,
  no.~6, pp. 3564--3571, June 2020.

\bibitem{palash_qcss_tit_24}
P.~Sarkar, C.~Li, S.~Majhi, and Z.~Liu, ``New correlation bound and
  construction of quasi-complementary sequence sets,'' \emph{IEEE Trans. Inf.
  Theory}, 2024.

\bibitem{dfan}
{X. Deng} and {P. Fan}, ``Spreading sequence sets with zero correlation zone,''
  \emph{Electro. Lett.}, vol.~36, no.~11, pp. 993--994, 2000.

\bibitem{ltsu}
Y.~{Liu}, C.~{Chen}, and Y.~T. {Su}, ``New constructions of zero-correlation
  zone sequences,'' \emph{IEEE Trans. Inf. Theory}, vol.~59, no.~8, pp.
  4994--5007, 2013.

\bibitem{appus}
R.~{Appuswamy} and A.~K. {Chaturvedi}, ``A new framework for constructing
  mutually orthogonal complementary sets and {ZCZ} sequences,'' \emph{IEEE
  Trans. Inf. Theory}, vol.~52, no.~8, pp. 3817--3826, 2006.

\bibitem{Tang2010}
X.~Tang, P.~Fan, and J.~Lindner, ``{Multiple binary ZCZ sequence sets with good
  cross-correlation property based on complementary sequence sets},''
  \emph{IEEE Trans. Inf. Theory}, vol.~56, no.~8, pp. 4038--4045, Aug. 2010.

\bibitem{Liu_ITW2014}
Z.~Liu, Y.~L. Guan, and U.~Parampalli, ``A new construction of zero correlation
  zone sequences from generalized reed-muller codes,'' in \emph{Proc. 2014 IEEE
  Inf. Theory Workshop (ITW'2014)}.

\bibitem{golay1961}
M.~J.~E. Golay, ``Complementary series,'' \emph{IRE Trans. Inf. Theory},
  vol.~7, no.~2, pp. 82--87, Apr. 1961.

\bibitem{Budisin90a}
S.~Z. Budi\u{s}in, ``New complementary pairs of sequences,'' \emph{Electro.
  Lett.}, vol.~26, no.~13, pp. 881--883, June 1990.

\bibitem{Budisin90b}
------, ``New multilevel complementary pairs of sequences,'' \emph{Electro.
  Lett.}, vol.~26, no.~22, pp. 1861--1863, Oct. 1990.

\bibitem{Schmid2007}
K.~U. Schmidt, ``Complementary sets, generalized {Reed-Muller} codes, and power
  control for {OFDM},'' \emph{IEEE Trans. Inf. Theory}, vol.~53, no.~2, pp.
  808--814, Feb. 2007.

\bibitem{palcs}
P.~Sarkar, S.~Majhi, and Z.~Liu, ``A direct and generalized construction of
  polyphase complementary sets with low pmepr and high code-rate for ofdm
  system,'' \emph{IEEE Trans. Commun.}, vol.~68, no.~10, pp. 6245--6262, 2020.

\bibitem{chentit}
C.~{Chen}, ``Complementary sets of non-power-of-two length for peak-to-average
  power ratio reduction in {OFDM},'' \emph{IEEE Trans. Inf. Theory}, vol.~62,
  no.~12, pp. 7538--7545, Dec. 2016.

\bibitem{chencommlett}
C.~{Chen}, C.~{Wang}, and C.~{Chao}, ``Complete complementary codes and
  generalized {Reed-Muller} codes,'' \emph{IEEE Commun. Lett.}, vol.~12,
  no.~11, pp. 849--851, Nov. 2008.

\bibitem{swuc}
S.-W. Wu, C.-Y. Chen, and Z.~Liu, ``How to construct mutually orthogonal
  complementary sets with non-power-of-two lengths?'' \emph{IEEE Trans. Inf.
  Theory}, vol.~67, no.~6, pp. 3464--3472, 2021.

\bibitem{nwmocs1}
B.~Shen, Y.~Yang, Y.~Feng, and Z.~Zhou, ``A generalized construction of
  mutually orthogonal complementary sequence sets with non-power-of-two
  lengths,'' \emph{IEEE Trans. Commun.}, vol.~69, no.~7, pp. 4247--4253, 2021.

\bibitem{xiao2023new}
H.~Xiao and X.~Cao, ``New constructions of mutually orthogonal complementary
  sets and z-complementary code sets based on extended {Boolean} functions,''
  \emph{Cryptogr. Commun.}, 2023.

\bibitem{pku1}
P.~Kumar, S.~Majhi, and S.~Paul, ``A direct construction of golay complementary
  pairs and binary complete complementary codes of length non-power of two,''
  \emph{IEEE Trans. commun.}, vol.~71, no.~3, pp. 1352--1363, 2023.

\bibitem{Budisin_QAM}
S.~Budisin and P.~Spasojevi\'{c}, ``Paraunitary generation/correlation of {QAM}
  complementary sequence pairs,'' \emph{Proc. Cryptogr. Commun.}, vol.~6,
  no.~1, pp. 59--102, Oct. 2014.

\bibitem{Wang_SETA2016}
Z.~Wang, G.~Wu, and D.~Ma, ``A new method to construct {Golay} complementary
  set by paraunitary matrices and {Hadamard matrices},'' in \emph{Proc.
  Sequences and Their Applications (SETA)}, Sept. 2016, pp. 252--264.

\bibitem{sdas}
S.~Das, S.~Budi\v{s}in, S.~Majhi, Z.~Liu, and Y.~L. Guan, ``A multiplier-free
  generator for polyphase complete complementary codes,'' \emph{IEEE Trans.
  Signal Process.}, vol.~66, no.~5, pp. 1184--1196, Mar. 2018.

\bibitem{Sdas_lett}
S.~Das, S.~Majhi, and Z.~Liu, ``A novel class of complete complementary codes
  and their applications for {APU} matrices,'' \emph{IEEE Signal Process.
  Lett.}, vol.~25, no.~9, pp. 1300--1304, Sept. 2018.

\bibitem{shibu2}
S.~{Das}, S.~{Majhi}, S.~{Budišin}, and Z.~{Liu}, ``A new construction
  framework for polyphase complete complementary codes with various lengths,''
  \emph{IEEE Trans. Signal Process.}, vol.~67, no.~10, pp. 2639--2648, May
  2019.

\bibitem{wang2020new}
Z.~Wang, D.~Ma, G.~Gong, and E.~Xue, ``New construction of complementary
  sequence (or array) sets and complete complementary codes,'' \emph{IEEE
  Trans. Inf. Theory}, vol.~67, no.~7, pp. 4902--4928, 2021.

\bibitem{zhao2007survey}
Q.~Zhao and B.~M. Sadler, ``A survey of dynamic spectrum access,'' \emph{IEEE
  Signal Process. Mag.}, vol.~24, no.~3, pp. 79--89, 2007.

\bibitem{haykin2005cognitive}
S.~Haykin, ``Cognitive radio: brain-empowered wireless communications,''
  \emph{IEEE journal on selected areas in communications}, vol.~23, no.~2, pp.
  201--220, 2005.

\bibitem{haykin2006cognitive}
------, ``Cognitive radar: a way of the future,'' \emph{IEEE Signal Process.
  Mag.}, vol.~23, no.~1, pp. 30--40, 2006.

\bibitem{Shen2022construction}
B.~Shen, Y.~Yang, P.~Fan, and Z.~Zhou, ``Constructions of non-contiguous
  complementary sequence sets and their applications,'' \emph{IEEE Trans.
  Wireless Commun.}, vol.~21, no.~7, pp. 4871--4882, 2022.

\bibitem{aplphan_snc_aapl}
A.~Șahin and R.~Yang, ``An uplink control channel design with complementary
  sequences for unlicensed bands,'' \emph{IEEE Trans. Wireless Commun.},
  vol.~19, no.~10, pp. 6858--6870, 2020.

\bibitem{csahin2021generic}
A.~{\c{S}}ahin and R.~Yang, ``A generic complementary sequence construction and
  associated encoder/decoder design,'' \emph{IEEE Trans. Commun.}, vol.~69,
  no.~11, pp. 7691--7705, 2021.

\bibitem{zhou2020new}
Y.~Zhou, Y.~Yang, Z.~Zhou, K.~Anand, S.~Hu, and Y.~L. Guan, ``New complementary
  sets with low papr property under spectral null constraints,'' \emph{IEEE
  Trans. Inf. Theory}, vol.~66, no.~11, pp. 7022--7032, 2020.

\bibitem{ipanov2018radar}
R.~N. Ipanov, A.~I. Baskakov, N.~Olyunin, and M.-H. Ka, ``Radar signals with
  zacz based on pairs of d-code sequences and their compression algorithm,''
  \emph{IEEE Signal Process. Lett.}, vol.~25, no.~10, pp. 1560--1564, 2018.

\bibitem{li2022spectrally}
Y.~Li, L.~Tian, and Y.~Zeng, ``Spectrally-null-constrained zcz sequences for
  mimo-ofdm channel estimation over non-contiguous carriers,'' \emph{IEEE
  commun. Lett.}, vol.~27, no.~2, pp. 442--446, 2022.

\bibitem{shen2023}
B.~Shen, Y.~Yang, Z.~Zhou, and S.~Mesnager, ``Constructions of spectrally null
  constrained complete complementary codes via the graph of extended {Boolean}
  functions,'' \emph{IEEE Trans. Inf. Theory}, vol.~69, no.~9, pp. 6028--6039,
  2023.

\bibitem{sarkar2021multivariable}
P.~Sarkar, Z.~Liu, and S.~Majhi, ``Multivariable function for new complete
  complementary codes with arbitrary lengths,'' 2021.

\bibitem{wangong}
Z.~Wang and G.~Gong, ``Constructions of complementary sequence sets and
  complete complementary codes by ideal two-level autocorrelation sequences and
  permutation polynomials,'' \emph{IEEE Trans. Inf. Theory}, vol.~69, no.~7,
  pp. 4723--4739, 2023.

\end{thebibliography}
\end{document}